\documentclass[aps,prb,preprint,onecolumn,amsmath,amssymb,superscriptaddress,eqsecnum,floatfix,scrartcl]{revtex4-1}
\pdfoutput=1      
\usepackage{amssymb}  
\usepackage{color,soul,colortbl} 
\usepackage{graphicx}  
\usepackage[normalem]{ulem}
\usepackage[tight]{subfigure}
\usepackage{mathrsfs}
\usepackage[pdftex,pdfusetitle,bookmarks=true,colorlinks=true,citecolor=blue,urlcolor=blue,linkcolor=magenta]{hyperref}
\usepackage{hypcap}
\usepackage{multirow}
\usepackage{float}
\usepackage{array}
\usepackage{bm}

\definecolor{Gray}{gray}{0.9}
\DeclareMathOperator{\Tr}{Tr}
\definecolor{forestgreen}{rgb}{0.13, 0.55, 0.13}

\begin{document}

	\title{Generalized Gibbs Ensemble Description of Real Space Entanglement Spectra of (2+1)-dimensional Chiral Topological Systems with $\rm{SU}(2)$ Symmetry}
	\author{Mark J.~Arildsen}
	\email{arildsen@physics.ucsb.edu}
	\author{Andreas W.~W.~Ludwig} 
	\affiliation{Department of Physics, University of California, Santa Barbara, California 93106}
	
	\date{\today}

	\begin{abstract} 
    	We provide a quantitative analysis of the splittings in low-lying numerical entanglement spectra (ES), at given momentum, of a number of quantum states that can be identified, based on ``Li-Haldane state-counting", as ground states of (2 + 1)-dimensional chiral topological phases with global \rm{SU}(2) symmetry. The ability to account for numerical ES splittings solely within the context of conformal field theory (CFT) is an additional diagnostic of the underlying topological theory, of finer sensitivity than ``state-counting". We use the conformal boundary state description of the ES, which can be viewed as a quantum quench. In this language, the ES splittings arise from local conservation laws in the chiral CFT besides the energy, which we view as a Generalized Gibbs Ensemble (GGE). Global \rm{SU}(2) symmetry imposes strong constraints on the number of such conservation laws, so that only a small number of parameters can be responsible for the splittings. We work out these conservation laws for chiral \rm{SU}(2) Wess-Zumino-Witten CFTs at levels one and two, and for the latter we notably find that some of the conservation laws take the form of local integrals of operators of fractional dimension, as proposed by Cardy for quantum quenches. We analyze numerical ES from systems with \rm{SU}(2) symmetry including chiral spin-liquid ground states of local 2D Hamiltonians and two chiral Projected Entangled Pair States (PEPS) tensor networks, which exhibit the ``state-counting" of the \rm{SU}(2)-level-one and -level-two theories. We find that the low-lying ES splittings can be well understood by the lowest of our conservation laws, and we demonstrate the importance of accounting for the fractional conservation laws at level two. Thus the states we consider, including the PEPS, appear chiral also under our more sensitive diagnostic.
	\end{abstract}
	
	\maketitle   
		
	\tableofcontents
		
	\newpage

	\section{Introduction}
	\label{sec:introduction}
		
		Topological states of matter exhibit a new kind of robust ``nonlocal'' order that is of great interest to condensed matter physics.\cite{Wen1991,Kitaev2003,Wen2004,Kitaev2006,Nayak2008}
		Some types of topological states in two spatial dimensions, including fractional quantum Hall states, lack time-reversal symmetry and are known as chiral topological states, which possess gapless topologically-protected edge states at their boundaries with universal finite-size spectra governed by (1+1)D chiral conformal field theories (CFTs) particular to the specific bulk topological order.\cite{Halperin1982,Witten1989,Wen1990,Nayak2008} 
		The searches for physical Hamiltonians that give rise to two-dimensional quantum states of this type, as well as investigations of the states themselves, make up a very active area of research.
		
		A useful method to help identify these states in numerical simulations is the entanglement spectrum (ES). Bipartitioning the Hilbert space into two disjoint halves $A$ and $\bar{A}$, we can compute the reduced density matrix of $A$ as $\rho_A = \Tr_{\bar{A}}~\rho$, where $\rho$ is the (global, pure-state) density matrix of the topological quantum state, and $\Tr_{\bar{A}}$ indicates that we trace out the degrees of freedom associated to $\bar{A}$.
		Then we can define the entanglement Hamiltonian as $H_{\text{entanglement}} = -\log \rho_A$. Its spectrum will be the ES. If the bipartition corresponds to degrees of freedom in two regions of real space, we call this spectrum the real space entanglement spectrum (RSES), and we refer to the division (interface) between the two spatial regions $A$ and $\bar{A}$ as the entanglement cut. 
		The crucial realization by Li and Haldane was that, for chiral topological systems, the 
		low-energy 
		states of the theory on a physical edge in real space, known to be described by a chiral CFT, are in a one-to-one correspondence with the low-lying eigenstates of the entanglement Hamiltonian computed across the entanglement cut placed at the same location as that physical edge.\cite{Li2008,Qi2012,Peschel2011,Chandran2011,Dubail2012,Swingle2012}.
		
\begin{figure}
    	    \centering
    	    \includegraphics{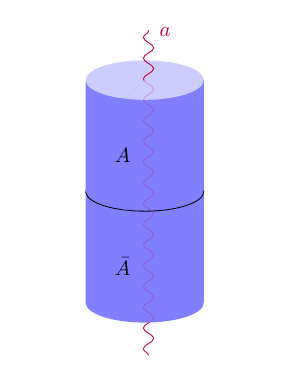}
            \caption{
                An infinite cylinder is bipartitioned into two sections $A$ and $\bar{A}$ by a circumferential (virtual) entanglement cut (in black). If the cylinder consists of a chiral topological bulk state, then by the result of Li and Haldane, the low-lying eigenstates of the entanglement Hamiltonian computed across the depicted entanglement cut are in a one-to-one correspondence with the low-energy states of the theory on the physical edge in real space that would be present if a physical, separating cut was made along the entanglement cut. An anyon flux of type $a$ is shown threaded through the cylinder.
            }
	        \label{fig:cylinder}
	    \end{figure}

		This wonderful fact, judiciously applied, allows the ES to become a diagnostic for identifying the presence of chiral topological states in numerical simulations. Typically the entanglement spectrum is computed across the entanglement cut 
		(see Fig.~\ref{fig:cylinder}, e.g., for an entanglement cut that is a circle, which appears when the surface of a cylinder is cut into two parts), and for the low-(entanglement)-energy
		part of the spectrum, the number of states at each momentum (along the circular cut) and of each type is counted and compared to the corresponding number in the relevant CFT. This agreement is taken as evidence that the correct topological state (or something similar  to it) has been produced by the computation.\cite{Bauer2014,Hickey2016,Hackenbroich2018,Chen2018}
		
		Numerical results at finite size, however, typically show splitting of the energy levels of entanglement spectra at a given momentum (see, e.g., Refs.~\onlinecite{Zaletel2012,Dubail2012,Davenport2015,Bauer2014,Hickey2016,Hackenbroich2018,Chen2018,Huang2021}, among many others), energy levels that would be degenerate considering the corresponding CFT Hamiltonian alone. 
		We would like to gain a more complete understanding of the detailed characteristics of these splittings in the low-energy levels for real space entanglement spectra, as these may provide information (going beyond the state-counting alone) about the underlying \mbox{$(2+1)$}-dimensional topological field theory, which is directly reflected in the structure of the resulting CFT, as well as the performance of the numerical methods used. 
		In particular, regarding the former point, if the underlying $(2+1)$-dimensional bulk topological field theory has been correctly identified, it must be possible to account for the splittings in the ES entirely within the context of the resulting CFT, i.e.,~without recourse to any other principles. If this is not possible, then the nature of the bulk state was not correctly identified.
		
		Various papers\cite{Zaletel2012, Dubail2012, Davenport2015} have successfully endeavored to study splittings in the entanglement spectra for Laughlin (or Pfaffian) quantum Hall states in a number of cases by irrelevant and/or dispersive terms, or composite fermion descriptions.
		We focus our attention on systematically investigating the ability of CFT to describe and characterize these splittings in a variety of different numerically generated chiral topological states, including Projected Entangled Pair States (PEPS)---see below.
		We use the conformal boundary state description\cite{Qi2012,Cardy2016,ChoLudwigRyu-PRB2017,Cardy2017} to do this.
	    The specific chiral topological states we consider are those where the bulk, and consequently also the edge theory (and therefore the chiral CFT describing the entanglement spectrum), possess {\it global $\rm{SU}(2)$ symmetry}: in particular, those where the topological properties of the bulk are described by $\rm{SU}(2)$-level-$k$ ($\rm{SU}(2)_k$) Chern-Simons theory\cite{Witten1989}, and thus where a physical edge is described by a chiral $\rm{SU}(2)_k$ Wess-Zumino-Witten (WZW) CFT (reviewed in Sec.~\ref{sec:wzwreview}), as is the case for some types of topological state.\cite{Nayak2008} 
	    The Kalmeyer-Laughlin spin liquid\cite{Kalmeyer1987} is one example\cite{Bauer2014} of such a chiral topological state that hosts, at a physical edge, edge modes described by a chiral $\rm{SU}(2)_1$ WZW CFT, and the non-Abelian $\rm{SU}(2)_2$ chiral spin liquid has also been investigated
	    \footnote{See for instance Ref.~\onlinecite{Huang2021} for recent numerical work on real-space entanglement spectra of this chiral spin liquid, as well as Ref.~\onlinecite{Chen2018}, a PEPS study that we consider in further detail below. Both works present the real-space entanglement spectrum in the spin-1/2 sector of $\rm{SU}(2)_2$, which is of particular interest in our present paper.}. 
		In this work specifically, we consider chiral spin liquids described by 
		(chiral) $\rm{SU}(2)_1$ and $\rm{SU}(2)_2$ Chern-Simons theories, and indeed we account for special features appearing in the entanglement spectrum of the half-integer spin sector of the $\rm{SU}(2)_2$ theory, stressing in particular the importance of fractional conservation laws\cite{Cardy2016} in understanding that entanglement spectrum. As compared to the quantum Hall states mentioned above, global $\rm{SU}(2)$ symmetry turns out to strongly constrain the number of parameters that can be responsible for the splittings. We additionally consider the effect of a discrete symmetry, the composition of spatial reflection and time reversal ($\mathcal{RT}$), which further constrains the number of parameters responsible for the splittings in some cases.
		
		For chiral topological states, the splittings can be understood in terms of CFT, as outlined below, by considering certain conserved quantities in the chiral CFT describing the ES, which contribute to a Generalized Gibbs Ensemble (GGE) form for the reduced density matrix. Thus the entanglement Hamiltonian will incorporate not only the data of the CFT Hamiltonian itself, but also data on the locally conserved quantities of the theory of the CFT that obey the $\rm{SU}(2)$ symmetry, as well as relevant discrete symmetries. 
		These locally conserved quantities can be thought of as arising physically in the process\cite{Qi2012} of generating the reduced density matrix along the real-space entanglement cut and are a property of the underlying two-dimensional topological quantum state itself. 
		(See Sec.~\ref{sec:locconquant} below for a review of this process using the ``conformal boundary state" formulation.)
		Further, in Ref.~\onlinecite{Cardy2016}, Cardy describes, in the context of quantum quenches, ``semi-local" conserved quantities, integrals of operators with noninteger dimension, that should in general be present in the GGE. These are the aforementioned fractional conservation laws. In sectors with twisted boundary conditions (such as, in a cylindrical geometry, those arising from threading the cylinder with topological flux, as shown in Fig.~\ref{fig:cylinder}), these quantities take the form of integrals of local operators, commute with the Hamiltonian, and belong to the GGE along with other locally conserved quantities. We detail the contribution by these conserved quantities described by Cardy to the splittings in the half-integer spin sector of the chiral $\rm{SU}(2)_2$ theory (which can be viewed as the Ramond sector of a theory with $N=1$ supersymmetry). This discussion, of how to understand the entanglement spectrum splittings with conserved quantities from CFT, and which conserved quantities will contribute to the splittings, is found in Secs.~\ref{sec:locconquant} and \ref{sec:locconquant2}.
		
		In Sec.~\ref{sec:results}, we 
		fit this description of the ES (which by construction is capable of describing the entanglement spectrum of any such chiral topological state) to numerically computed entanglement spectrum data of a variety of different states, all of which were observed to obey Li-Haldane counting.
		What we find from these fits is that we can understand the numerical spectra very well using our approach, including the novel fractional dimension conserved quantities, which we show to be essential to the understanding of the entanglement spectra of chiral topological states described by $\rm{SU}(2)_2$ Chern-Simons theory.
		As mentioned above, this provides confirmation that the description of the bulk topological state by the corresponding topological field theory is correct for the systems considered, as the ES, {\it including the splittings}, can be described entirely in terms of quantities of the CFT associated with that bulk topological field theory.
		
		This approach also gives insight into the success of the various computational techniques used by the works considered here in capturing the chiral topological nature of the states in question. For instance, Refs.~\onlinecite{Bauer2014} and \onlinecite{Hickey2016}, whose work we discuss in Sec.~\ref{sec:results}, use Density Matrix Renormalization Group (DMRG)-based techniques, and we verify that they exhibit detailed splitting behavior in the entanglement spectrum that is consistent with the possibilities allowed by the construction of the entanglement Hamiltonian out of available terms in the GGE from the chiral CFT, thereby supporting the ability of these DMRG-based techniques to capture chiral topological behavior. 
		The other two numerical entanglement spectra discussed in Sec.~\ref{sec:results}, Refs.~\onlinecite{Hackenbroich2018} and \onlinecite{Chen2018}, are generated instead through the use of chiral projected entangled pair states (PEPS) tensor network techniques. There, too, our results support the ability of these (PEPS) techniques to capture chiral topological behavior.
		
		PEPS tensor network techniques are a powerful general tool and the subject of considerable interest in their own right.\cite{Cirac2021}
        Here, we consider in particular {\it chiral} interacting PEPS\cite{Poilblanc2016,Hackenbroich2018,Chen2018}. However, noninteracting fermionic chiral topological PEPS (that is, noninteracting fermionic PEPS possessing a nontrivial Chern number) are known to obey a ``no-go theorem", stating that they cannot be ground states of local Hamiltonians gapped in the bulk.\cite{Dubail2015,Wahl2013} Thus, if a noninteracting fermionic chiral topological PEPS is the ground state of a local parent Hamiltonian, that Hamiltonian must be gapless in the bulk. Whether this ``no-go theorem", or a variation thereof, generalizes to interacting chiral topological PEPS is, to date, an open question.
        Our work can provide insight into this topic. For example, we consider the entanglement spectrum of the spin-1/2 PEPS investigated in Ref.~\onlinecite{Hackenbroich2018} (shown, with our fit to the splittings, 
        in Fig.~\ref{fig:hackenbroichfit}, corresponding to those for an Abelian $\rm{SU}(2)_1$ chiral spin liquid). The observed finite size splittings of the entanglement spectrum can be very well understood by our approach, only using the information from the CFT. This analysis lends substantial support to the claim of chirality for the PEPS of Ref.~\onlinecite{Hackenbroich2018}, consistent with earlier work pointing toward chirality of this PEPS.
        \footnote{
        Ref.~\onlinecite{Haegeman2017} sees a gapless chiral entanglement spectrum (``dispersion law'')
        of the PEPS of Ref.~\onlinecite{Hackenbroich2018} in the thermodynamic limit, as depicted in their Figure 27.\cite{SchuchPriv}} 
        Our approach also bolsters the evidence for chirality in the spin-1 PEPS of Ref.~\onlinecite{Chen2018}, where we find splittings characteristic of the non-Abelian chiral $\rm{SU}(2)_2$ theory, consistent with the observed Li-Haldane counting. In Ref.~\onlinecite{Chen2018}, as well as in other interacting PEPS that appear to be chiral based on Li-Haldane counting (including the PEPS of Ref.~\onlinecite{Poilblanc2016}, which is similar to the PEPS found in Ref.~\onlinecite{Hackenbroich2018}, discussed above), it has been observed numerically that equal-time correlations of local operators in the PEPS quantum state appear to have long-range correlations.\cite{Poilblanc2015,Poilblanc2016,Poilblanc2017,Chen2018} 
        It is expected (see, e.g., Refs.~\onlinecite{Hastings2004,Hastings2010}) that this implies that local parent Hamiltonians of these PEPS would be gapless, which would be further evidence in support of an interacting ``no-go theorem". Therefore, by providing additional evidence for chiral topological behavior in interacting PEPS, our approach shows the ability to help determine whether such a ``no-go theorem" holds for interacting chiral topological PEPS. 
        
    \section{Structure of Chiral $\rm{SU}(2)_k$ Wess-Zumino-Witten CFTs}
    \label{sec:wzwreview}
		
		The CFTs we will encounter in the entanglement spectra we look at in this work are chiral $\rm{SU}(2)_k$ WZW CFTs.~\cite{Knizhnik1984} The Hilbert space of such a CFT consists of $k+1$ primary states and their descendants under the actions of elements of an affine $\rm{SU}(2)$ current algebra. 
		Each primary state and its (affine) descendants comprise a separate topological sector of the theory. Each descendant of a primary state has a particular (descendant) level associated to it based on the elements of the current algebra used to specify the state. The descendant states, ordered by increasing level above the primary state, form ``conformal towers" for each topological sector of the theory.
		
\begin{figure}[hbt]
			\centering
			\subfigure[]{\label{fig:baueretalplot_integer} \includegraphics[width=.50625\textwidth]{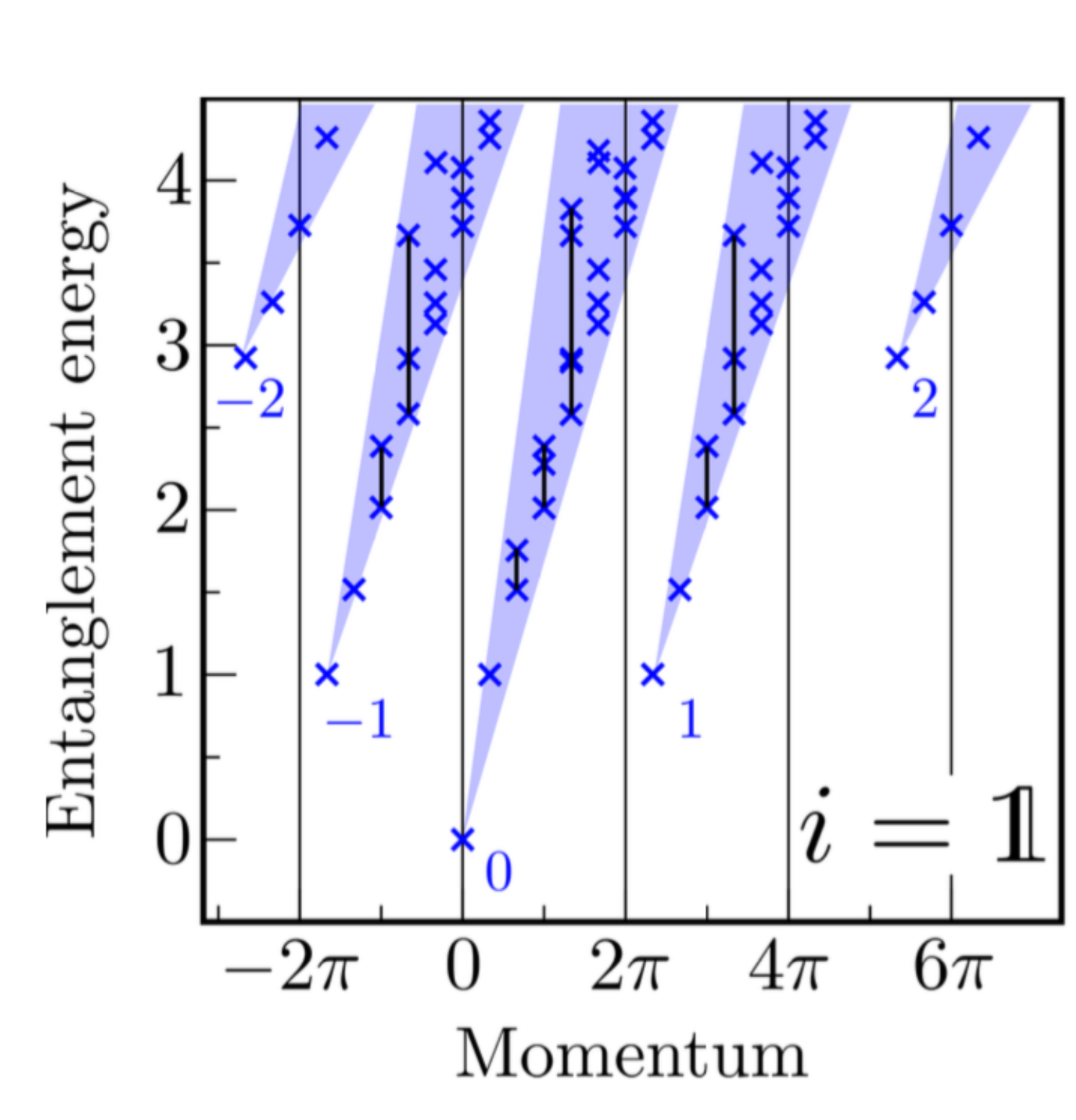}}
			\subfigure[]{\label{fig:baueretalplot_halfinteger} \includegraphics[width=.4\textwidth]{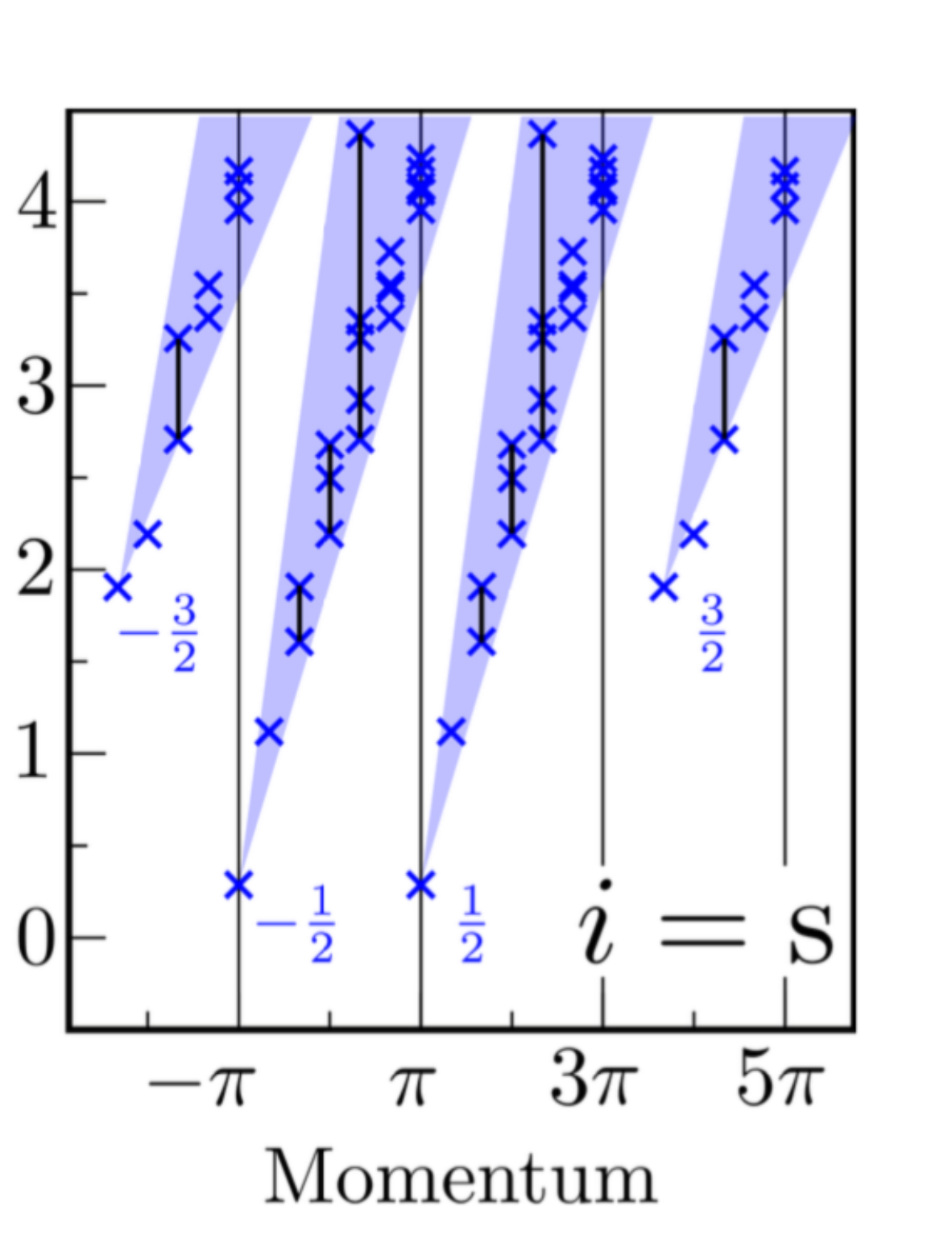}}   
			\caption{
				A reproduction of the entanglement spectrum of Figure 4\footnote{Our Fig.~\ref{fig:baueretalplot} is taken from Figure 4 of the arXiv version:  \href{https://arxiv.org/abs/1401.3017}{\tt arXiv:1401.3017 [cond-mat.str-el].}} from Bauer et al.~(Ref.~\protect\onlinecite{Bauer2014}). Two topological sectors are manifest, the integer in (a) and half-integer in (b), with conformal towers rooted, respectively, in the $|j=0\rangle$ and $|j=1/2\rangle$ primary states (the lowest markers at momenta 0 and $\pm\pi$, respectively). Extending above these primary states in each blue-shaded subtower of fixed $j^z$ are their descendant states, all of which display $\rm{SU}(2)$ symmetry across the indicated $j^z$ written at the bottom of the subtowers. 
				[In the Figure, these different subtowers are horizontally offset from each other by $2\pi$ to make them clearly visible (i.e., they actually have the same momenta).]
				Within each subtower, indicated by vertical black lines up to the fourth descendant level, we have the 1-1-2-3-5 counting of states that are degenerate in momentum and under the pure CFT Hamiltonian. This state-counting pattern is characteristic of the chiral $\rm{SU}(2)_1$ CFT. (See Appendix \ref{app:su2kintro} for more detail on the counting of states here.)
			}
			\label{fig:baueretalplot}
		\end{figure}

		We can denote each of the $k+1$ primary states of $\rm{SU}(2)_k$ by its global $\rm{SU}(2)$ spin quantum number $j$, with the notation $|j=i/2\rangle$ for some integer $i = 0,\ldots,k$. The state $|j\rangle$ takes the form of the $(2j+1)$-dimensional spin-$j$ representation of
		global $\rm{SU}(2)$ (with $j^z$ ranging from $-j$ to $+j$ in integral increments in the usual way). The descendant states in the conformal tower also obey the global $\rm{SU}(2)$ symmetry, and we can therefore view all of them as $\rm{SU}(2)$ multiplets of various dimensions. The pattern of multiplicities of these multiplets is characteristic of the theory. In the $\rm{SU}(2)_1$ theory, the two topological sectors, which correspond to $|j=0\rangle$ and $|j=1/2\rangle$ primary states (of the affine current algebra), are often referred to as the integer and half-integer sectors, respectively. This structure is illustrated for the chiral $\rm{SU}(2)_1$ WZW CFT in Fig.~\ref{fig:baueretalplot}.
		
		Our analysis will be done in a spatially circular geometry with fixed time coordinate in the CFT, in which we take the affine $\rm{SU}(2)$ currents $J^a(x)$ (where $a = 1,2,3$ is the index of the generator in the adjoint representation of $\rm{SU}(2)$) to be periodic in the spatial coordinate $x$ around the circle: 
		$J^a(x) = J^a(x+\ell)$, where $\ell$ is the circumference of the circle. 
		The energy-momentum tensor $T(x)$ of the CFT can be related to the $J^a(x)$ by expressing $T(x)$ in the Sugawara form\cite{Knizhnik1984}
		\begin{equation}
			\label{eq:sugawara}
			T(x) = \frac{1}{k+2}\sum_{a=1}^3 (J^a J^a)(x),
	    \end{equation}
		where we use brackets $()$ around $J^a J^a$ to indicate normal ordering. 
		We note that $T(x)$ will inherit the periodicity of $J^a(x)$: $T(x) = T(x+\ell)$. We can then write $T(x)$ in terms of modes $L_{-n}$ as
		\begin{equation}
			\label{eq:lmodeexpansion}
			T(x) =\left(\frac{2\pi}{\ell}\right)^2 \left(-\frac{c}{24}+\sum_{n=-\infty}^\infty L_{-n}e^{2\pi i n x/\ell}\right),
		\end{equation}
		where the central charge $c = \frac{3k}{k+2}$ for an $\rm{SU}(2)_k$ WZW CFT. 
		Thus we see that the Hamiltonian of this theory, which describes the left-moving states, will be
		\begin{equation}
			\label{eq:cfth}
			H_L = \frac{v}{2\pi}\int_0^{\ell} T(x) dx = 
			\frac{2\pi v}{\ell} \left( L_0 - \frac{c}{24}\right)
		\end{equation}
		where $v$ is a (nonuniversal) velocity. From now on we will suppose $v = 1$ for simplicity, since the conclusions we draw will not depend on an overall scale factor.
		The eigenvalue of $L_0$ (or $\overline{L}_0$) is the overall conformal dimension $h+K$ of the state on which it acts: the sum of the conformal weight $h$ of the WZW primary state at the base of the relevant conformal tower and, if the state in question is a descendant of a WZW primary state, the level $K$ of that descendant state above the corresponding primary state.
		
		The left-moving momentum $k_L$ acts uniformly on all the states $|h,K\rangle$ at descendant level $K$ above a primary state with conformal weight $h$ to give
		\begin{equation}
		\label{eq:hkstate}
		    k_L|h,K\rangle = \frac{2\pi}{\ell}L_0|h,K\rangle =\frac{2\pi}{\ell} \left(h + K\right)|h,K\rangle.
		\end{equation}We can also write down a corresponding set of quantities for right-moving states (in the Hilbert space of the chiral right-moving CFT): a Hamiltonian $H_R$ that is instead an integral of the conjugate energy-momentum tensor $\overline{T}(x)$, $\overline{L}_0$ the zero-mode of $\overline{T}(x)$, and the right-moving momentum $k_R$. 
		Therefore, all of the states in a given conformal tower at the same descendant level should be degenerate in both momentum, and energy, the eigenvalue of the CFT Hamiltonian $H_L$ (or $H_R$). The counts of these degeneracies are a characteristic marker of the CFT. These degeneracies and their description in terms of $\rm{SU}(2)$ multiplets are described in a more detailed discussion of the structure of the chiral $\rm{SU}(2)_k$ WZW CFT Hilbert space in Appendix \ref{app:su2kintro}.
	    
	    We can observe these degeneracies in numerical results. If this (1+1)D CFT on a circular spatial geometry is in fact the theory of one of the
	    two decoupled circular edges that arises from cutting, along its circumference, an infinite cylinder home to a (2+1)D bulk topological quantum field theory, then the left- and right-moving momenta $k_L$ and $k_R$, and Hamiltonians $H_L$ and $H_R$, govern the left- and right-moving states along the respective physical edges created by that cut of the cylinder. The aforementioned degeneracies of the momenta and energy spectra will also be present.
	    
	    The geometry of this setting is depicted in Fig.~\ref{fig:cylinder}. As discussed in the introduction, the Li-Haldane correspondence means that these degeneracies will then appear in the low-lying (entanglement) energy levels of the real space entanglement spectrum computed with an infinite cylinder bipartitioned with a (virtual) entanglement cut in the same location as the physical cut determining the edge. Crucially, though, when this computation is done, as in the numerical work, for cylinders 
	    of finite circumference for the entanglement cut, the degeneracies in question appear only for momentum, but are split for the entanglement energies. To illustrate this with a particular example, we can consider Fig.~\ref{fig:baueretalplot}, an entanglement spectrum exhibiting the characteristic pattern of degeneracies associated with the chiral $\rm{SU}(2)_1$ WZW CFT. The degeneracies are present in momentum, but split in the entanglement energy levels of the spectrum. We will consider this particular entanglement spectrum in greater detail in Sec.~\ref{sec:results}.
	
	    \vspace{1.5cm}
		
	\section{Explaining Entanglement Spectrum Splitting with Locally Conserved Quantities}
	\label{sec:locconquant}
	
		As we saw in Fig.~\ref{fig:baueretalplot}, in the real space entanglement spectra that we study, numerically computed at finite size, the states at the same descendant level are not all degenerate. Instead, we observe splitting of such states despite their common conformal dimension.
		Our goal is to gain control over the RSES, splittings and all. To do this, we need to re-examine the origin of the entanglement/edge correspondence that allows us to find the edge CFT data in the entanglement spectra in the first place. 
		
		Ref.~\onlinecite{Qi2012} (as briefly summarized here below) understands the entanglement/edge correspondence in a bipartitioned cylindrical geometry (Fig.~\ref{fig:cylinder}) from the point of view of a quantum quench, that for time $t>0$ decouples the two Hamiltonians $H_L$ and $H_R$ of the counter-propagating left- and right-moving physical edges produced by physically cutting the cylinder along the entanglement cut. 
		Then, the actual topological ground state $|{\bf G}\rangle$ of the system on the surface of the cylinder will serve as the $t = 0$ boundary condition of the quench, i.e.,~as the {\it initial} state. 
		That short-range entangled (and short-range correlated) initial state $|{\bf G}\rangle$ can be represented as the state resulting from a coupling between the two counter-propagating edges above (e.g., by relevant or marginally relevant operators), a process which generates a gap. 
		This initial state plays a dual role (see Refs.~\onlinecite{Qi2012,ChoLudwigRyu-PRB2017,Cardy2017}): (i)~on the one hand, as described above, it can be viewed as the ground state of a CFT gapped by coupling left- and right-movers, and (ii)~at the same time, it is a boundary condition on a gapless CFT describing the initial state of the quantum quench after Wick-rotation to imaginary time. 
		That boundary condition itself undergoes a renormalization group~(RG) flow, and is controlled at large scales by a scale-invariant boundary fixed point. When formulated in the language of a state in the Hilbert space of the bulk CFT, this fixed point is described by what is known as a conformally invariant boundary state $|{\bf G}_*\rangle$. That fixed point boundary state is itself not immediately useful for representing the physical state $|{\bf G}\rangle$ because, representing a fixed point, it turns out to be not normalizable.
		$|{\bf G}_*\rangle$ can be given a finite norm by slightly ``deforming" this state by slightly moving away from the fixed point in the RG sense by irrelevant boundary operators. The initial work by Cardy and Calabrese on quantum quenches\cite{Calabrese2006,Calabrese2007} made a particular (special) choice for an irrelevant boundary operator, which they chose to be the energy momentum tensor of the CFT at the boundary; thus, they represented  the physical state $|{\bf G}\rangle$ in terms of the fixed point conformal boundary state as
		\begin{equation}
			\label{eq:qig}
			|{\bf G}\rangle \propto e^{-\tau_0 (H_L + H_R)}|{\bf G}_*\rangle,
		\end{equation}
		where $H_L$ and $H_R$ are the CFT Hamiltonians for left- and right-movers, which are precisely the spatial integrals of the left- and right-moving energy-momentum tensors $T(x)$ and $\overline{T}(x)$, respectively. 
		The so-obtained state $|{\bf G}\rangle$ turns out to be a purification of a thermal density matrix (as reviewed below), where the ``extrapolation length'' $\tau_0$ plays the role of an inverse temperature, which sets the scale for the finite correlation length of all ``equal time'' correlation functions of local operators in this state.
		\footnote{That correlation length goes to zero as the fixed point is approached when $\tau_0\to 0$, in line with the expectation of the absence of a scale at a fixed point.}
		
		But Eq.~\eqref{eq:qig} is not the whole story, since the energy-momentum tensor is only one of the possible irrelevant operators that may be used to deform the boundary fixed point state $|{\bf G}_*\rangle$ in order to make it normalizable. 
		The integral of the energy-momentum tensor alone is not general enough to represent the actual ground state $|{\bf G}\rangle$ of the topological system on the surface of the cylinder in terms of the fixed point boundary state $|{\bf G}_*\rangle$. The procedure that allows for a representation of a general ground state on the cylinder surface will be described in Eq.~\eqref{eq:gboundarystate}.
		Before describing this, we need to also introduce the different topological flux sectors. We do that first using the state in Eq.~\eqref{eq:qig}, and generalize this subsequently.
		
		The topological sector of the theory on the surface of the cylinder depends on the topological flux through the cylinder (Fig.~\ref{fig:cylinder}). 
		This will select, following Ref.~\onlinecite{Qi2012}, a corresponding sector in the entanglement spectrum we are interested in describing. In any so-called rational CFT, such as those under consideration, all conformally invariant boundary states $|{\bf G}_*\rangle$ turn out to be finite linear combinations of states $|{\bf G}_{*,a} \rangle$, where $a$ denotes topological flux.
		\footnote{These states are maximally entangled states between the left- and the right-moving descendants of a primary state in the bulk CFT Hilbert space\cite{Ishibashi1989}.} 
		This leads to the obvious generalization of Eq.~\eqref{eq:qig}, namely, $|{\bf G}_a\rangle~\propto~e^{-\tau_0 (H_L + H_R)}|{\bf G}_{*,a}\rangle$, where $|{\bf G}_a\rangle$ describes the ground state of the system in the topological flux sector $a$.
		We are interested in the reduced density matrix $\rho_{L,a}$ of the ground state $|{\bf G}_a\rangle$ of the topological system in question, for which the half of the cylinder with the right-moving edge has been traced out. It turns out that under the assumption of Eq.~\eqref{eq:qig}, tracing out the half of the cylinder with the right-moving edge degrees of freedom yields\cite{Qi2012} 
		\begin{equation}
		    \label{LabelEq-rho-L-a}
			\rho_{L,a} = \Tr_R \left(|{\bf G}_a\rangle\langle {\bf G}_a|\right) \propto P_a e^{-4\tau_0 H_L} P_a,
		\end{equation}
		where $P_a$ is a projector onto the sector $a$. 
		That is, the reduced density matrix has a thermal form with effective inverse temperature $\beta = 4\tau_0$. Then we would have an entanglement Hamiltonian simply proportional to the Hamiltonian $H_L$ of the left-moving chiral edge CFT Eq.~\eqref{eq:cfth} projected onto the sector $a$. 
		As previously noted, though, under this Hamiltonian, all the states of a given conformal dimension are degenerate, so there must be additional terms in the entanglement Hamiltonian. Cardy has more recently argued\cite{Cardy2016} in the context of quantum quenches that one can generalize the assumption of Eq.~\eqref{eq:qig} to include also conservation laws which are integrals of irrelevant local boundary operators $\Phi_i$ and $\overline{\Phi}_i$ (left- and right-moving,
		respectively, in the bulk, with coinciding boundary limits
		\footnote{I.e., the additional boundary operators should be such that they are boundary limits of both purely chiral (holomorphic $\Phi_i(z)$ or anti-holomorphic $\bar{\Phi}_i(\bar{z})$) bulk operators.
		(This is the case, e.g., for the energy momentum tensor:
		$\Phi_{i=1}=T(x)$ and $\bar{\Phi}_{i=1}=\bar{T}(x)$ in the notation of Table \ref{table:modereps} of Appendix \ref{app:details}. These are equal, i.e.,~$T(x)=\bar{T}(x)$, at a conformally invariant boundary such as the one under consideration.)},
		$\Phi_i(x)=\overline{\Phi}_i(x)$, corresponding to the boundary operator) other than the energy-momentum tensor $T(x)$ and $\overline{T}(x)$, leading to a Generalized Gibbs Ensemble (GGE):
		\begin{equation}
				\label{eq:gboundarystate}
				|{\bf G}\rangle \propto e^{-(\beta/4)(H_L+H_R)}\prod_i e^{-(\beta_i/4)  \int \left[\Phi_i(x)+\overline{\Phi}_i(x) \right]dx}|{\bf G}_*\rangle
		\end{equation}
		That is to say, we now introduce a more fine-grained ansatz to guide $|{\bf G}\rangle$ closer toward the actual topological ground state on the surface of the cylinder by deforming $|{\bf G}_*\rangle$ with a more complete set of irrelevant boundary operators
		\footnote{The actual ground state on the surface of the cylinder of any such two-dimensional chiral topological system can be fully represented in this manner, upon inclusion of a sufficient number of operators $\Phi_i(x)$ and $\overline{\Phi}_i(x)$.}.
		When we calculate the reduced density matrix $\rho_{L,a}$ by tracing out the right-moving parts of the above expression, we obtain a GGE form for $\rho_{L,a}$,
		\begin{equation}
			\label{eq:realrhol}
			\rho_{L,a} = \Tr_R \left(|{\bf G}_a\rangle\langle {\bf G}_a|\right) 
			\propto P_a e^{-\beta H_L} \prod_i e^{-\beta_i  \int \Phi_i(x) dx} P_a,
		\end{equation}
		where $\Phi_i(x)$ can be viewed as operators acting only on the left-moving boundary
		\footnote{As an aside, we remark that in the usual quantum quench problem one wishes to describe the expectation value of a product of a finite number of local bulk operators of the CFT, each of which consists of a left- and a right-moving part, after waiting a sufficiently long (real) time. Such information is contained in the reduced density matrix for a bipartition of position space into a compact spatial interval $A$ containing the locations of the finite number of operators, and its complement $\bar{A}$. Under (real) time-evolution, such a density matrix relaxes\cite{Cardy2016,WenRyuLudwigJStatMech2018} into the thermal density matrix or GGE density matrix on the interval $A$. 
		The bipartition in the situation we consider here, in contrast, is that between left- and right-moving degrees of freedom (in contrast to that between the two spatial regions $A$ and $\bar{A}$ above), and the reduced density matrix in our Eqs.~\eqref{LabelEq-rho-L-a} and \eqref{eq:realrhol} is obtained by performing a trace over the right-moving degrees of freedom. This density matrix reflects the left-right entanglement of the boundary state $|{\bf G}_a\rangle$.}.
		This clarifies the cause of the splittings: additional locally conserved quantities $H^{(i)} = \frac{1}{2\pi}\int_0^\ell  \Phi_i(x) dx$ in the theory, besides just the Hamiltonian $H_L$, are entering into the form of the reduced density matrices $\rho_{L,a}$. These may include integrals of such irrelevant operators as powers of the energy-momentum tensor $T(x)$, e.g., $\frac{1}{2\pi} \int dx (TT)(x)$, a hierarchy of integrals of motion that can be written down for a wide range of CFTs.\cite{Sasaki1988,Bazhanov1996} 
		Integrals of other available local operators of even integer conformal dimension will come in as well, however, and for the case of the chiral $\rm{SU}(2)_2$ theory (in the spin-1/2 sector, where the boundary conditions allow them) we will also have integrals of operators with half-integer conformal dimension, a case discussed in Ref.~\onlinecite{Cardy2016} in the context of quantum quenches. 
		One example of these in our case is the integral $G_0 = \frac{1}{2\pi} \int dx G(x)$ of a chiral superconformal current operator $G(x)$, analogous to that included in the hierarchy of integrals of motion in an $N = 1$ superconformal field theory.\cite{Kulish2005} We will find that this same conservation law can be written down in the chiral $\rm{SU}(2)_2$ theory, since it is known\cite{Mussardo1988,FriedanQiuShenkerPhysLettB-SUSY-1985,DixonGinspargHarvey-SUSY-NPB1988} to possess $N = 1$ superconformal invariance.
		
		The entanglement Hamiltonian is then a sum involving all these conserved quantities, and we have
		\footnote{The constant below satisfies ``${\rm const.}=
	    \ln Z_a $'', where $1/Z_a$ is the factor by which the right hand side of \eqref{eq:realrhol} has to be multiplied to ensure the proper normalization of the density matrix on the left hand side. 
	    In general, this constant has a complicated dependence on the coefficients $\beta_i$. However, if in the limit of large system size $\ell$ the entanglement Hamiltonian is dominated by the Hamiltonian ${\tilde H}_L$ of the CFT [compare \eqref{eq:lincomb-SizeDependence}], then this constant is known\cite{KitaevPreskillPRL2006,Qi2012,AffleckLudwig1991} to take the form ``${\rm const.} =\alpha \ell - \gamma_a$'', where $\alpha$ is nonuniversal and $\gamma_a$ is related to the quantum dimension of the topological excitation $a$.}
		(within a given sector, so we can suppress the subscripts $a$)
		\begin{equation}
			\label{eq:lincomb}
			H_{\text{entanglement}} - {\rm const.} =
			-\log \rho_L - {\rm const.}=
			\sum_{i}^{\infty} \beta_i H^{(i)}= \beta H_L + \sum_{i \neq 1}^{\infty} \beta_i H^{(i)},
		\end{equation}
		where $\beta_i$ are parameters chosen to properly match the initial conditions of the quench,  and we take $\beta_1 = \beta$ and $H^{(1)} = H_L$ (the Hamiltonian of the CFT, which is just one of the conservation laws).
		Thus, we explain the splittings by some linear combination of the locally conserved quantities $H^{(i)}$.
		
		We end this section by highlighting an aspect of the fact that the entanglement Hamiltonian Eq.~\eqref{eq:lincomb} is described by a linear combination of conservation laws involving irrelevant boundary operators, the particular linear combination being dictated by the wave function of the topological quantum state on the cylinder surface (the initial condition of the quench). When expressed in terms of system size $\ell$-independent conservation laws ${\tilde H}^{(i)}:=$
		$\left(\frac{\ell}{2\pi}\right)^{\Delta_i-1}H^{(i)}$ (where $\Delta_i >1$ is the associated conformal dimension ---  these are summarized in Table \ref{table:modereps} of Appendix \ref{app:details}),
		\begin{equation}
			\label{eq:lincomb-SizeDependence}
			H_{\text{entanglement}} - {\rm const.} = 
			\beta 
			\left (\frac{2\pi}{\ell}\right )
			{\tilde H}_L + \sum_{i \neq 1}^{\infty} \beta_i 
			\left (\frac{2\pi}{\ell}\right )^{\Delta_i -1}
			{\tilde H}^{(i)},
		\end{equation}
		the entanglement Hamiltonian in the limit of large system size $\ell$ (cylinder circumference) ought to be dominated by the least irrelevant conservation law (the ${\tilde H}^{(i)}$ with the smallest $\Delta_i$). 
		While we find that for most systems (and topological sectors) whose numerical ES data  we investigated, this least irrelevant conservation law is the Hamiltonian ${\tilde H}_L$ of the CFT, the integral of the energy momentum tensor (i.e.,~all $\Delta_i>2$, for $i\neq 1$), we find instead in Sec.~\ref{sec:su22results} below that, in contrast, the entanglement spectrum of the spin-1/2 sector of the non-Abelian chiral $\rm{SU}(2)_2$ spin liquid requires a fractional conservation law  ${\tilde H}^{(0)}\propto G_0$ with scaling dimension ($\Delta_{i=0}=3/2$) smaller than that of the energy momentum tensor (and any other conservation laws).
		More generally, this conservation law is allowed to occur in this sector on symmetry grounds\footnote{see also Appendix \ref{LabelSubsectionRTSymatryConservedQuantities}}, so it will in general be present in the entanglement Hamiltonian. It would appear, then, that this fractional conservation law, arising (as already mentioned above) from an underlying supersymmetry of the corresponding CFT, will dominate the entanglement spectrum at large system size (far beyond the small finite sizes of the spectra shown in Sec.~\ref{sec:results}). 
		As a consequence of this, the entanglement spectrum in the limit of large system size would therefore not be expected to approach the spectrum of a CFT Hamiltonian [which would describe the spectrum at a physical edge as in \eqref{eq:cfth}], but rather the spectrum of $G_0$,
		\begin{eqnarray}
		\label{eq:g0asymptote}
		&&
		H_{\text{entanglement}} \sim \beta_0
			\left (\frac{2\pi}{\ell}\right )^{1/2}
			G_0,
			\qquad (\ell \to \infty).
		\end{eqnarray}
	    Since the Hamiltonian of the CFT is related to the square of $G_0$ due to the space-time supersymmetry\cite{FriedanQiuShenkerPhysLettB-SUSY-1985}, the eigenvalues of $G_0$ are related to those of the energy (and the momentum) via
	    \footnote{The modes $G_n$ of the superconformal current, together with the Virasoro modes $L_n$ of the energy-momentum tensor, obey the $N = 1$ superconformal algebra\cite{Cohn1988,FriedanQiuShenkerPhysLettB-SUSY-1985} at central charge $c=3/2$.
	    In particular, $\{G_m, G_n\} =2 L_{m+n} +\frac{c}{3} \left(m^2-\frac{1}{4}\right) \delta_{m, -n}$, so in particular, $G_0^2 = \frac{\{G_0,G_0\}}{2} = L_0 - \frac{c}{24}$. Therefore the eigenvalues of $G_0$ must in every case be one of $\pm\sqrt{L_0 - \frac{c}{24}}$.} 
		\begin{eqnarray}
		G_0=\pm \sqrt{L_0 - \frac{c}{24}}.
		\end{eqnarray}
		Given this relation, it would seem that in this sector the entanglement spectrum would take values equal to the positive and negative square-root of the (left-moving) momenta $k_L$ [compare \eqref{eq:cfth}]:
		\begin{eqnarray}
		&&
		H_{\text{entanglement}} \sim \pm \beta_0
			\sqrt{v k_L} + O(k_L)
			\qquad (\ell \to \infty).
		\end{eqnarray}
		It would be interesting to try to understand this in future work, 
		e.g., through numerical investigation by accessing the thermodynamic limit of the entanglement spectrum using an excitation ansatz approach, as discussed in Ref.~\onlinecite{Haegeman2017} for the simpler, Abelian Kalmeyer-Laughlin chiral spin liquid\cite{Note2}.
	
	\section{Choosing the Locally Conserved Quantities}
	\label{sec:locconquant2}
		
		A crucial question in this effort is how to properly choose the locally conserved quantities $H^{(i)}$ that go into our parametrization Eq.~\eqref{eq:lincomb} of the entanglement Hamiltonian.
		These quantities $H^{(i)}$ must both commute with the Hamiltonian $H_L$ of Eq.~\eqref{eq:cfth} and, crucially, must preserve both the $\rm{SU}(2)$ symmetry of the CFT and any applicable discrete symmetries. The requirement of global $\rm{SU}(2)$ symmetry serves as a strong constraint on the irrelevant local operators whose integrals we can use for the $H^{(i)}$. For the systems we consider, the principal discrete symmetry in question is the $\mathcal{RT}$ symmetry of the (1+1)D theory found in the entanglement spectrum that results from the symmetry of the (2+1)D theory under the composition of a spatial reflection through a plane parallel to the axis of the cylinder and time reversal. Further, in the case of each of the PEPS we consider, the $\mathcal{RT}$ symmetry has been built into the PEPS wavefunction by construction. The details of the $\mathcal{RT}$ symmetry can be found in Appendix \ref{app:rtsymmetry}.
		
		We will not require our conserved quantities to commute among themselves as long as they commute with $H_L$. One might rightly worry about the issue of noncommutativity of the $H^{(i)}$ among themselves when taking the logarithm of Eq.~\eqref{eq:realrhol} to obtain Eq.~\eqref{eq:lincomb}, but since our list of $H^{(i)}$ will be exhaustive of quantities that commute with $H_L$ and preserve $\rm{SU}(2)$ and $\mathcal{RT}$ symmetry, we can simply reassign the $\beta_i$ as needed to account for the commutators. 
		
		To simplify our calculations and to avoid a surfeit of parameters, we only consider integrals of local operators up to conformal dimension $\Delta = 6$. (Operators of higher conformal dimension, more irrelevant, will have less significant contributions to the spectral splittings, and could of course be incorporated if necessary.) 
		At a given conformal dimension $\Delta$, we begin by considering the complete list of all the independent $\rm{SU}(2)$-invariant operators of dimension $\Delta$ in the theory. The operators of this type available will exactly correspond to the $\rm{SU}(2)$ singlet descendant states of the primary states of the theory.
		For the $\rm{SU}(2)_1$ WZW CFT, the number of operators in that list is the number of $\rm{SU}(2)$ singlet descendant operators of the identity with dimension $\Delta$, equal to the number of singlet descendant states of the primary state $|j=0\rangle$ at descendant level $K = \Delta$.
		Since the level $k$ of the $\rm{SU}(2)_{k=1}$ theory is $k=1$, we can take advantage of Abelian bosonization to construct these singlet descendants from $|j=0\rangle$ using only the Virasoro modes of the energy-momentum tensor, and therefore these operators will consist only of combinations of the energy-momentum tensor $T(x)$ and its derivatives.
		We do, however, exclude total derivatives from the list of operators we consider, as their spatial integrals will not contribute given the periodicity of the cylinder. 
		[Indeed, while the number of singlet descendant states of $|j=0\rangle$ can be found in Table \ref{table:su21countings} of Appendix \ref{app:su2kintro} as the number of singlets ($\bm{1}$'s) in the ``Multiplet content" column for the $|j=0\rangle$ primary sector, a further calculation is required to get the actual number of operators we consider at each dimension $\Delta = K$: one must account for the exclusion of the derivatives of the operators of dimension $\Delta - 1$, corresponding to the subtraction of the number of singlet states at descendant level $K = 1$.] Additionally, the $\mathcal{RT}$ symmetry requires that we only include operators with even dimension $\Delta$, as it can be observed that operators with an odd number of derivatives will be odd under $\mathcal{RT}$, while all factors of $T(x)$ are even under $\mathcal{RT}$. In $\rm{SU}(2)_1$, though, this does not exclude any further operators, as all such operators of odd dimension $\Delta$ in the $\rm{SU}(2)_1$ theory up to $\Delta = 6$ turn out to be total derivatives as well. 
		A list of the operators $\Phi_i(x)$ we consider for the $\rm{SU}(2)_1$ theory can be found in the leftmost column of Table \ref{table:operators}, arranged by their corresponding conformal dimension $\Delta_i$.
		\begin{table}[hbt]
			\centering
			\begin{tabular}{c|c|c|c}
					$\Delta_i$ & $\Phi_i(x)$ in $\rm{SU}(2)_{k\geq 1}$ & $\Phi_i(x)$ in $\rm{SU}(2)_2$ ($\Delta_i \in \mathbb{Z}$) & $\Phi_i(x)$ in $\rm{SU}(2)_2$ ($\Delta_i \in \mathbb{Z}+\frac12$) \\
					\hline
					3/2 & --- & --- & $G(x)$ \\
					\hline
					2 & $T(x)$ & --- & --- \\
					\hline
					7/2 & --- & --- & $(TG)(x)$ \\
					\hline
					4 & $(TT)(x)$ & $i(G\partial G)(x)$ & ---  \\
					\hline
					9/2 & --- & --- & \cellcolor{Gray} $(T\partial G)(x)$ \\ 
					\hline
					11/2 & --- & --- & $(G(TT))(x)$, $(\partial T\partial G)(x)$\\
					\hline
					6 & $(T(TT))(x)$, $(\partial T \partial T)(x)$ & $i(T(G \partial G))(x)$, $i(\partial G \partial^2 G)(x)$ & --- \\
					\hline
			\end{tabular}
			\caption{
    			An enumeration of the lower-dimensional (of conformal dimensions $\Delta_i \leq 6$) irrelevant operators $\Phi_i(x)$ we will use to fit the splittings of the numerical spectra within each descendant level for (leftmost column) the chiral $\rm{SU}(2)_1$ and (all columns) chiral $\rm{SU}(2)_2$ WZW theories we consider. The shaded cell indicates an irrelevant operator that we choose to exclude due to consideration of $\mathcal{RT}$ symmetry.
    		}
			\label{table:operators}
		\end{table}
		
		For the $\rm{SU}(2)_2$ WZW CFT, the picture is somewhat more complex. As we did for $\rm{SU}(2)_1$, we can use the operator-state correspondence to find the operators available to us in the $\rm{SU}(2)_2$ theory. For $\rm{SU}(2)_2$, the $\rm{SU}(2)$ singlet states come in two sets: descendants of the $|j=0\rangle$ primary state and descendants of the $|j=1\rangle$ primary state\footnote{As will be seen later, the $\rm{SU}(2)_2$ entanglement spectrum data available to us from Ref.~\onlinecite{Chen2018} only includes clearly observable countings for the $|j=0\rangle$ and $|j=1/2\rangle$ primary sectors, so those will be the only sectors of $\rm{SU}(2)_2$ we fit in Sec.~\ref{sec:su22results}. This does not in any way preclude our use of the integrals of \textit{operators} that correspond to descendant states in the $|j=1\rangle$ primary sector.}. (This is because these are the two sectors that have integer spin multiplets, and therefore contain $\rm{SU}(2)$ singlet states.) Descendants of the $|j=0\rangle$ primary state correspond to operators that have integer conformal dimension, since the conformal weight of the $|j=0\rangle$ primary state is $h_{j=0} = 0$. This set of operators includes as a subset the operators we considered for the $\rm{SU}(2)_1$ theory, all of which had integer conformal dimension. Descendants of the $|j=1\rangle$ primary state, on the other hand, correspond to operators that have fractional (half-integer) conformal dimension, as the conformal weight of the $|j=1\rangle$ primary state is $h_{j=1} = 1/2$. [The number of singlet descendant states at each descendant level $K$ for both the $|j=0\rangle$ and $|j=1\rangle$ primary sectors, corresponding to the number of available operators of dimension $\Delta = K$ (for $|j=0\rangle$) or $\Delta = K+1/2$ (for $|j=1\rangle$), can be found in the respective ``Multiplet content" columns in Table \ref{table:su22countings} of Appendix \ref{app:su2kintro} 
		(where singlets are denoted by $\bm{1}$'s), 
		though as in the $\rm{SU}(2)_1$ case, one must account for the exclusion of total derivatives when comparing to the enumeration of operators in Table \ref{table:operators}.]

		It turns out that we will only need to consider the action of the integrals of the operators with half-integer conformal dimension on the states of the $|j=1/2\rangle$ sector, in which the operators possess periodicity around the cylinder. 
		One way to see this is to express\cite{ZamolodchikovFateevSovJNuclPhys1986} the chiral $\rm{SU}(2)_2$ WZW CFT as a theory of three free real Majorana fermions, which we will denote by $\psi^a(x)$ for $a = 1,2,3$. (See e.g., Ref.~\onlinecite{Goddard1986}.) We can relate the 3-fermion theory to the chiral $\rm{SU}(2)_2$ WZW CFT as described in Sec.~\ref{sec:wzwreview} by writing the $\rm{SU}(2)$ current $J^a(x)$ as the fermion bilinear
		\begin{equation}
			\label{eq:jpsi}
			J^a(x) = -\frac{i}{2} \epsilon_{abc} :\psi^b \psi^c:(x),
		\end{equation}
		where the $::$ indicates normal ordering. 
		The 3-fermion theory possesses $N=1$ supersymmetry. The sectors of the $|j=0\rangle$ and $|j=1\rangle$ primary states of the chiral $\rm{SU}(2)_2$ theory correspond to the Neveu-Schwarz sector of the 3-fermion theory, while the sector of the $|j=1/2\rangle$ primary state corresponds to the Ramond sector. $\psi^a(x)$ has conformal dimension 1/2, so in the 3-fermion theory the half-integer dimensional operators are exactly the fermionic operators. On the cylinder, fermionic operators have periodic boundary conditions only in the Ramond sector, with anti-periodic boundary conditions in the Neveu-Schwarz sector. Thus integrals of fermionic operators will be non-trivial only in the Ramond sector. And indeed, this result holds in general for the half-integer dimensional operators of the $\rm{SU}(2)_2$ theory we consider, so we will only take them into account in our set of conserved quantities for the $|j=1/2\rangle$ primary state sector, which corresponds to the Ramond sector, of $\rm{SU}(2)_2$.

		We will not work with the three fermions $\psi^a(x)$ per se, however, but rather with bosonic spin-1 Kac-Moody (affine) primary operators $\phi^a(x)$ of conformal dimension $\Delta = 1/2$ that possess identical ``anti-commutation relations" within the chiral theory
		\footnote{In contrast to the $\psi^a(x)$ and their antichiral counterparts $\bar{\psi}^a(x)$, which anticommute with each other, $\phi^a(x)$ and $\bar{\phi}^a(x)$ actually commute with each other. Within the chiral (antichiral) theory, though, $\phi^a(x)$ and $\psi^a(x)$ ($\bar{\phi}^a(x)$ and $\bar{\psi}^a(x)$) will behave the same way in correlation functions (see, e.g., Ref.~\onlinecite{Maldacena1997}.)}
		and satisfy 
		\begin{equation}
				\label{eq:jphi}
				J^a(x) = -\frac{i}{2} \epsilon_{abc} :\phi^b \phi^c:(x).
		\end{equation}
		\noindent We can then additionally write down a current operator $G(x)$ in terms of the $\phi^a(x)$ operators, which has the same operator product expansion relations within the chiral theory as the superconformal current operator $G(x)$ 
		(see, e.g., Refs.~\onlinecite{Cohn1988,FriedanQiuShenkerPhysLettB-SUSY-1985,Mussardo1988}):
		\begin{equation}
			\label{eq:gdef}
			G(x) = \frac{i}{6} \epsilon_{abc} :\phi^a \phi^b \phi^c:(x).
		\end{equation}
		$G(x)$ has conformal dimension $\Delta_{i=0} = 3/2$ (as it is composed of three of the $\Delta = 1/2$ operators $\phi^a(x)$). Furthermore, $G(x)$ is an $\rm{SU}(2)$ singlet. $G(x)$ corresponds to the lowest-level $\rm{SU}(2)$ singlet descendant in the $|j=1\rangle$ primary sector, and is thus one of the fractional dimension operators available in $\rm{SU}(2)_2$. 
		The operators $T(x)$, $G(x)$, and combinations of both and their derivatives will comprise the set of all of the operators $\Phi_i(x)$ we consider in the $\rm{SU}(2)_2$ theory. 
		These are explicitly listed up to conformal dimension $\Delta_i = 6$ in all the columns of Table \ref{table:operators}. The left two columns comprise the integer dimensional operators, those which contain an even number of half-integer dimensional factors ($G(x)$ or its derivatives) and hence are periodic in, and therefore found in, all sectors. The right column contains the half-integer dimensional operators, which contain an odd number of factors of $G(x)$ or its derivatives, and are only available to us in the $|j=1/2\rangle$ sector as discussed above. Note that we again exclude total derivatives. The $\mathcal{RT}$ symmetry also requires that we again exclude all operators with odd integer dimension $\Delta$, though as was the case for $\rm{SU}(2)_1$, all operators of odd integer dimension below $\Delta = 6$ for $\rm{SU}(2)_2$ will again be total derivatives of even integer dimensional operators as well, and thus they are already excluded.
		$G(x)$, however, which has $\Delta_{i=0} = 3/2$, is invariant under $\mathcal{RT}$ in the Ramond sector as discussed in Appendix~\ref{LabelSubsectionRTSymatryConservedQuantities}. 
		Thus, of the half-integer dimensional operators we consider, only $(T\partial G)(x)$, where an odd number of derivatives are not multiplied by an $i$ (which will become $-i$ under the anti-unitary $\mathcal{RT}$), will actually be wholly excluded from the entanglement Hamiltonian due to the $\mathcal{RT}$ symmetry. This is indicated in Table \ref{table:operators} by a shaded cell.
		
		For each of these operators $\Phi_i(x)$, we compute the mode-expanded form of the system size $\ell$-independent integral $\tilde{H}^{(i)}$. (Recall that $\tilde{H}^{(i)} = \left(\frac{\ell}{2\pi}\right)^{\Delta_i - 1}H^{(i)}$, where $\Delta_i$ is the conformal dimension of $\Phi_i(x)$.)
		The mode-expanded forms of the $\tilde{H}^{(i)}$ can be found in Table \ref{table:modereps} of Appendix \ref{app:details}.
		Using the mode-expanded forms for the $\tilde{H}^{(i)}$, $H_{\text{entanglement}}$ of Eq.~\eqref{eq:lincomb} is diagonalized on the space of descendant states of each of the $k+1$ CFT primary states of $\rm{SU}(2)_k$, a description also found in Appendix \ref{app:details}. 
		For each primary state, this gives a method for finding the entanglement spectrum of the corresponding sector of the CFT in terms of the set of parameters $\{\beta_i\}$ of Eq.~\eqref{eq:lincomb}. As can be seen from Table \ref{table:operators}, this will give 4 free parameters $\beta_i$ for the $\rm{SU}(2)_1$ case and 11 free parameters $\beta_i$ for the $\rm{SU}(2)_2$ case. To test the ability of this method to fit the RSES data, we use a least-squares method to find the set of parameters that lead to splittings that best fit the numerical data. 
		Details on the fitting procedure are found in Appendix \ref{app:fitting}. The results of this fitting procedure are shown for a number of data sets in the plots of the next section. Precise values of the parameters and statistics of individual fits are found in Tables \ref{table:su21fitparams} and \ref{table:su22fitparams} of Appendix \ref{app:parameters}.
	
	\section{Results}
	\label{sec:results}
	
		We present fits to four different sets of numerical entanglement spectrum data, of which the first three exhibit the chiral $\rm{SU}(2)_1$ WZW CFT, characteristic of the Kalmeyer-Laughlin spin liquid, while the last exhibits the chiral $\rm{SU}(2)_2$ WZW CFT, characteristic of a non-Abelian chiral $\rm{SU}(2)_2$ spin liquid.
	    In each case, we consider a certain number of low-lying descendant levels of the numerical entanglement spectrum data, as typically, numerical results may be less reliable as we get to higher entanglement energies.
		
		\subsection{Entanglement spectra containing an $\rm{SU}(2)_1$ WZW CFT}
		\label{sec:su21results}
		
\begin{figure}[H]
			\centering
			\subfigure[]{\label{fig:bauerfit_integer} \includegraphics[width=.45\textwidth]{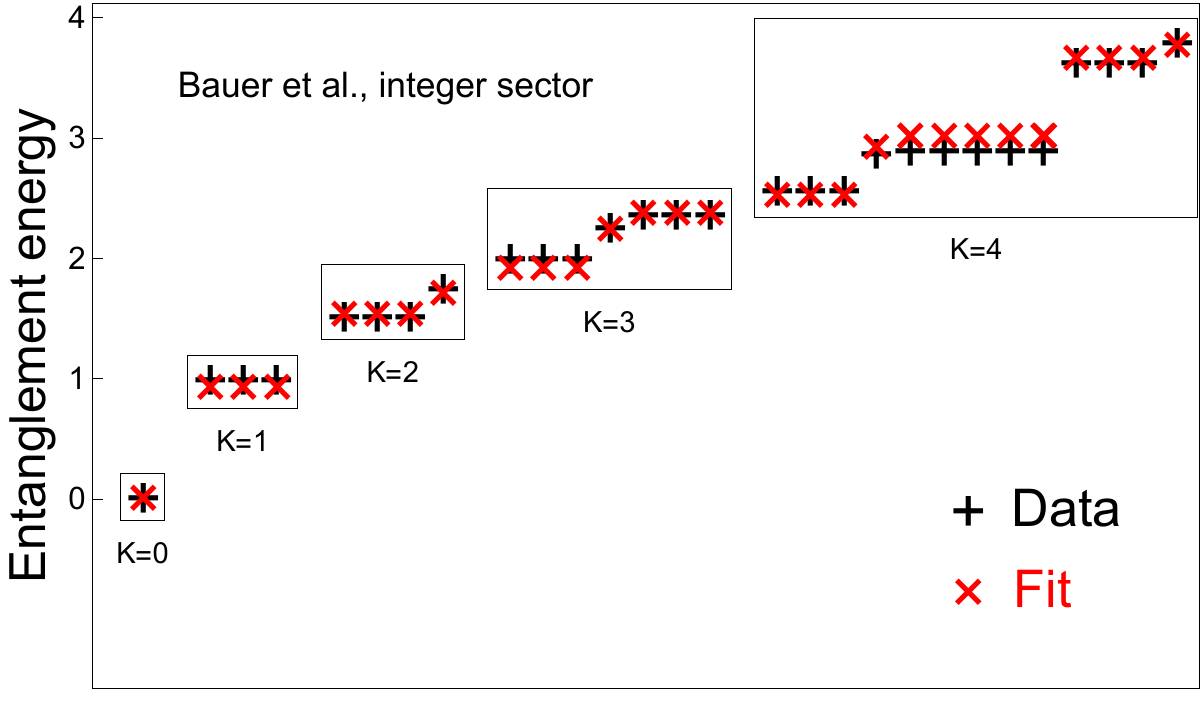}}
			\subfigure[]{\label{fig:bauerfit_halfinteger} \includegraphics[width=.45\textwidth]{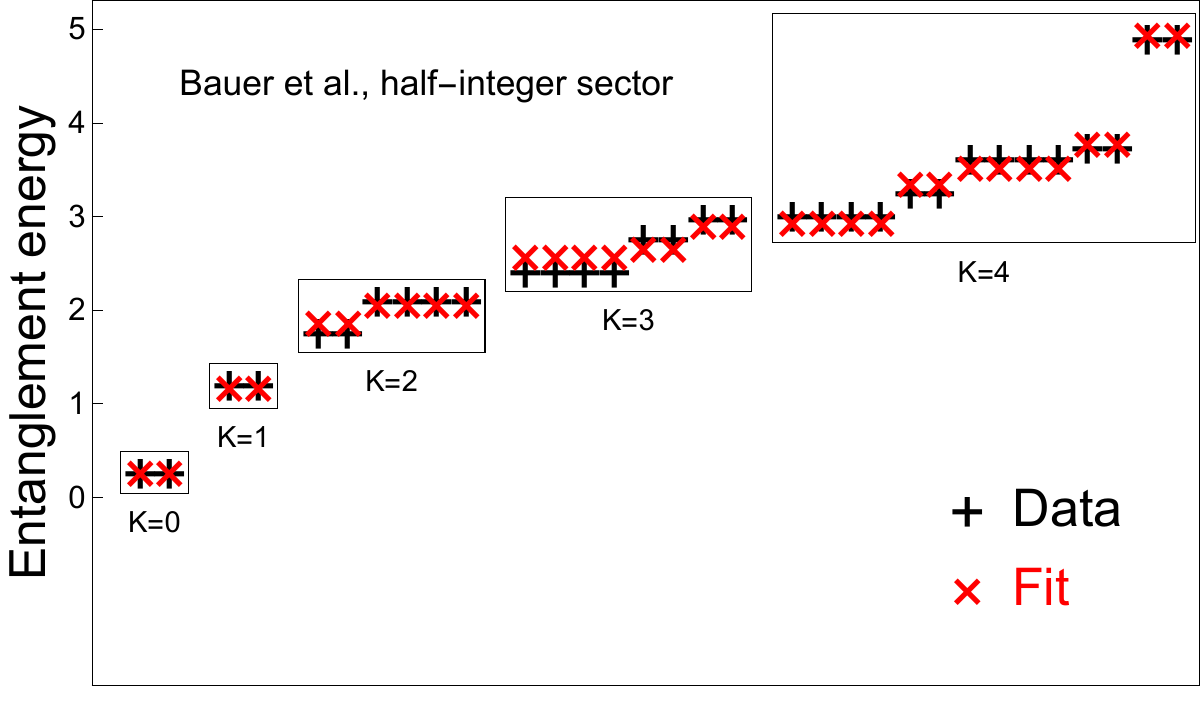}}   
			\caption{
				Our fit to the $\rm{SU}(2)_1$ entanglement spectrum data of Bauer et al.~(Ref.~\protect\onlinecite{Bauer2014}) is shown in (a) the integer sector ($|j=0\rangle$ primary state and descendants) and (b) the half-integer sector ($|j=1/2\rangle$ primary state and descendants). The original data is indicated by black +'s, while red $\times$'s mark the fit produced by our approach. The black boxes indicate states with the same momentum, and hence the same descendant level $K$ above the corresponding primary state, while $\rm{SU}(2)$ multiplets are grouped within each box. (The multiplet content of each box may be compared to Table \ref{table:su21countings} in Appendix \ref{app:su2kintro}.) We attempt to fit 11 differences between multiplets in the integer sector and 11 differences between multiplets in the half-integer sector.
				For each sector, our approach uses 4 parameters: 4 coefficients $\beta_i$ in Eq.~\eqref{eq:lincomb} for the conserved quantities corresponding to the 4 distinct operators of $\Delta \leq 6$ available in $\rm{SU}(2)_1$ found in the left column of Table \ref{table:operators}. The data was computed with a cylinder of circumference $\ell = 12$.\protect\cite{Bauer2014} The scales of the vertical entanglement energy axes are normalized such that $\beta (2\pi/\ell) = 1$ [$\beta (2\pi/\ell)$ being the coefficient of $\tilde{H}_L$ in Eq.~\eqref{eq:lincomb-SizeDependence}], with the zero point appropriate to the conformal weight of the primary state for each sector.
			}
			\label{fig:bauerfit}
		\end{figure}

		The first fit is to the numerical entanglement spectrum of the Kalmeyer-Laughlin chiral spin liquid found in a Mott insulator on the kagome lattice with broken time-reversal symmetry by Bauer et al.~in Figure 4 of their 2014 paper, Ref.~\onlinecite{Bauer2014}. Bauer et al.~employed the method of Ref.~\onlinecite{Cincio2013} to compute their spectrum, using infinite DMRG to optimize a variational MPS state on a cylinder of circumference $\ell = 12$ sites. 
	    The authors cite the observed degeneracies and multiplets of the global $\rm{SU}(2)$ symmetry as evidence that the entanglement spectrum is described by a chiral $\rm{SU}(2)_1$ CFT.\cite{Bauer2014} Indeed, we discussed the chiral $\rm{SU}(2)_1$ countings in Sec.~\ref{sec:wzwreview}, accompanied by the depiction of this particular numerical entanglement spectrum in Fig.~\ref{fig:baueretalplot}. Our fit of the splittings of the first five descendant levels of the spectrum is seen in Fig.~\ref{fig:bauerfit}. Note that the depiction of the $\rm{SU}(2)$ multiplets in Fig.~\ref{fig:bauerfit}, as well as our subsequent plots of fits, differs from the depiction in Fig.~\ref{fig:baueretalplot}. In Figure \ref{fig:baueretalplot}, the individual states of each multiplet at fixed $j^z$ are depicted in each blue-shaded subtower. By contrast, in Fig.~\ref{fig:bauerfit}, each spin-$j$ $\rm{SU}(2)$ multiplet of dimension $d=(2j+1)$ is depicted as a horizontal row of $d$ markers at the vertical coordinate corresponding to the entanglement energy of the multiplet. Multiplets are grouped by descendant level $K$. The associated countings of various dimensions of multiplets at a given $K$ in each of the integer ($|j=0\rangle$) and half-integer ($|j=1/2\rangle$) sectors may be compared with Table \ref{table:su21countings} in Appendix \ref{app:su2kintro}. For the Bauer et al.~data set, the fits are performed independently in both the integer and half-integer sectors.
		\footnote{Attempts to fit both sectors simultaneously resulted in some of the very highest-energy multiplets, in the highest descendant level considered, not agreeing with our expectations based on the set of parameters that successfully fit the low-energy part of the spectrum.}
		On the whole, while not perfect, we see that the fits are fairly successful in explaining the 11 splittings between the 12 multiplets in those levels with the 4 parameters available by considering the coefficients of the integrals of the 4 operators up to dimension $\Delta = 6$ in the $\rm{SU}(2)_1$ theory (see Table \ref{table:operators}). In particular, the fits match the relative positioning of the different multiplets within each descendant level of the spectrum.
		
\begin{figure}[H]
			\centering
			\subfigure[]{\label{fig:hickeyfit_integer} \includegraphics[width=.45\textwidth]{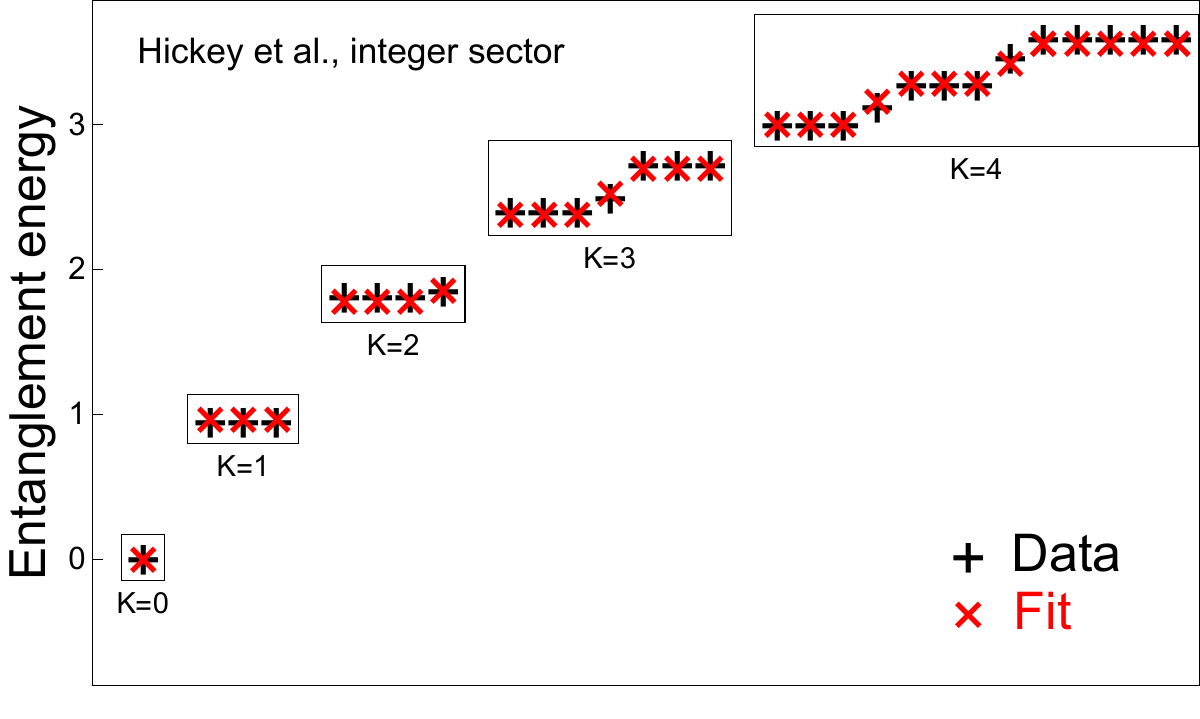}}
			\subfigure[]{\label{fig:hickeyfit_halfinteger} \includegraphics[width=.45\textwidth]{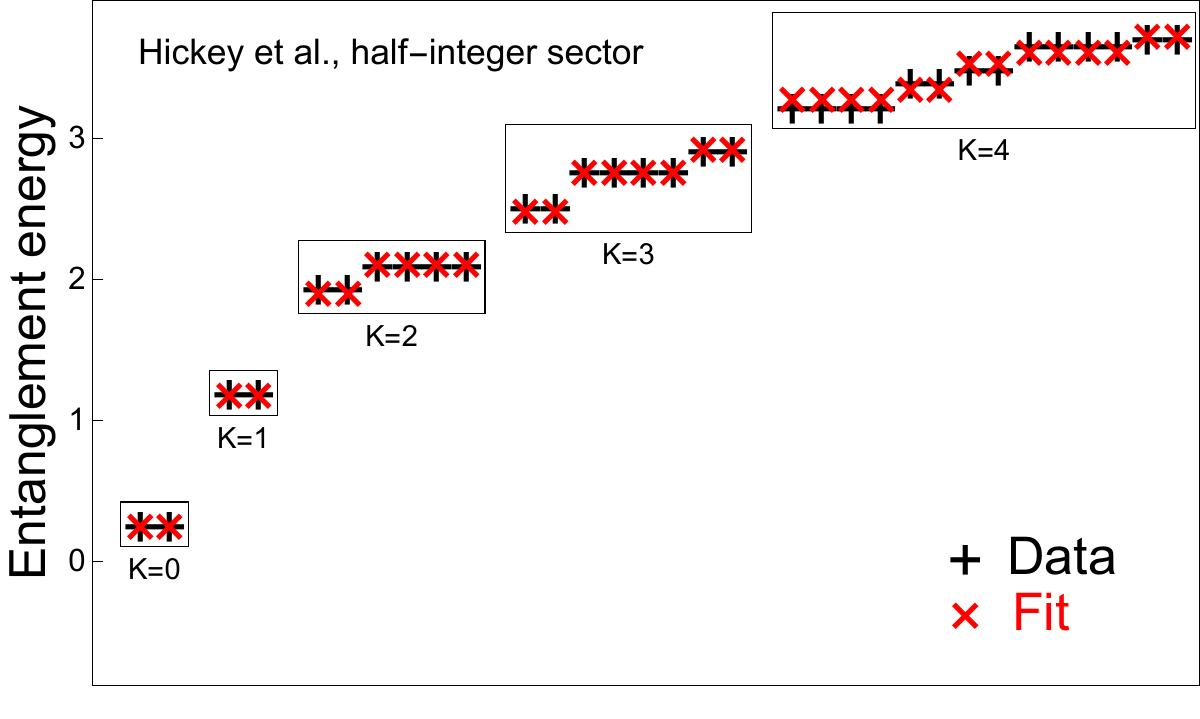}} 
			\caption{
				Our fit to the $\rm{SU}(2)_1$ entanglement spectrum data of Hickey et al.~(Ref.~\protect\onlinecite{Hickey2016}) is shown in (a) the integer sector ($|j=0\rangle$ primary state and descendants) and (b) the half-integer sector ($|j=1/2\rangle$ primary state and descendants). The original data is indicated by black +'s, while red $\times$'s mark the fit produced by our approach. The black boxes indicate states with the same momentum, and hence the same descendant level $K$ above the corresponding primary state, while $\rm{SU}(2)$ multiplets are grouped within each box. (The multiplet content of each box may be compared to Table \ref{table:su21countings} in Appendix \ref{app:su2kintro}.) We attempt to fit 11 differences between multiplets in the integer sector and 11 differences between multiplets in the half-integer sector, i.e.,~a total of 22 differences.
				Fitting both sectors simultaneously, our approach uses 4 parameters: 4 coefficients $\beta_i$ in Eq.~\eqref{eq:lincomb} for the conserved quantities corresponding to the 4 distinct operators of $\Delta \leq 6$ available in $\rm{SU}(2)_1$ found in the left column of Table \ref{table:operators}. The data was computed with a cylinder of circumference $\ell = 8$.\protect\cite{Hickey2016} The scales of the vertical entanglement energy axes are normalized such that $\beta (2\pi/\ell) = 1$ [$\beta (2\pi/\ell)$ being the coefficient of $\tilde{H}_L$ in Eq.~\eqref{eq:lincomb-SizeDependence}], with the zero point appropriate to the conformal weight of the primary state for each sector.
			}
			\label{fig:hickeyfit}
		\end{figure}

		The second fit is to the numerical entanglement spectrum of the chiral spin liquid found in a Haldane-Hubbard Mott Insulator on the honeycomb lattice by Hickey et al.~in Figure 3 of their 2016 paper, Ref.~\onlinecite{Hickey2016}. Hickey et al.~used infinite DMRG as well, on a cylinder with a circumference of $\ell=8$ sites. Also here, the degeneracies and multiplets of global $\rm{SU}(2)$ symmetry were cited as evidence of a description of the observed entanglement by the $\rm{SU}(2)_1$ WZW CFT.\cite{Hickey2016} Our fit of the splittings of the first five descendant levels of this spectrum is seen in Fig.~\ref{fig:hickeyfit}. 
		For the Hickey et al.~data set, we are able to very successfully fit both integer and half-integer sectors simultaneously, with one set of 4 parameters. In particular, the fit captures the relative positioning of the multiplets within each descendant level of the spectrum, in both the integer and half-integer sectors.
		The 22 splittings of the 24 multiplets of the first five levels in both the integer and half-integer sectors are explained with the 4 parameters available by considering the coefficients of the integrals of the 4 operators up to dimension $\Delta = 6$ in the $\rm{SU}(2)_1$ theory (see Table \ref{table:operators}).
		
		The third fit is to the entanglement spectrum of a particular PEPS on an infinite cylinder by Hackenbroich et al.~in their 2018 paper, Ref.~\onlinecite{Hackenbroich2018}. We fit the spectrum found in their Figure 14.
		Hackenbroich et al.~worked with a cylinder of circumference $\ell = 8$. In this spectrum the authors cited the degeneracies and relative computed conformal weight of the $|j=0\rangle$ and $|j=1/2\rangle$ primary states as evidence of a description of the entanglement spectrum by a chiral $\rm{SU}(2)_1$ WZW CFT.\cite{Hackenbroich2018} 
		
\begin{figure}[H]
			\centering
			\subfigure[]{\label{fig:hackenbroichfit_integer} \includegraphics[width=.45\textwidth]{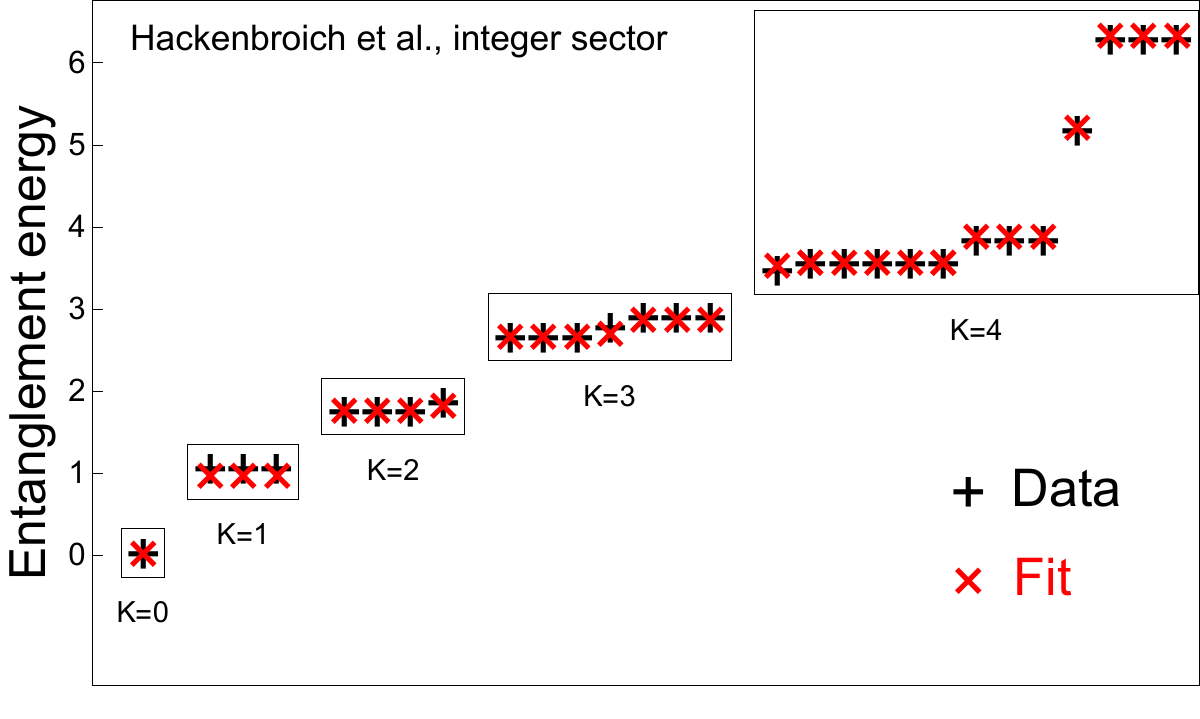}}
			\subfigure[]{\label{fig:hackenbroichfit_halfinteger} \includegraphics[width=.45\textwidth]{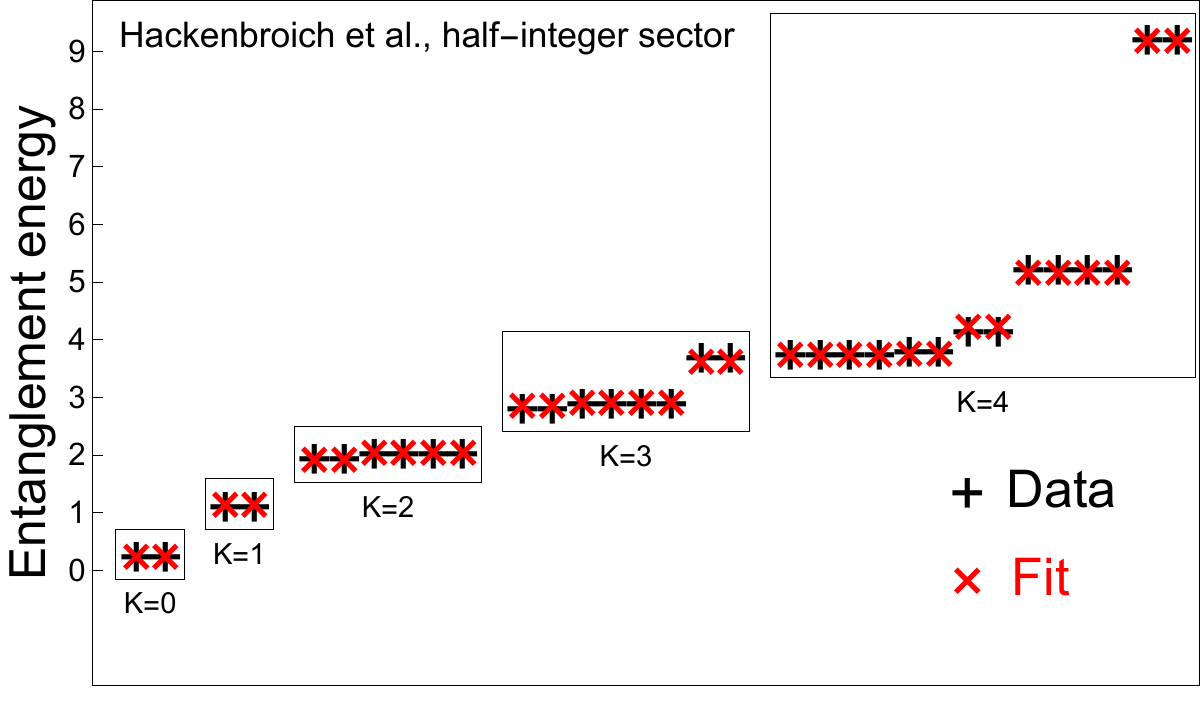}}
			\caption{
				Our fit to the $\rm{SU}(2)_1$ entanglement spectrum data of Hackenbroich et al.~(Ref.~\protect\onlinecite{Hackenbroich2018}) is shown in (a) the integer sector ($|j=0\rangle$ primary state and descendants) and (b) the half-integer sector ($|j=1/2\rangle$ primary state and descendants). The original data is indicated by black +'s, while red $\times$'s mark the fit produced by our approach. The black boxes indicate states with the same momentum, and hence the same descendant level $K$ above the corresponding primary state,
				while $\rm{SU}(2)$ multiplets are grouped within each box. (The multiplet content of each box may be compared to Table \ref{table:su21countings} in Appendix \ref{app:su2kintro}.) We attempt to fit 11 differences between multiplets in the integer sector and 11 differences between multiplets in the half-integer sector, i.e.,~a total of 22 differences.
				Fitting both sectors simultaneously up to a scale factor, our approach uses 5 parameters: 4 coefficients $\beta_i$ in Eq.~\eqref{eq:lincomb} for conserved quantities corresponding to the 4 distinct operators of $\Delta \leq 6$ available in $\rm{SU}(2)_1$ found in the left column of Table \ref{table:operators}, plus a relative scale factor between the two sectors. The data was computed with a cylinder of circumference $\ell = 8$.\cite{Hackenbroich2018} The scales of the vertical entanglement energy axes are normalized such that $\beta (2\pi/\ell) = 1$ [$\beta (2\pi/\ell)$ being the coefficient of $\tilde{H}_L$ in Eq.~\eqref{eq:lincomb-SizeDependence}], with the zero point appropriate to the conformal weight of the primary state for each sector.
			}
			\label{fig:hackenbroichfit}
		\end{figure}

		Fig.~\ref{fig:hackenbroichfit} is our fit of the spectrum. We are able to fit both sectors of the Hackenbroich et al.~dataset simultaneously with great success, though a relative scale factor between the two sectors is used due to their possibly differing velocities $v$ in Eq.~\eqref{eq:cfth}.\cite{Hackenbroich2018} We thus end up using 5 parameters---the 4 parameters available by considering the coefficients of the integrals of the 4 operators up to dimension $\Delta = 6$ in the chiral  $\rm{SU}(2)_1$ theory (see Table \ref{table:operators}), plus the relative scale factor, to fit the 22 differences between the multiplets in the first five descendant levels (with 12 multiplets each) in both the integer and half-integer sectors. The fit is quite good, and certainly captures the relative positioning of the multiplets within each descendant level of the spectrum in both sectors. 
		
		The power of our approach was demonstrated by one phenomenon in the Hackenbroich et al.~data: the highest-energy multiplet in the highest descendant level that we fit in the half-integer sector is not simply the next higher doublet at the same value of momentum, as the Brillouin zone of the spectrum is ``folded" in the half-integer sector (as compared to the integer sector---momenta are multiples of $2\pi/4$ instead of $2\pi/8$).\cite{Hackenbroich2018} This leads data from higher descendant levels to overlap with the data of the level we are trying to fit. Yet, by fitting the integer sector, we were able to determine an estimate of the correct parameters $\{\beta_i\}$ for the half-integer sector, as well. These parameters predicted where the final highest-energy doublet in the half-integer sector could be found. Indeed, there was a doublet present at that point in the entanglement spectrum, and so we were able to show (in Fig.~\ref{fig:hackenbroichfit}) the success of our fit for the Hackenbroich et al.~dataset.
		\newpage
		
		\subsection{Entanglement spectra containing an $\rm{SU}(2)_2$ WZW CFT}
		\label{sec:su22results}
		
		The fourth and final set of fits is to the entanglement spectrum of a PEPS aimed at representing a spin-1 chiral Heisenberg antiferromagnet defined on a square lattice by Chen et al.~in their 2018 paper, Ref.~\onlinecite{Chen2018}. The spectra we fit are those of their Figures~11~(c)~and~(d), which display numerical entanglement spectra computed on a cylinder of circumference $\ell = 6.$\cite{Chen2018} 
		The authors obtained degeneracies and multiplet content consistent with a chiral $\rm{SU}(2)_2$ WZW CFT, as may be seen by comparison with the data of Table \ref{table:su22countings} of Appendix \ref{app:su2kintro}. Recall from Sec.~\ref{sec:wzwreview} that the chiral $\rm{SU}(2)_2$ WZW CFT will have three (affine) primary states, $|j=0\rangle$, $|j=1/2\rangle$, and $|j=1\rangle$, each with an associated topological sector of descendant states.
		The Chen et al.~data does not contain a clear representation of the $|j=1\rangle$ sector
		\footnote{The states in this sector can be thought of possessing odd fermion parity in $\rm{SU}(2)_2$, as opposed to the states in the $|j=0\rangle$ sector (see, e.g., the discussion in Sec.~\ref{sec:locconquant2}). Perhaps this might be related to the fact that they do not seem to be visible in the data.}.
		We perform several fits to the Chen et al.~data. 
			
\begin{figure}[H]
			\centering
			\subfigure[]{\label{fig:chenmmfit_bosons} \includegraphics[width=.45\textwidth]{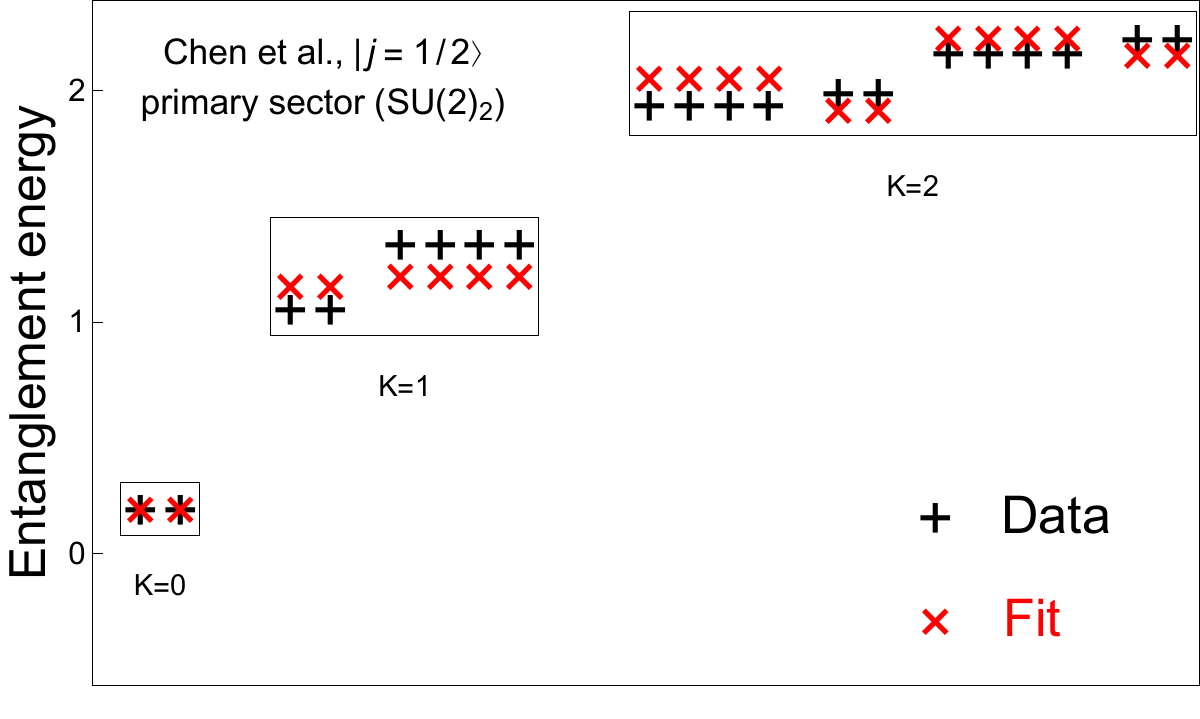}}
			\subfigure[]{\label{fig:chenmmfit_fermions} \includegraphics[width=.45\textwidth]{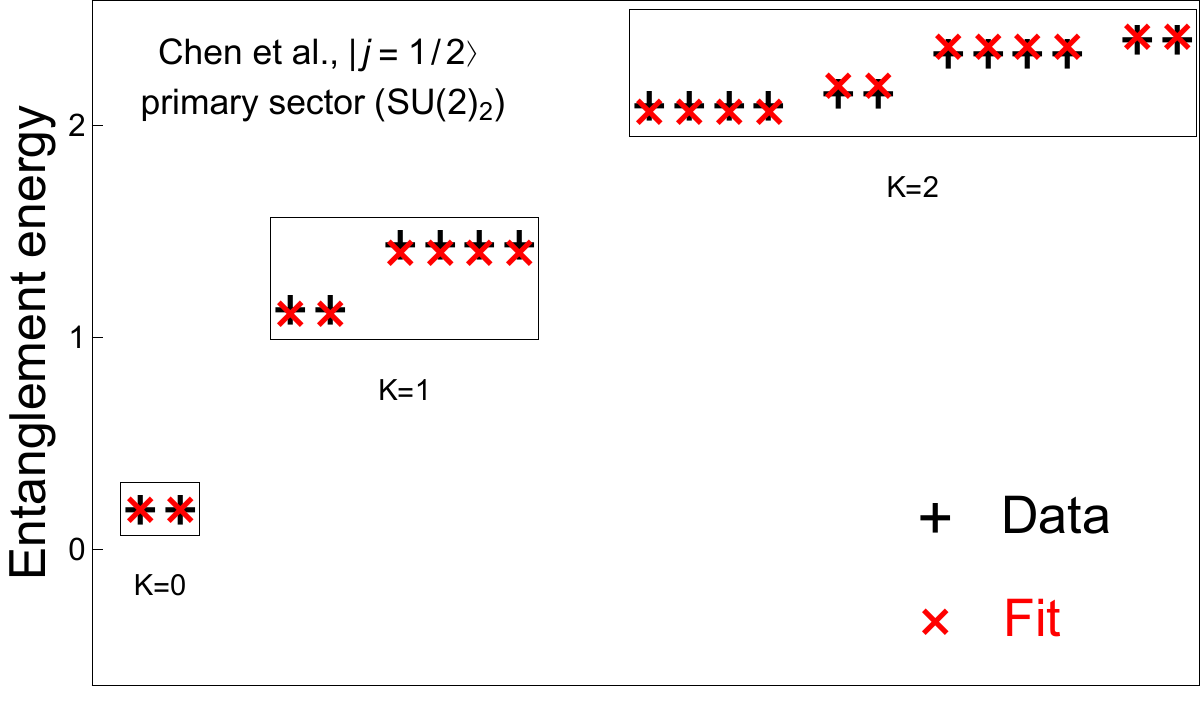}}
			\caption{
				Two fits to the $\rm{SU}(2)_2$ entanglement spectrum data of Chen et al.~(Ref.~\protect\onlinecite{Chen2018}) in the sector of the $|j=1/2\rangle$ primary state and descendants are shown. The original data is indicated by black +'s, while red $\times$'s mark the fits produced by our approach. The black boxes indicate states with the same descendant level $K$ above the corresponding primary state, while states in the same $\rm{SU}(2)$ multiplet are grouped within each box. (The multiplet content of each box may be compared to Table \ref{table:su22countings} in Appendix \ref{app:su2kintro}.) We attempt to fit the 6 differences between multiplets in the 3 depicted levels of the $|j=1/2\rangle$ sector, up to $K = 2$. 
				The fits use 3 parameters corresponding to the coefficients $\beta_i$ in Eq.~\eqref{eq:lincomb} for the conserved quantities corresponding to, in (a), the 3 integer dimensional operators with $\Delta \leq 4$ available in $\rm{SU}(2)_2$, ($T(x)$, $(TT)(x)$, and $i(G\partial G)(x)$), and, in (b), the 3 integer and half-integer dimensional operators with $\Delta \leq 7/2$ available 
				in $\rm{SU}(2)_2$, ($G(x)$, $T(x)$, and $(TG)(x)$). (See Table \ref{table:operators}.) The scales of the vertical entanglement energy axes are normalized such that $\beta (2\pi/\ell) = 1$ [$\beta (2\pi/\ell)$ being the coefficient of $\tilde{H}_L$ in Eq.~\eqref{eq:lincomb-SizeDependence}], with the zero point appropriate to the conformal weight of the primary state for the sector.
		    }
			\label{fig:chenmmfit}
		\end{figure}

		In the first pair of fits, depicted in Fig.~\ref{fig:chenmmfit} in the same format as the figures of Sec.~\ref{sec:su21results}, we illustrate the necessity of including the conserved quantities corresponding to the half-integer dimensional operators (found in the rightmost column of Table \ref{table:operators}) in the fit. 
		To show this, we fit only the three lowest descendant levels of the $|j = 1/2\rangle$ sector, by two different approaches. In one approach, shown in Fig.~\ref{fig:chenmmfit_bosons}, we use the conserved quantities corresponding to the first three integer dimensional operators from Table \ref{table:operators} available in $\rm{SU}(2)_2$, up to $\Delta = 4$, which are $T(x)$, $(TT)(x)$, and $i(G\partial G)(x)$, to fit the numerical spectrum. 
		In the other approach, shown in Fig.~\ref{fig:chenmmfit_fermions}, we use the conserved quantities corresponding to the first three operators available of both integer and half-integer dimension, up to $\Delta = 7/2$, which are $G(x)$, $T(x)$, and $(TG)(x)$, to fit the numerical spectrum. In both approaches, there are 3 parameters available corresponding to the coefficients of the conserved integrals of the operators in question, which are used to fit the 6 differences present in the data for the first three levels of the $|j=1/2\rangle$ sector. Yet it is clear from Fig.~\ref{fig:chenmmfit} that the second approach, which makes use of the two half-integer dimensional operators, produces a far better fit.~\footnote{Observe that, for example, the first method is not able to lift the degeneracy of even the lowest excited momentum state at descendant level $K=1$, while the second method achieves that goal with ease.}
		
\begin{figure}[H]
			\centering
			\subfigure[]{\label{fig:chen5opsfit_integer} \includegraphics[width=.45\textwidth]{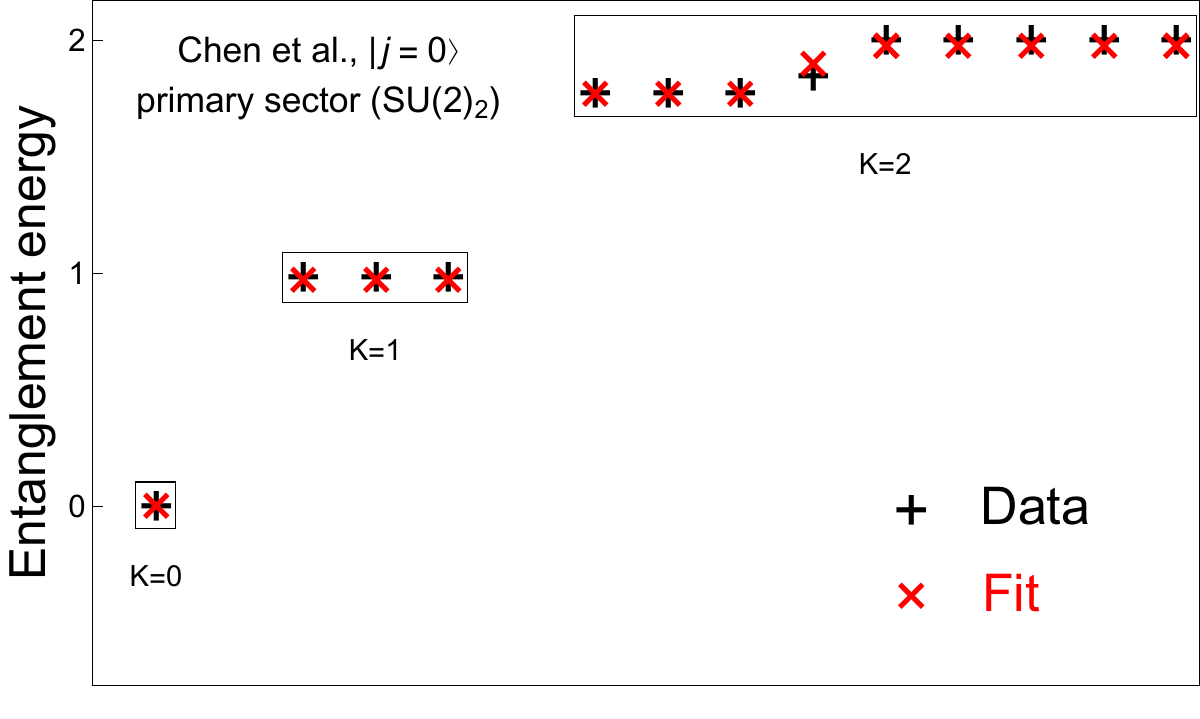}}
			\subfigure[]{\label{fig:chen5opsfit_halfinteger} \includegraphics[width=.45\textwidth]{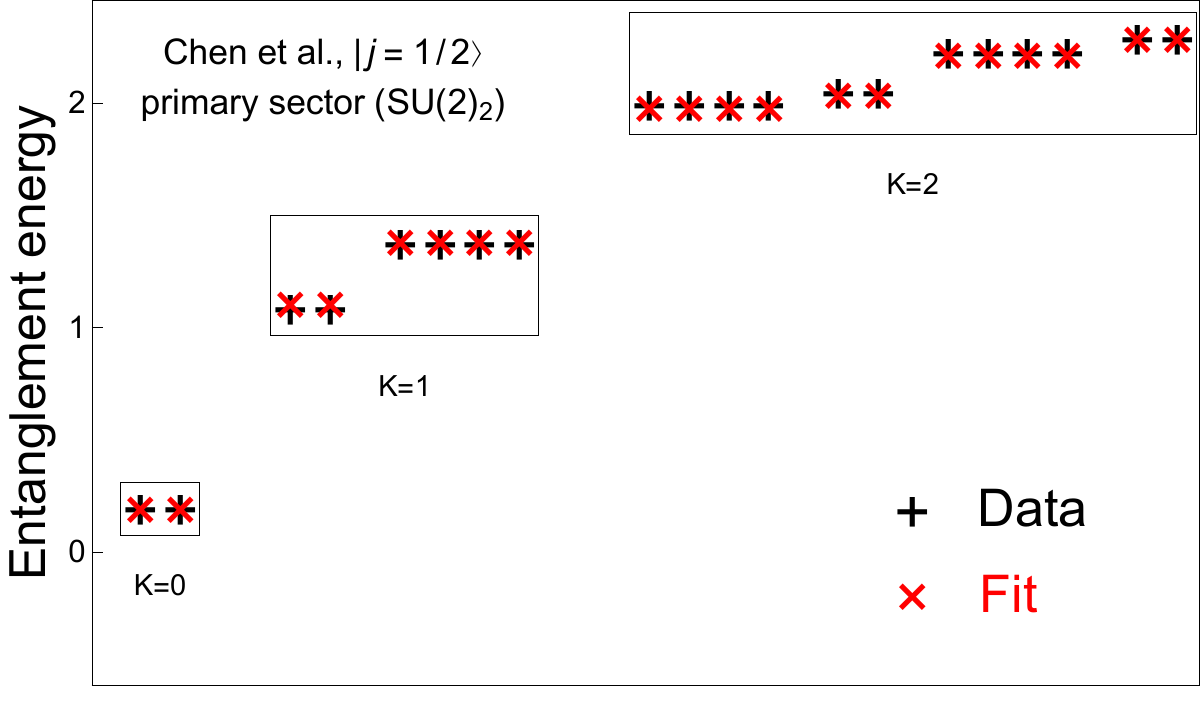}}
			\caption{
				A fit to the $\rm{SU}(2)_2$ entanglement spectrum data of Chen et al.~(Ref.~\protect\onlinecite{Chen2018}) is shown for the first three levels (up to $K=2$) in (a) the sector of the $|j=0\rangle$ primary state and descendants and (b) the sector of the $|j=1/2\rangle$ primary state and descendants. The original data is indicated by black +'s, while red $\times$'s mark the fit produced by our approach. The black boxes indicate states with the same descendant level $K$ above the corresponding primary state, while $\rm{SU}(2)$ multiplets are grouped within each box. (The multiplet content of each box may be compared to Table \ref{table:su22countings} in Appendix \ref{app:su2kintro}.) We attempt to fit 4 differences in the $|j=0\rangle$ sector and 6 differences in the $|j=1/2\rangle$ sector. 
				Fitting both sectors simultaneously up to a relative scale factor, our approach uses 6 parameters: 5 coefficients $\beta_i$ in Eq.~\eqref{eq:lincomb} for conserved quantities corresponding to the 5 distinct operators of $\Delta \leq 4$ available in $\rm{SU}(2)_2$ (see Table \ref{table:operators}), plus a scale factor, which corresponds to the relative scale of the two sectors. The scales of the vertical entanglement energy axes are normalized such that $\beta (2\pi/\ell) = 1$ [$\beta (2\pi/\ell)$ being the coefficient of $\tilde{H}_L$ in Eq.~\eqref{eq:lincomb-SizeDependence}], with the zero point appropriate to the conformal weight of the primary state for each sector.
			}
			\label{fig:chen5opsfit}
		\end{figure}

		In the second fit, depicted in Fig.~\ref{fig:chen5opsfit}, we consider the first three descendant levels of both (in Fig.~\ref{fig:chen5opsfit_integer}) the $|j=0\rangle$ and (in Fig.~\ref{fig:chen5opsfit_halfinteger}) the $|j=1/2\rangle$ sectors in tandem. We employ the conserved integrals of all integer and half-integer dimensional operators in the $\rm{SU}(2)_2$ theory (all three columns of Table \ref{table:operators}) up to $\Delta = 4$ to fit the numerical spectra. The 2 half-integer dimensional operators will contribute to the fit in the $|j=1/2\rangle$ sector only, while the 3 integer dimensional operators will contribute in both sectors. We thus fit the 4 differences between the multiplets of the first three levels of the $|j=0\rangle$ sector and the 6 differences between the multiplets of the first three levels of the $|j=1/2\rangle$ sector, for a total of 10 differences simultaneously, with the 6 parameters available by considering the coefficients of conserved integrals of the 5 operators up to dimension $\Delta = 4$ in Table \ref{table:operators}, plus an additional relative scale factor between the two sectors. The fit is quite good in both sectors and correctly captures the relative positioning of multiplets within each level of both sectors. 
		
\begin{figure}[H]
			\centering
			\subfigure[]{\label{fig:chenallopsfit_integer} \includegraphics[width=.45\textwidth]{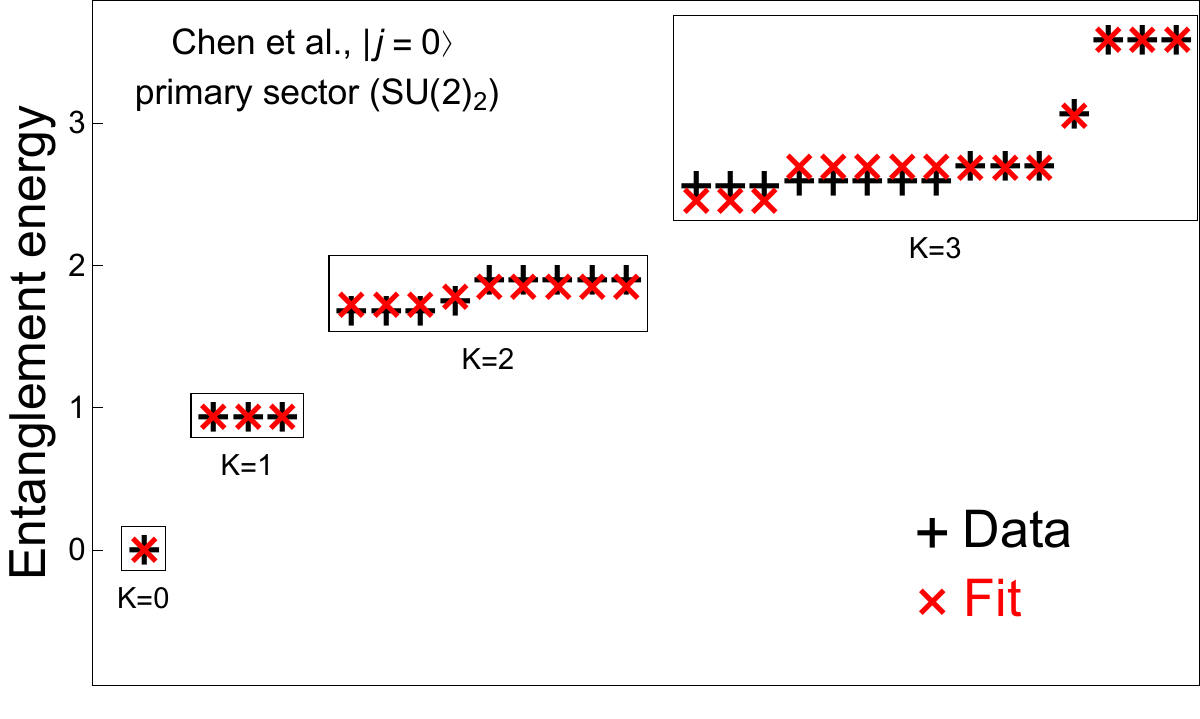}}
			\subfigure[]{\label{fig:chenallopsfit_halfinteger} \includegraphics[width=.45\textwidth]{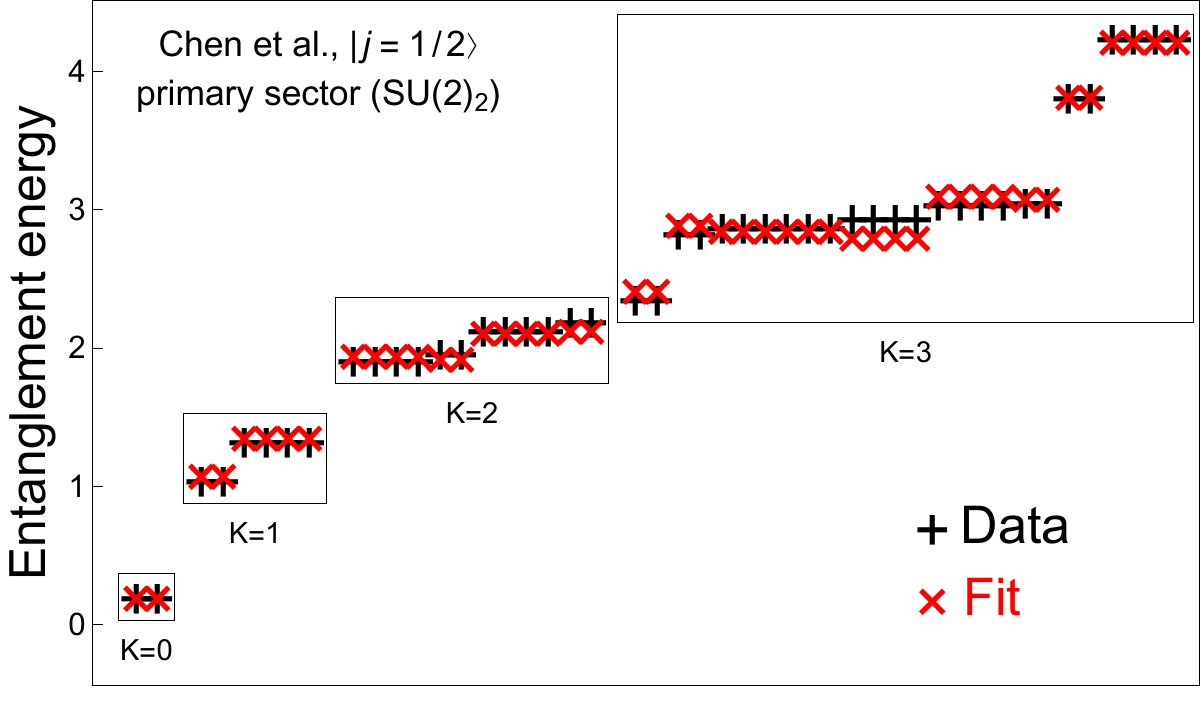}}
			\caption{
				A fit to the $\rm{SU}(2)_2$ entanglement spectrum data of Chen et al.~(Ref.~\protect\onlinecite{Chen2018}) is shown for the first four levels (up to $K=3$) in (a) the sector of the $|j=0\rangle$ primary state and descendants and (b) the sector of the $|j=1/2\rangle$ primary state and descendants. The original data is indicated by black +'s, while red $\times$'s mark the fit produced by our approach. The black boxes indicate states with the same descendant level $K$ above the corresponding primary state, while $\rm{SU}(2)$ multiplets are grouped within each box. (The multiplet content of each box may be compared to Table \ref{table:su22countings} in Appendix \ref{app:su2kintro}.) We attempt to fit 9 differences in the $|j=0\rangle$ sector and 14 differences in the $|j=1/2\rangle$ sector.
				Fitting both sectors simultaneously up to a relative scale factor, our approach uses 12 parameters: 11 coefficients $\beta_i$ in Eq.~\eqref{eq:lincomb} for included conserved quantities corresponding to the 11 distinct operators of $\Delta \leq 6$ available in $\rm{SU}(2)_2$ (see Table \ref{table:operators}), plus a scale factor, which corresponds to the relative scale of the two sectors. The scales of the vertical entanglement energy axes are normalized such that $\beta (2\pi/\ell) = 1$ [$\beta (2\pi/\ell)$ being the coefficient of $\tilde{H}_L$ in Eq.~\eqref{eq:lincomb-SizeDependence}], with the zero point appropriate to the conformal weight of the primary state for each sector.
			}
			\label{fig:chenallopsfit}
		\end{figure}

		In the third and final fit, found in Fig.~\ref{fig:chenallopsfit}, we consider a simultaneous fit of the first four descendant levels of both (in Fig.~\ref{fig:chenallopsfit_integer}) the $|j=0\rangle$ and (in Fig.~\ref{fig:chenallopsfit_halfinteger}) the $|j=1/2\rangle$ sector with all 11 unshaded operators of Table \ref{table:operators} (up to $\Delta = 6$). As before, the half-integer dimensional operators, of which there are now 4, will contribute in the $|j=1/2\rangle$ sector only, while the 7 integer dimensional operators will contribute to the fit in both sectors. This fit attempts to account for 9 differences between multiplets in the $|j=0\rangle$ sector and 14 differences between multiplets in the $|j=1/2\rangle$ sector, for a total of 23 differences between multiplets, using the 12 parameters available by considering the coefficients of conserved integrals of the 11 included operators up to dimension $\Delta = 6$ in Table \ref{table:operators}, plus an additional relative scale factor between the two sectors. The resulting fit is reasonable, especially in the lower levels. In each of the sectors, though, there are instances where the relative ordering of the multiplets has been changed in a few places in the higher levels.
		
		That this occurs despite the use of all 11 operators with $\Delta \leq 6$ indicates that, at higher entanglement energies, the numerical entanglement spectrum for the particular system of Chen et al.~begins to differ slightly from our expectations for a topological state with an entanglement spectrum exhibiting the chiral $\rm{SU}(2)_2$ WZW CFT.
		(This may be a consequence of numerical limitations on the accuracy of the data at higher entanglement energies.)
		We did also perform a slight modification of this last fit that included the integral of $(T\partial G)(x)$ (the operator from Table \ref{table:operators} otherwise excluded due to the $\mathcal{RT}$ symmetry). The inclusion of this additional integral did not result in a significant improvement of the quality of the fit, consistent with the fact that the integral ought to be excluded. Taken together, the results of the three fits in this subsection do demonstrate the necessity of including the fractional conserved quantities, and in particular $G_0$, in the chiral $\rm{SU}(2)_2$ entanglement spectrum.
		
		\newpage
		
		
	\section{Conclusion and Outlook}
	\label{sec:conclusion}
		
		The results of Sec.~\ref{sec:results} demonstrate the success of our approach to quantitatively understanding splittings in low-lying numerical entanglement spectra entirely within the framework of CFT, and thereby further support the chiral nature of the considered quantum states.
		We also note that such a success of the fits applies even more so to the chiral topological PEPS data of Hackenbroich et al.~(Fig.~\ref{fig:hackenbroichfit}). The results for the PEPS data from Chen et al., as well, are close to what we expect (Figs.~\ref{fig:chen5opsfit} and \ref{fig:chenallopsfit}), and these, as well as the results shown in Fig.~\ref{fig:chenmmfit}, clearly illustrate the necessity of including the integrals of the {\it half-integer dimensional} operators of Table \ref{table:operators}. 
		Because we are able to reproduce the splittings of the respective entanglement spectra at low energies, we are able to confirm, with more confidence than based on the characteristic Li-Haldane countings of the chiral $\rm{SU}(2)_1$ or $\rm{SU}(2)_2$ WZW CFT alone, that the PEPS entanglement spectra of Hackenbroich et al.~and Chen et al., as well as the non-PEPS entanglement spectra we consider from Bauer et al.~and Hickey et al., reflect the presence of an underlying (2+1)-dimensional chiral topological theory.
		Where the PEPS we considered are concerned, our analysis thus provides substantial support to the claim of the chiral topological nature of the corresponding PEPS wavefunctions. As mentioned, our analysis and approach also show the ability to help determine whether a ``no-go theorem" holds for interacting topological PEPS.
	
		As numerical methods for calculating the entanglement spectra of chiral topological states develop further, more data will become amenable to analysis by the approach developed in this work. 
		The next target could perhaps be the chiral $\rm{SU}(2)_3$ spin-liquid state\cite{Read1999}, known to possess non-Abelian anyonic excitations capable of supporting universal quantum computation\cite{Nayak2008,Wang2010}. This system will require the understanding of generalizations of the conserved quantities in this work, including those of half-integer dimension introduced in the context of $\rm{SU}(2)_2$. The CFT of $\rm{SU}(2)_3$, however, is known to possess fractional conservation laws\cite{AhnBernardLeClairNPB1990} (also referred to as ``fractional supersymmetry'') which are a direct generalization of the (actual, nonfractional) $N=1$ supersymmetry used in the present work to handle the $\rm{SU}(2)_2$ case. Based on the present work, it is presumably to be expected that these more unusual fractional conservation laws will be needed to explain the splittings in the entanglement spectrum of the chiral topological $\rm{SU}(2)_3$ spin-liquid. We leave the discussion of this to future work. 
		Beyond that, it may also become worthwhile to extend our approach to more general $\rm{SU}(N)_k$ spin-liquids such as those with $N > 2$, or $\rm{SU}(2)_k$ spin-liquids with $k > 3$. Most such extensions will also require understanding generalizations of the conserved quantities discussed here, including those of fractional dimension, and potentially again including those found in Ref.~\onlinecite{AhnBernardLeClairNPB1990}.

		\begin{acknowledgments}
    		We thank Bela Bauer for useful discussions regarding the application of our method to chiral PEPS, as well as Norbert Schuch and Didier Poilblanc for interesting discussions regarding the spectra.
    		We thank the authors of Refs.~\onlinecite{Hickey2016,Hackenbroich2018,Chen2018} for permitting us to use the relevant numerical entanglement spectra presented in their work for the analysis of those spectra performed in this paper.
    		This work was supported in part by the National Science Foundation under Grant No.~DMR-1309667 (AWWL).\hphantom{\cite{null}}
		\end{acknowledgments}
		
	\appendix
	
	\section{Brief Review of Properties of $\rm{SU}(2)_k$}
	\label{app:su2kintro}
	    Here we expand somewhat on the structure of the chiral $\rm{SU}(2)_k$ WZW theory that we first discussed in Sec.~\ref{sec:wzwreview}. We wish to understand the structure of the chiral $\rm{SU}(2)_k$ WZW Hilbert space in more detail. That Hilbert space is built up from the primary states. The $k+1$ primaries of the $\rm{SU}(2)_k$ theory can be thought of as $\rm{SU}(2)$ multiplets, so we can write down states of each $|j=i/2\rangle$ primary (where $i = 0,\ldots,k$) as states $|j,j^z\rangle$ that correspond to the $2j+1$ individual states with a particular $j^z$ within the primary spin-$j$ multiplet. These states of the primary spin-$j$ multiplet will all share the same conformal weight given by \begin{equation}
	        h_j = \frac{j(j+1)}{k+2}.
	    \end{equation}

	    In the WZW theory, we can go beyond the modes $L_n$ of $T(x)$ defined in Eq.~\eqref{eq:lmodeexpansion} and define modes $J^a_{-n}$ of the (``affine'') $\rm{SU}(2)$ Noether current $J^a(x)$ as well, writing
    	\begin{equation}
    		\label{eq:jmodeexpansion}
    		J^a(x) = \frac{2\pi}{\ell}\sum_{n=-\infty}^\infty J^a_{-n}e^{2\pi i n x/\ell}.
    	\end{equation} These modes can be used to build up the chiral $\rm{SU}(2)_k$ Hilbert space from the $k+1$ primary multiplets. Descendant states can then be written down, of the form 
	    \begin{equation}
    	\label{eq:jsnonabelian}
    	    J^{a_1}_{-n_1}\cdots J^{a_m}_{-n_m}|j,j^z\rangle.
    	\end{equation}
	    Such a state will have descendant level $K = \sum_{i=1}^m n_m$, and the spin-$j$ primary of which it is a descendant will have conformal weight $h = h_j$. With these values of $h$ and $K$, the expression of Eq.~\eqref{eq:jsnonabelian} then provides a more concrete realization of the state $|h,K\rangle$ discussed in Eq.~\eqref{eq:hkstate}. Note that such a realization is not unique, leading to the degeneracies of momentum and energy (considered as eigenvalues of $k_L$ and $H_L$ from Sec.~\ref{sec:wzwreview}) present at each descendant level in each primary sector of the $\rm{SU}(2)_k$ theory.
	    Even the states of Eq.~\eqref{eq:jsnonabelian} ought not to be considered distinct in general.\cite{Knizhnik1984}

	    The distinct such states can be organized into $\rm{SU}(2)$ representations, or multiplets, of various dimensions. The $\rm{SU}(2)$ multiplet content in the cases of the chiral $\rm{SU}(2)_1$ and $\rm{SU}(2)_2$ WZW CFTs can be calculated\cite{Kass1990} and is given up to $K = 4$ in Tables \ref{table:su21countings} and \ref{table:su22countings}, respectively. These multiplets are preserved despite the spectral splittings by $\rm{SU}(2)$-invariant conservation laws studied in this work. The countings of the ``Multiplet content" column of Table \ref{table:su21countings} can be observed in low-lying entanglement spectra from the studies of chiral $\rm{SU}(2)_1$ spin liquids (Refs. \onlinecite{Bauer2014,Hickey2016,Hackenbroich2018}) in Figs.~\ref{fig:baueretalplot}--\ref{fig:hackenbroichfit}. The multiplets are written as representations of $\rm{SU}(2)$, with the dimension $d=(2j+1)$ of the spin-$j$ representation shown in bold. For example, the $\bm{1}$ representation is the singlet ($j=0$). 
	    The ``\# at $j^z = 0$" and ``\# at $j^z = +1/2$" columns simply describe the number of multiplets at each descendant level $K$, since every multiplet, even singlets (in the $|j=0\rangle$ primary sector, wherein the multiplets have integer spin) or doublets (in the $|j=1/2\rangle$ primary sector, wherein the multiplets have half-integer spin), has a single state at that value of $j^z$. In Fig.~\ref{fig:baueretalplot} this can be seen by looking at the central subtowers, which exhibit the 1-1-2-3-5 degeneracy in momentum, consistent with these columns for Table \ref{table:su21countings}. In Figs.~\ref{fig:bauerfit}--\ref{fig:hackenbroichfit}, these numbers are simply the number of multiplets depicted in each box of level $K$. Since we fit the entanglement energies of the multiplets, these are also the numbers of data points involved in the fit at each level $K$.

    	We have written down the same columns for the chiral $\rm{SU}(2)_2$ WZW CFT in Table \ref{table:su22countings}. The countings of the ``Multiplet content" column of Table \ref{table:su22countings} can be observed in low-lying entanglement spectra from the study of the chiral $\rm{SU}(2)_2$ spin liquid (Ref.~\onlinecite{Chen2018}) in Figs.~\ref{fig:chenmmfit}--\ref{fig:chenallopsfit}, at least for the $|j=0\rangle$ and $|j=1/2\rangle$ primary sectors found in these spectra. The countings of the $|j=1\rangle$ sector are also included in Table \ref{table:su22countings}.
    	The data of the ``\# at $j^z = 0$" and ``\# at $j^z = +1/2$" columns (for the $|j=0\rangle$ and $|j=1/2\rangle$ primary sectors) is represented in the number of multiplets depicted in each box of level $K$ in Figs.~\ref{fig:chenmmfit}--\ref{fig:chenallopsfit}, and again these data represent the numbers of data points involved in the fit at each level $K$. 
	    \begin{table}[hbt]
		    \centering
		    \begin{tabular}{c | c | c | c | c}
		        \multirow{2}{*}{$K$} & \multicolumn{2}{c|}{$|j=0\rangle$ primary sector} & \multicolumn{2}{c}{$|j=1/2\rangle$ primary sector} \\
		        & Multiplet content & \# at $j^z = 0$ & Multiplet content & \# at $j^z = +1/2$\\
		        \hline
                0 & $\bm{1}$ & 1 & $\bm{2}$ & 1 \\
                 1 & $\bm{3}$ & 1 & $\bm{2}$ & 1 \\
                 2 & $\bm{1}+\bm{3}$ & 2 & $\bm{2}+\bm{4}$ & 2 \\
                 3 & $\bm{1}+2(\bm{3})$ & 3 & $2(\bm{2})+\bm{4}$ & 3 \\
                 4 & $2(\bm{1})+2(\bm{3})+\bm{5}$ & 5 & $3(\bm{2})+2(\bm{4})$ & 5 \\
		    \end{tabular}
		    \caption{
			    $\rm{SU}(2)$ multiplet content of the chiral $\rm{SU}(2)_1$ WZW CFT, in both the $|j=0\rangle$ and $|j=1/2\rangle$ primary sectors, listed by descendant level $K$. The multiplets are written as representations of $\rm{SU}(2)$, with the dimension 
			    $d=(2j+1)$
			    of the spin-$j$ representation shown in bold. For example the primary state multiplets are shown at $\Delta = 0$ as $\bm{1}$ and $\bm{2}$ (i.e. singlet and doublet), respectively. The ``\#" columns indicate the total number of multiplets at each level, which corresponds to the number of states at the indicated central $j^z$ value.
		    }
		    \label{table:su21countings}
	    \end{table} 
	    \begin{table}[hbt]
		    \centering
		    \begin{tabular}{c | c | c | c | c | c | c}
		        \multirow{2}{*}{$K$} & \multicolumn{2}{c|}{$|j=0\rangle$ primary sector} & \multicolumn{2}{c|}{$|j=1/2\rangle$ primary sector} & \multicolumn{2}{c}{$|j=1\rangle$ primary sector} \\
		        & Multiplet content & \# at $j^z = 0$ & Multiplet content & \# at $j^z = +1/2$ & Multiplet content & \# at $j^z = 0$ \\
		        \hline
                0 & $\bm{1}$ & 1 & $\bm{2}$ & 1 & $\bm{3}$ & 1 \\
                1 & $\bm{3}$ & 1 & $\bm{2}+\bm{4}$ & 2 & $\bm{1}+\bm{3}$ & 2 \\
                2 & $\bm{1}+\bm{3}+\bm{5}$ & 3 & $2(\bm{2})+2(\bm{4})$ & 4 & $\bm{1}+2(\bm{3})+\bm{5}$ & 4 \\
                3 & $\bm{1}+3(\bm{3})+\bm{5}$ & 5 & $4(\bm{2})+3(\bm{4})+\bm{6}$ & 8 & $2(\bm{1})+3(\bm{3})+2(\bm{5})$ & 7 \\
                4 & $3(\bm{1})+4(\bm{3})+3(\bm{5})$ & 10 & $6(\bm{2})+6(\bm{4})+2(\bm{6})$ & 14 & $3(\bm{1})+6(\bm{3})+3(\bm{5})+\bm{7}$ & 13 \\
		    \end{tabular}
		    \caption{
			    $\rm{SU}(2)$ multiplet content of the chiral $\rm{SU}(2)_2$ WZW CFT, in the $|j=0\rangle$, $|j=1/2\rangle$, and $|j=1\rangle$ primary sectors, listed by descendant level $K$. The multiplets are written as representations of $\rm{SU}(2)$, with the dimension $d=(2j+1)$
			    of the spin-$j$ representation shown in bold. For example the primary state multiplets are shown at $\Delta = 0$ as $\bm{1}$, $\bm{2}$, and $\bm{3}$ (i.e. singlet, doublet, and triplet), respectively. The ``\#" columns indicate the total number of multiplets at each level, which corresponds to the number of states at the indicated central $j^z$ value.
		    }
		    \label{table:su22countings}
	    \end{table}

	\section{Calculation of the Conserved Quantities}
	\label{app:details}
		
		To compute the values of the conserved quantities we will use in the GGE, we first write them in terms of the modes $L_n$ of $T(x)$ defined by Eq.~\eqref{eq:lmodeexpansion}, and the modes $G_n$ defined by a similar mode expansion for the superconformal currents $G(x)$ of Eq.~\eqref{eq:gdef}:
		\begin{equation}
			\label{eq:gmodeexpansion}
			G(x) = \left(\frac{2\pi}{\ell}\right)^{3/2}\sum_{m=-\infty}^\infty G_{-m}e^{2\pi imx/\ell},
		\end{equation}
		where we will choose $m \in \mathbb{Z}+1/2$ if the conserved quantity is to be used in the Neveu-Schwarz sector (the sector of descendants of the $|j=0\rangle$ and $|j=1\rangle$ primary states), and $m \in \mathbb{Z}$ if the conserved quantity is to be used in the Ramond sector (the sector of descendants of the $|j=1/2\rangle$ primary state).
		The expressions we get are collected in Table \ref{table:modereps}. For completeness, the table includes $\tilde{H}^{(5)}$, which is excluded from the entanglement Hamiltonian by $\mathcal{RT}$ symmetry (indicated by the shading of the row), though we take the associated coefficient $\beta_5 = 0$ in the expression of the entanglement Hamiltonian as a linear combination of conserved quantities in Eq.~\eqref{eq:lincomb}.
		
		\begin{table}[hbt]
			\centering
			\begin{tabular}{c | c | c | c}
				$i$ & $\Delta_i$ & $\Phi_i(x)$ & $\tilde{H}^{(i)} = \left(\frac{\ell}{2\pi}\right)^{\Delta_i-1}\frac{1}{2\pi}\int_0^\ell \Phi_i(x) dx$ \\
				\hline
				\hline
				0 & 3/2 & $G(x)$ & $G_0$ \\
				\hline
				1 & 2 & $T(x)$ & $L_0 - \frac{c}{24}$\\
				\hline
				2 & 7/2 & $(TG)(x)$ & $\sum_{n>0} (L_{-n}G_n + G_{-n}L_n) + L_0 G_0 - \frac{c}{24}G_0 $ \\
				\hline
			    3 & 4 & $(TT)(x)$ & $2\sum_{n>0} L_{-n} L_n + L_0^2 - \frac{c}{12}L_0 + \frac{c^2}{576} $  \\
				\hline
				4 & 4 & $i(G\partial G)(x)$ & $2 \sum_{m>0} m G_{-m}G_m$ \\
				\hline
				\rowcolor{Gray}
				5 & 9/2 & $(T\partial G)(x)$ & $i \sum_{n > 0}n (- L_{-n}G_n + G_{-n}L_n)$ \\
				\hline
				\multirow{3}{*}{6} & \multirow{3}{*}{11/2} & \multirow{3}{*}{$(G(TT))(x)$} & $\sum_{{n_1},n_2\leq 0}G_{n_2} L_{n_1} L_{-{n_2}-{n_1}} + \sum_{{n_1}>0,{n_2}\leq 0}G_{n_2} L_{-{n_2}-{n_1}}L_{n_1} $\\
				& & & $+\sum_{{n_1}\leq 0, {n_2}> 0}L_{n_1} L_{-{n_2}-{n_1}}G_{n_2} + \sum_{{n_1},{n_2}>0} L_{-{n_2}-{n_1}}L_{n_1} G_{n_2}$ \\
				& & & $-\frac{c}{12}\sum_{n> 0}(G_{-n} L_n + L_{-n}G_n) - \frac{c}{12}G_0 L_0 + \frac{c^2}{576}G_0$\\
				\hline
				7 & 11/2 & $(\partial T\partial G)(x)$ & $\sum_{n>0} n^2(L_{-n}G_n + G_{-n}L_n)$ \\
				\hline
				\multirow{2}{*}{8} & \multirow{2}{*}{6} & \multirow{2}{*}{$(T(TT))(x)$} & $\sum_{n_1+n_2+n_3=0}:L_{n_1} L_{n_2} L_{n_3}: + \frac{3}{2}\sum_{n>0} n^2 L_{-n} L_n + \frac{3}{2}\sum_{n>0} L_{1-2n} L_{2n-1}$ \\
				& & & $-\frac{c}{4}\sum_{n>0} L_{-n}L_n - \frac{c}{8}L_0^2 + \frac{c^2}{192}L_0 -\frac{c^3}{13824}$ \\
				\hline
				9 & 6 & $(\partial T \partial T)(x)$ & $2 \sum_{n>0} n^2 L_{-n} L_n$ \\
				\hline 
				\multirow{3}{*}{10} & \multirow{3}{*}{6} & \multirow{3}{*}{$i(T(G\partial G))(x)$} & $-\sum_{m,n\leq 0}(n+m)L_n G_m G_{-n-m} + \sum_{m>0,n\leq 0}(n+m)L_n G_{-n-m}G_m $\\
				& & & $-\sum_{m\leq 0, n> 0}(n+m)G_m G_{-n-m}L_n + \sum_{m,n>0} (n+m) G_{-n-m}G_m L_n$ \\
				& & & $-\frac{c}{12}\sum_{m> 0} m G_{-m} G_m$\\
				\hline
				11 & 6 & $i(\partial G \partial^2 G)(x)$ & $2 \sum_{m>0} m^3 G_{-m}G_m$ \\
				\hline
			\end{tabular}
			\caption{
				Expressions for the size-independent parts $\tilde{H}^{(i)}$ of the corresponding locally conserved quantities $H^{(i)}$ in terms of the Fourier modes $L_n$ and $G_n$ of the energy-momentum tensor $T(x)$ and the superconformal current $G(x)$, respectively. 
				The shaded row indicates that we will exclude $\tilde{H}^{(5)}$ from fits on the basis of the $\mathcal{RT}$ symmetry. The index\footnote{The index $i$ should not be confused with the imaginary number $i$ found in some entries.} 
				$i$ denotes the quantity the parameter $\beta_i$ refers to, useful for comparison to Tables \ref{table:su21fitparams} and \ref{table:su22fitparams}. 
				$\Delta_i$ indicates the conformal dimension of the operator $\Phi_i(x)$, which is integrated to give $H^{(i)}$. 
				The symbols $::$ in the $i = 6$ row indicate normal ordering by increasing subscripts $n_1,n_2,n_3$. 
				$c$ is the central charge: $c = 1$ for $\rm{SU}(2)_1$, while $c = 3/2$ for $\rm{SU}(2)_2$. 
				Note that while $\tilde{H}^{(i)}$ with associated half-integer $\Delta_i$ occur only in the Ramond ($|j = 1/2\rangle$ primary) sector, leading to modes $G_m$ with integer $m$, the $\tilde{H}^{(i)}$ with associated integer $\Delta_i$ can occur in both the Neveu-Schwarz ($|j = 0\rangle$ and $|j = 1\rangle$ primary) and Ramond sectors, leading to modes $G_m$ with half-integer $m$ in the Neveu-Schwarz sector and integer $m$ in the Ramond sector. Indices that are always integers have been denoted by $n$ (or $n_1$, etc.) above, while indices that vary between integers and half-integers depending on the sector have been denoted by $m$.
			}
			\label{table:modereps}
		\end{table}

		To find the spectral levels with our approach, we diagonalize the expression Eq.~\eqref{eq:lincomb} using these mode representations.
		\footnote{Strictly, we ignore the constant terms in the expressions of Table \ref{table:modereps}, because each will shift every state in the spectrum by the exact same amount, and therefore none of these terms will affect the splittings.}
		This requires finding a basis to represent the descendant states in each level of the conformal tower. 
		For the $\rm{SU}(2)_1$ case, in particular, where we can make use of Abelian bosonization, we can then represent the central $j^z = 0$ or $j^z = +1/2$ (depending on the sector) state of every descendant multiplet by a unique linear combination of states with the form
		\begin{equation}
			\label{eq:abelianbasis}
			J^3_{-n_1}\cdots J^3_{-n_\ell}\left|j,+j\right\rangle
		\end{equation}
		where $|j,+j\rangle$ is the highest-weight state in the $|j=0\rangle$ or $|j=1/2\rangle$ primary $\rm{SU}(2)$ multiplets, for some choice of positive integers $n_1 \leq \ldots \leq n_\ell$.\cite{DiFrancesco1997}
		The eigenvalue of $L_0$ on the state Eq.~\eqref{eq:abelianbasis} will be equal to $h_j+K$, where $h_j$ is the eigenvalue of $L_0$ on the primary state $|j,+j\rangle$. ($h_{j=0} = 0$, while $h_{j=1/2} = 1/4$.) The descendant level $K$ of the state is then given by $K = \sum_{i=1}^\ell n_i$.
		We then use the Virasoro and affine Lie commutation relations along with the Sugawara form for $L_n$ [the mode-expanded form of Eq.~\eqref{eq:sugawara}] to evaluate the mode expressions of Table \ref{table:modereps} corresponding to the integrals of the operators of the leftmost column of Table \ref{table:operators} on the basis of states Eq.~\eqref{eq:abelianbasis}. This is possible because these mode expressions $\tilde{H}^{(i)}$ are exactly those in Table \ref{table:modereps} that include only Virasoro modes $L_n$. The primary states of the chiral $\rm{SU}(2)_1$ WZW CFT are also Virasoro primary, in the sense that $L_n|j,+j\rangle = 0$ for all $n > 0$, which simplifies this process greatly. Diagonalizing an arbitrary linear combination of the $\tilde{H}^{(i)}$ evaluated in that basis, we obtain an expression for the splittings of the $\rm{SU}(2)_1$ conformal tower that we can fit to entanglement spectra. 
		
		It remains to identify each eigenvalue of the combined operator with a particular dimension of multiplet in the conformal tower. In the Abelian case, we take advantage of the symmetry of the conformal tower, and the fact that the mode expressions we use are expressed solely in terms of Virasoro modes. It turns out that in the $\rm{SU}(2)_1$ theory, in addition to the WZW primary states, the lowest-level descendant $\rm{SU}(2)$ multiplet of a given dimension is also a Virasoro primary state.
		\footnote{We can see this by considering the effect of acting with any Virasoro mode $L_n$ for $n > 0$ on a state $|\phi\rangle$ of the multiplet. Because the mode $L_n$ is a singlet under $\rm{SU}(2)$, $L_n|\phi\rangle$ must be a state with the same $j^z$ quantum number in a multiplet of the same dimension, but at a lower level. As there is no such state, it must be the case that $L_n|\phi\rangle = 0$. In the $\rm{SU}(2)_1$ theory, the 1-1-2-3-5 counting of states in the subtowers of the highest-weight states of these Virasoro primary multiplets and their descendants also guarantees the uniqueness of such multiplets within the descendant levels in which they appear (see, e.g., Fig.~\ref{fig:baueretalplot}.)}
		We can then write down an additional basis similar to Eq.~\eqref{eq:abelianbasis} for the states of descendant multiplets that have $j^z = j+i$, for an integer $i \geq 0$: 
		\begin{equation}
			\label{eq:szabelianbasis}
			J^3_{-n_1}\cdots J^3_{-n_\ell}\left|j+i,j+i \right\rangle,
		\end{equation}
		where the notation $|j+i,j+i\rangle$ now denotes the $\rm{SU}(2)$ highest-weight state in the lowest-level descendant spin-$(j+i)$ multiplet.
		Because the $|j+i,j+i\rangle$ states are Virasoro primary, acting on the elements of this basis with the mode expressions of the conserved quantities we need to evaluate in the $\rm{SU}(2)_1$ theory (composed solely of Virasoro modes) does not require knowledge of the explicit form of the $|j+i,j+i\rangle$ state in terms of the modes of the affine $\rm{SU}(2)$ current acting on the underlying WZW primary state $|j\rangle$.
		Each basis Eq.~\eqref{eq:szabelianbasis} at fixed $i$ spans a section of the full $\rm{SU}(2)_1$ conformal tower with fixed $j^z = j+i$. For $i = 0$, $j^z = j$, so the basis Eq.~\eqref{eq:szabelianbasis} is simply the original Abelian basis Eq.~\eqref{eq:abelianbasis}. The number of such bases at fixed $i$ in which an eigenvalue of the conserved quantities occurs then determines the dimension of the $\rm{SU}(2)$ multiplet associated to that eigenvalue: if the eigenvalue occurs in $s$ such bases in addition to Eq.~\eqref{eq:abelianbasis}, the multiplet is a spin-$(j+s)$ multiplet.
		
		For the calculation in the $\rm{SU}(2)_2$ theory, we have to use non-Abelian bosonization, and so we can no longer use the Abelian basis Eq.~\eqref{eq:abelianbasis}. Instead of building a similar non-Abelian basis from the modes $J^a_{-n}$, we take the 3-fermion theory point of view, though instead making use of the $\phi^a(x)$ of Eq.~\eqref{eq:jphi}. We can define modes $\phi^a_{-m}$ by
		\begin{equation}
			\label{eq:phimodeexpansion}
			\phi^a(x) = \left(\frac{2\pi}{\ell}\right)^{1/2}\sum_{m=-\infty}^\infty \phi^a_{-m}e^{2\pi i m x/\ell}.
		\end{equation}
		One non-Abelian basis is then 
		\begin{equation}
			\label{eq:nonabelianbasis}
			\phi^{a_1}_{-m_1}\cdots \phi^{a_\ell}_{-m_\ell}\left|\sigma\right\rangle,
		\end{equation}
		where $\sigma \in \{\text{NS},\text{R}\}$ corresponds to the ground state of the Neveu-Schwarz or Ramond sectors, respectively. If $\sigma=\text{NS}$, $m_i \in \mathbb{Z}+1/2$, whereas if $\sigma=\text{R}$, $m_i \in \mathbb{Z}$. In either case, $a_i \in \{1,2,3\}$.
		The Neveu-Schwarz sector of this theory corresponds to the $|j=0\rangle$ and $|j=1\rangle$ primary sectors of the $\rm{SU}(2)_2$ WZW CFT: states with an even number of $\phi^a_{-m}$ represent the descendants (by action of the affine current algebra) of the $|j=0\rangle$ primary state, while states with an odd number of $\phi^a_{-m}$ represent the descendants of the $|j=1\rangle$ primary state. The Ramond sector corresponds to the descendants of the $|j=1/2\rangle$ primary state. 
		The eigenvalue of $L_0$ on a state in the basis Eq.~\eqref{eq:nonabelianbasis} is given by $h_\sigma + \sum_{i=1}^\ell m_i$, where $h_{\sigma} = h_{\sigma=\text{NS}} = 0$ for states in the Neveu-Schwarz sector, and $h_\sigma = h_{\sigma=\text{R}} = 3/16$ for states in the Ramond sector. The descendant level $K$ of that state is then found by subtracting off the eigenvalue $h_j$ of $L_0$ of the corresponding primary state $|j\rangle$. For states in the $|j=0\rangle$ and $|j=1/2\rangle$ primary state sectors, this is simply $K = \sum_{i=1}^\ell m_i$, since $h_{j=0} = h_{\sigma=\text{NS}}$ and $h_{j=1/2} = h_{\sigma=\text{R}}$. For the $|j=1\rangle$ primary state sector, we have $h_{j=1} = 1/2 = h_{\text{NS}}+1/2$, so the level $K$ of the descendant state is given by $K = \sum_{i=1}^\ell m_i-1/2$.

		To obtain a basis that more directly reflects the multiplet structure of the theory, however, we diagonalize the operator $J^3_0$ [which can be expressed in terms of $\phi^a_{m}$ modes by the relation Eq.~\eqref{eq:jphi}] on the basis Eq.~\eqref{eq:nonabelianbasis} and use those eigenstates as the basis. The eigenvalues of $J^3_0$ can be thought of as $j^z$ quantum numbers. We can then diagonalize linear combinations Eq.~\eqref{eq:lincomb} of the conserved quantities corresponding to the mode representations of Table \ref{table:modereps} (which can be algorithmically re-written in terms of the $\phi^a_{m}$ modes) in each of the fixed-$j^z$ sectors formed by that basis. The presence of a given eigenvalue across $2s+1$ fixed-$j^z$ sectors is used to associate that eigenvalue with a spin-$s$ multiplet. 
		\newpage
	\section{$\mathcal{RT}$ Symmetry Mechanism}
	\label{app:rtsymmetry}
	
		\begin{figure}[H]
			\centering
			\includegraphics[]{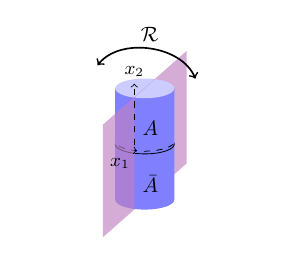}
			\caption{
    			A truncated representation of the infinite cylindrical geometry we consider is shown in blue, along with a diagram of the spatial reflection $\mathcal{R}$. $\mathcal{R}$ reflects the cylinder about the violet plane, mapping regions $A$ and $\bar{A}$ to themselves. The coordinate $x_1$ is the compactified coordinate around the cylinder, while the coordinate $x_2$ denotes spatial position along the cylinder. The entanglement cut lies along the spatial circle $x_2 = 0$.
			}
			\label{fig:cylindergeometry}
		\end{figure}
		
		\subsection{The $\mathcal{RT}$ Symmetry on a Cylinder}
		
		We consider a chiral topological state arranged in the cylindrical geometry of Fig.~\ref{fig:cylindergeometry}. The infinite cylinder is bipartitioned into two regions $A$ and $\bar{A}$, with an entanglement cut between them. The coordinate $x_1$ is the compactified spatial coordinate around the cylinder, which has circumference $\ell$, while the coordinate $x_2$ denotes spatial position along the cylinder. The region $A$ is thus the half-cylinder $x_2 \geq 0$, while the region $\bar{A}$ is the half-cylinder $x_2 < 0$. The entanglement cut lies along $x_2 = 0$. 
		We can then define an orientation-reversing spatial transformation that preserves the position of the entanglement cut:
		\begin{align}
			\mathcal{R} &: (x_1,x_2) \mapsto (-x_1,x_2)
		\end{align}
		$\mathcal{R}$ reflects the cylinder about the violet plane in Fig.~\ref{fig:cylindergeometry}, mapping region $A$ into region $A$ and region $\bar{A}$ into region $\bar{A}$. 
		
		This reflection reverses the orientation of the cylinder, so it inverts the chirality of the chiral topological state, in the sense that, among other things, the associated chiral edge modes, if one were to physically cut the cylinder along the entanglement cut, would reverse their direction of flow. We can restore the system to its original state if we follow spatial reflection with a time-reversal transformation $\mathcal{T}$, which will also reverse the direction of flow of the chiral edge modes. This gives rise to the discrete symmetry we will consider, $\mathcal{R T}$, under which we consider our system to be invariant.
		
		\subsection{The $\mathcal{RT}$ Symmetry in the PEPS}
		
		At the level of the square-lattice PEPS wavefunctions we consider in the $\rm{SU}(2)_1$ (Ref.~\onlinecite{Hackenbroich2018}) and $\rm{SU}(2)_2$ (Ref.~\onlinecite{Chen2018}) cases, we can see the $\mathcal{R T}$ invariance explicitly. The projectors onto the PEPS, which are the building blocks of the PEPS wavefunctions, take the form of either $\mathcal{A}_1 + i\mathcal{A}_2$ or $\mathcal{B}_1 + i\mathcal{B}_2$, where $\mathcal{A}_1$, $\mathcal{A}_2$, $\mathcal{B}_1$, and $\mathcal{B}_2$ are linear combinations of projectors that transform like the $A_1$, $A_2$, $B_1$, and $B_2$ irreducible representations, respectively, under the actions of the elements of $C_{4v}$ point group of the square lattice.\cite{Mambrini2016} The consequence of this is that under the action of $\mathcal{R}$, we have 
		\begin{align}
			\mathcal{A}_1 \mapsto \mathcal{A}_1 & & \mathcal{A}_2 \mapsto -\mathcal{A}_2 & &  \mathcal{B}_1 \mapsto \mathcal{B}_1 & & \mathcal{B}_2 \mapsto -\mathcal{B}_2. 
		\end{align}
		At the same time, $\mathcal{A}_1$, $\mathcal{A}_2$, $\mathcal{B}_1$, and $\mathcal{B}_2$ are all real, and therefore invariant under $\mathcal{T}$. Thus the only effect of the antiunitary $\mathcal{T}$ is to conjugate the $i$ in $\mathcal{A}_1 + i\mathcal{A}_2$ or $\mathcal{B}_1 + i\mathcal{B}_2$:
		\begin{align}
			\mathcal{A}_1 + i\mathcal{A}_2 \mapsto \mathcal{A}_1 - i\mathcal{A}_2 & & \mathcal{B}_1 + i\mathcal{B}_2 \mapsto \mathcal{B}_1 - i\mathcal{B}_2.
		\end{align}
		Thus $\mathcal{A}_1 + i\mathcal{A}_2$ and $\mathcal{B}_1 + i\mathcal{B}_2$ are both invariant under $\mathcal{RT}$.
		
		\subsection{The Action of $\mathcal{RT}$ Symmetry on the Conserved Quantities of the Entanglement Hamiltonian}
		\label{LabelSubsectionRTSymatryConservedQuantities}
		
		We thus see that whether we think of an abstract chiral topological state or the concrete PEPS we are working with, we will have invariance of the overall density matrix $\rho$ under $\mathcal{RT}$.
		Since the $\mathcal{R T}$ symmetry preserves $\rho$ and maps $A \mapsto A$ and $\bar{A} \mapsto \bar{A}$, we can see that the reduced density matrix $\rho_A = \Tr_{\bar{A}} \rho$, which may be compared to Eq.~\eqref{eq:realrhol}, will be preserved under $\mathcal{R T}$. Hence, the entanglement Hamiltonian $H_{\text{entanglement}} = -\ln \rho_A$, Eq.~\eqref{eq:lincomb} in our case, should likewise be invariant under $\mathcal{R T}$. As a consequence, we demand that the $H^{(i)}$ of Eq.~\eqref{eq:lincomb} satisfy 
		\begin{equation}
			\label{eq:hirtcondition}
			(\mathcal{R T}) H^{(i)} (\mathcal{R T})^{-1} = H^{(i)}.
		\end{equation}
		
		We can deduce the action of $\mathcal{R T}$ on the $H^{(i)}$ by considering its action on the modes of the various operators. First, we consider the effect of $\mathcal{R T}$ on the left-moving energy-momentum tensor $T(t,x)$, the energy-momentum tensor of our chiral theory, expressed as a function of the time coordinate $t$ and compact spatial coordinate $x$ ($x_1$ in the notation of Fig.~\ref{fig:cylindergeometry}, since $T(t,x)$ is an operator in the $(1+1)$-dimensional theory along the cut). Conjugating by spatial reflection $\mathcal{R}$ alone, we have that 
		\begin{align}
		    \label{eq:RLop}
			\mathcal{R} T(t,x) \mathcal{R}^{-1} &= \overline{T}(t, \ell - x) = \overline{T}(t,- x) \\
			\mathcal{R} \overline{T}(t,x) \mathcal{R}^{-1} &= {T}(t,\ell - x) = {T}(t,-x),
		\end{align}
		where $\overline{T}(t,x)$ indicates the energy-momentum tensor of the right-moving theory, and we have used the spatial periodicity of $T(t,x)$ and $\overline{T}(t,x)$ around the cylinder in the last equality.
		Likewise, time reversal gives 
	    \begin{align}
    	   \mathcal{T} T(t,x) \mathcal{T}^{-1} &= \overline{T}(-t,x) \\
    	   \mathcal{T} \overline{T}(t,x) \mathcal{T}^{-1} &= {T}( -t,x) .
	    \end{align}
        We can then use the scaling property of the $\Delta = 2$ operator $T(t,x)$ to see that 
        \begin{align}
           \label{eq:RTLop}
           (\mathcal{RT} )T(t,x) (\mathcal{RT})^{-1} &={T}(-t, -x) = (-1)^2T(t,x) = T(t,x) \\
           (\mathcal{RT}) \overline{T}(t,x) (\mathcal{RT})^{-1} &=  \overline{T}(-t,- x) =  (-1)^2\overline{T}(t,x) = \overline{T}(t,x).
        \end{align}
        Thus $T(t,x)$, and hence the $T(x)$ we have considered at fixed time, remains invariant under conjugation by $\mathcal{RT}$. By similar logic, we have for the affine $\rm{SU}(2)$ currents $J^a(t,x)$ that
        \begin{align}
           \label{eq:RTJop}
           (\mathcal{RT} )J^a(t,x) (\mathcal{RT})^{-1} &= J^a(-t, -x) = (-1)^1 J^a(t,x) = -J^a(t,x) \\
           (\mathcal{RT}) \overline{J}^a(t,x) (\mathcal{RT})^{-1} &=  \overline{J}^a(-t,- x) =  (-1)^1\overline{J}^a(t,x) = -\overline{J}^a(t,x),
           \label{eq:RTJbarop}
        \end{align}
        where in the last equalities of Eq.~\eqref{eq:RTJop} and Eq.~\eqref{eq:RTJbarop} we have used the scaling property of the $\Delta = 1$ operator $J^a(t,x)$.

        Now the invariance of $T(x)$ extends to its modes,
        \footnote{Expanding Eq.~\eqref{eq:RTLop} in terms of modes, we obtain $\sum_{n=-\infty}^\infty (\mathcal{R T})L_{n}(\mathcal{R T})^{-1}e^{2\pi inx/\ell}  =  \sum_{n=-\infty}^\infty L_{n}e^{2\pi inx/\ell}$, and since $(\mathcal{R T})e^{2\pi inx/\ell}(\mathcal{R T})^{-1} = e^{2\pi inx/\ell}$, this means that $(\mathcal{R T})L_{n}(\mathcal{R T})^{-1} =L_{n}$.} 
        and so we have 
		\begin{align}
			\label{eq:ltransformation}
			(\mathcal{R T}) L_n (\mathcal{R T})^{-1} &= L_n.
		\end{align}
		From Eq.~\eqref{eq:RTJop} we can likewise conclude that the modes $J^a_n$ of Eq.~\eqref{eq:jmodeexpansion} will obey
		\begin{align}
			\label{eq:jtransformation}
			(\mathcal{R T}) J^a_n (\mathcal{R T})^{-1} &= -J^a_n
		\end{align}
		under conjugation by $\mathcal{RT}$.
	    
        We observe that Eq.~\eqref{eq:ltransformation} and Eq.~\eqref{eq:jtransformation} are consistent with the commutation relations\cite{Knizhnik1984} of the $J^a_n$,
        \begin{equation}
	   	\label{eq:jcommutation}
	   	[J^a_n,J^b_m] = i\epsilon_{abc}J^c_{n+m} + \frac{k}{2}n\delta_{n+m,0},
        \end{equation}
        and the Sugawara relation Eq.~\eqref{eq:sugawara}.
        Eq.~\eqref{eq:sugawara} and Eq.~\eqref{eq:ltransformation} require that we must have one of $(\mathcal{R T}) J^a_n (\mathcal{R T})^{-1} = \pm J^a_n$, but only the minus sign of Eq.~\eqref{eq:jtransformation} is also consistent with Eq.~\eqref{eq:jcommutation}. This is due to the anti-unitarity of $\mathcal{RT}$, which causes $(\mathcal{R T}) i (\mathcal{R T})^{-1}= -i$ to flip the sign of the $i$ upon conjugation of both sides of Eq.~\eqref{eq:jcommutation} by $\mathcal{RT}$. In the $|j=1/2\rangle$ primary sector, $J^a_0$ acts\cite{Knizhnik1984} on the $|j=1/2\rangle$ doublet like $\sigma^a/2$, where $\sigma^a$ is the $a$th Pauli matrix,
        \footnote{The normalization can be found by comparing the commutation relations of the Pauli matrices $\sigma^a$, $[\sigma^a,\sigma^b] = 2i \epsilon_{abc}\sigma^c$, with Eq.~\eqref{eq:jcommutation}.} 
        and we have
        \begin{equation}
            \label{eq:j0action}
            J^a_0 |j =1/2\rangle^\alpha =  \frac{\sigma^a_{\beta \alpha}}{2} |j =1/2\rangle^\beta,
        \end{equation}
        where $|j =1/2\rangle^\alpha$, for $\alpha = \pm$, indicates the $|1/2,\pm 1/2\rangle$ state [in the notation of Eq.~\eqref{eq:abelianbasis}] within the $|j=1/2\rangle$ doublet. 
	    
	    We now examine the effect of $\mathcal{RT}$ on the $\phi^a_0$ modes. 
        Group theory guarantees that the $\phi^a_0$ modes act\cite{Knizhnik1984} on the $|j=1/2\rangle$ primary state as $\sigma^a/\sqrt{2}$.
        \footnote{The $\phi^a_n$ modes must satisfy the anti-commutation relations $\{\phi^a_n,\phi^a_m\} = \delta_{n+m,0}.$ The normalization of the action of the $\phi^a_0$ modes on the $|j=1/2\rangle$ primary state may be found by comparison with the Pauli anti-commutation relations $\{\sigma^a,\sigma^b\} = 2\delta_{ab}I$.}
        Up to normalization, this is the same as Eq.~\eqref{eq:j0action}, and so we see that we must also have 
		\begin{align}
		    \label{eq:phi0transformation}
			(\mathcal{R T}) \phi^a_0 (\mathcal{R T})^{-1} &= -\phi^a_0,
		\end{align}
		since the $J^a_0$ obey Eq.~\eqref{eq:jtransformation}. 
	    The $\phi^a_n$ satisfy commutation relations with the modes $L_n$ of the energy-momentum tensor,
	    \begin{equation}
	        \label{eq:lphicommutation}
	    [L_n,\phi^a_m] = -\left(\frac{n}{2}+m\right)\phi^a_{n+m}.
	    \end{equation}
	    When $m = 0$, this becomes
	    \begin{equation}
	        \label{eq:lphi0commutation}
	        [L_n,\phi^a_0] = -\frac{n}{2}\phi^a_{n}.
	    \end{equation}
	    From Eq.~\eqref{eq:ltransformation} and Eq.~\eqref{eq:phi0transformation}, we know how the left hand side of Eq.~\eqref{eq:lphi0commutation} transforms, and therefore it follows that for integer indices $n$,
		\begin{align}
			\label{eq:phiztransformation}
			(\mathcal{R T}) \phi^a_n (\mathcal{R T})^{-1} &= -\phi^a_n \text{, }(n \in \mathbb{Z}).
		\end{align} 
	    Note that the derivation of Eq.~\eqref{eq:phiztransformation} relies upon $n \in \mathbb{Z}$, i.e.,~that we are in the $|j=1/2\rangle$ primary sector, which is the Ramond sector for the $\phi^a_n$. This is because the indices of the $L_n$ in Eq.~\eqref{eq:lphi0commutation} can only take integer values. In the $|j=0\rangle$ and $|j=1\rangle$ primary sectors, which are the Neveu-Schwarz sector for the $\phi^a_n$, the indices of the $\phi^a_n$ will instead take half-integer values. There the fractional dimension operators will be anti-periodic in space, and so the conjugation by $\mathcal{R}$ will have the opposite sign.
	    \footnote{In terms of the operators $\phi^a(x)$, we have $\mathcal{R} \phi^a(x) \mathcal{R}^{-1} = \bar{\phi}^a(\ell - x)$ and $\mathcal{R} \bar{\phi}^a(x) \mathcal{R}^{-1} = \phi^a(\ell - x)$, analogous to Eq.~\eqref{eq:RLop}. In the Ramond sector, where $\phi^a(x) = \phi^a(\ell+x)$ and $\bar{\phi}^a(x) = \bar{\phi}^a(\ell+x)$ are periodic, this works out to $\mathcal{R} \phi^a(x) \mathcal{R}^{-1} = \bar{\phi}^a(- x)$ and $\mathcal{R} \bar{\phi}^a(x) \mathcal{R}^{-1} = \phi^a(-x)$, as in Eq.~\eqref{eq:RLop}. In the Neveu-Schwarz case, on the other hand, $\phi^a(x) = -\phi^a(\ell+x)$ and $\bar{\phi}^a(x) = -\bar{\phi}^a(\ell+x)$ are anti-periodic, so $\mathcal{R} \phi^a(x) \mathcal{R}^{-1} = -\bar{\phi}^a(- x)$ and $\mathcal{R} \bar{\phi}^a(x) \mathcal{R}^{-1} = -\phi^a(- x)$, leading to the relative minus sign of Eq.~\eqref{eq:phiz12transformation}.}
	    Thus we instead obtain 
		\begin{align}
			\label{eq:phiz12transformation}
			(\mathcal{R T}) \phi^a_n (\mathcal{R T})^{-1} &= \phi^a_n\text{, }(n \in \mathbb{Z} + 1/2).
		\end{align}
		From Eqs.~\eqref{eq:gdef}, \eqref{eq:phiztransformation}, and \eqref{eq:phiz12transformation}, we thus deduce that the superconformal current modes $G_n$ satisfy
		\begin{align}
			\label{eq:gztransformation}
			(\mathcal{R T}) G_n (\mathcal{R T})^{-1} &= G_n \text{, }(n \in \mathbb{Z})\text{, and} \\
			(\mathcal{R T}) G_n (\mathcal{R T})^{-1} &= -G_n\text{, }(n \in \mathbb{Z} + 1/2),
		\end{align}
		in the Ramond and Neveu-Schwarz sectors, respectively. 
		
		Taking Eqs.~\eqref{eq:ltransformation} and \eqref{eq:gztransformation} into account, we can see that of the conserved quantities in Table \ref{table:modereps}, only $H^{(5)}$ (considered, of necessity, in the Ramond sector) will fail to satisfy Eq.~\eqref{eq:hirtcondition}, due to the effect of the anti-unitarity of $\mathcal{RT}$ on the imaginary coefficient. This analysis is done in terms of the modes, useful for considering the conserved quantities themselves, but the translation to the corresponding local operators of which they are the integrals (i.e.,~the contents of Table \ref{table:operators}) is straightforward.

	\section{Discussion of Fitting Algorithm}
	\label{app:fitting}
	
		Above, we have discussed how we compute the splittings of entanglement spectra of chiral topological states that feature an $\rm{SU}(2)_1$ or $\rm{SU}(2)_2$ WZW CFT in the entanglement spectrum by incorporating the linear combination of terms with conserved integrals of irrelevant local operators Eq.~\eqref{eq:lincomb}. We now discuss how we determine the optimal values of the parameters $\beta_i$ in that linear combination.
		
		For a given set of numerical entanglement spectrum data, we choose the GGE parameters $\beta_i$ that best fit that spectrum.
		Within each topological sector of the entanglement spectrum, we calculate the eigenvalues of the linear combination of operators as a function of the $\beta_i$ and order those eigenvalues in increasing order of, first, the associated descendant level, then multiplet dimension, and finally value. We write the $\ell$th element of that ordered list as $\xi^{\text{fit}}_\ell\left(\{\beta_i\}\right)$.
		We order the actual entanglement spectrum data by the same criteria, and write the $\ell$th element of that ordered list as $\xi^{\text{data}}_\ell$.
		We can then write a fitting function of the chosen GGE parameters in the $|j\rangle$ primary sector:
		\begin{equation}
			\label{eq:fitfunc}
			R_j\left(\{\beta_i\}\right)=\sum_\ell \left[\xi^{\text{data}}_\ell-\xi^{\text{fit}}_\ell\left(\{\beta_i\}\right) \right]^2 W_\ell,
		\end{equation}
		where $W_\ell$ is a weight associated to the $\xi_\ell$. 
		We set the weights $W_\ell$ so that the states at each descendant level have, collectively, the same weight in the fit, with $W_\ell \propto \frac{1}{N_{\Delta_\ell}}$, where $N_\Delta$ is the number of states at descendant level $\Delta$ (i.e., the corresponding counting from the ``\#" column of Table \ref{table:su21countings} or \ref{table:su22countings} of Appendix \ref{app:su2kintro}), and $\Delta_\ell$ is the descendant level of the $\ell$th state from the spectral data. The weights $W_\ell$ are normalized, however, so that $\sum_\ell W_\ell = 1$.
		
		We then minimize either the individual $R_j\left(\{\beta_i\}\right)$ for each sector of descendants of each primary state $|j\rangle$, or, in the case of a simultaneous fit of multiple sectors, the sum $\sum_j R_j\left(\{\beta_i\}\right)$ of the $R_j\left(\{\beta_i\}\right)$ over all the relevant sectors. The method used for minimization is Mathematica's \texttt{NMinimize} function. The results of this minimization are the plotted results of Sec.~\ref{sec:results}, Figs.~\ref{fig:bauerfit}--\ref{fig:chenallopsfit}. The corresponding values of the $\tilde{\beta}_i$ can be found in Tables \ref{table:su21fitparams}--\ref{table:su22fitparams} of Appendix \ref{app:parameters}.
		
		We note that the \texttt{NMinimize} function may not always find the exact global minimum of the function that we attempt to minimize. We believe, however, that the minimizing sets of $\tilde{\beta}_i$ reported here reflect local minima which are representative, in the sense that the globally minimal fits would not be substantial improvements in fitting the data. In cases where we do obtain very good fits, this is necessarily true. But even in the more difficult case of the $\rm{SU}(2)_2$ data from Chen et al., e.g., in Fig.~\ref{fig:chenallopsfit}, the consistency of our results with the exact $\mathcal{RT}$ symmetry, as described in Sec.~\ref{sec:su22results}, gives us confidence in this conclusion. 
		
	\section{Fitting Parameter Data}
	\label{app:parameters}
		
		Tables \ref{table:su21fitparams} and \ref{table:su22fitparams} exhibit the numerical values of the parameters that were actually used in our approach to generate the best-fit results of the figures of Sec.~\ref{sec:results}. Since our calculations on the CFT Hilbert space are done using the size $\ell$-independent integrals of motion  $\tilde{H}^{(i)} = \left(\frac{\ell}{2\pi}\right)^{\Delta_i - 1}H^{(i)}$ (enumerated in Table \ref{table:modereps}) we end up computing correspondingly size-dependent parameters $\tilde{\beta}_i = \left(\frac{2\pi}{\ell}\right)^{\Delta_i - 1}\beta_i$. 
		Essentially, we rewrite Eq.~\eqref{eq:lincomb-SizeDependence} as
		\begin{equation}
			\label{eq:lincombtilde}
			H_{\text{entanglement}} - \textrm{const.} = \tilde{\beta} \tilde{H}^{(1)} + \sum_{i \neq 1}^{\infty} \tilde{\beta}_i \tilde{H}^{(i)}.
		\end{equation}
		
		To remove arbitrary factors of scale, the $\tilde{\beta}_i$, which determine the splitting within a descendant level, have been normalized by dividing by $\tilde{\beta}$, which is the parameter that determines the distance between descendant levels as the coefficient of $\tilde{H}^{(1)} = \frac{\ell}{2\pi} H_L$ in Eq.~\eqref{eq:lincombtilde}. Tables \ref{table:su21fitparams} and \ref{table:su22fitparams} also note the value of the corresponding fitting function $R_j\left(\{\beta_i\}\right)$ of Eq.~\eqref{eq:fitfunc} for the best fit that was achieved for the given data. Since $R_j\left(\{\beta_i\}\right)$ is a sum of the squares of the distances between the data and our fit, we have normalized $R_j\left(\{\beta_i\}\right)$ by dividing by $\tilde{\beta}^2$. 
		
		\vspace{1.5cm}
		
		\begin{table}[hbt]
			\centering
			\begin{tabular}{c|c|c|c|c|c|c|c}
				Data source & Figure & Size $\ell$ & Sector &  $R_j\left(\{\beta_i\}\right)$ & $\tilde{\beta}_3$ & $\tilde{\beta}_8$ & $\tilde{\beta}_9$ \\ 
				\hline
				\multirow{2}{*}{Bauer et al.~(Ref.~\onlinecite{Bauer2014})} & \multirow{2}{*}{Fig.~\ref{fig:bauerfit}} & \multirow{2}{*}{12} &  $|j=0\rangle$ & 0.00201 & -0.0733 & 0.00283 & 0.00401 \\
				\cline{4-8}
				& & & $\left|j=\frac{1}{2}\right\rangle$ & 0.00535 & -0.0423 & 0.000408 & 0.00452  \\
				\hline 
				\multirow{2}{*}{Hickey et al.~(Ref.~\onlinecite{Hickey2016})} & \multirow{2}{*}{Fig.~\ref{fig:hickeyfit}} & \multirow{2}{*}{8} & $|j=0\rangle$ & 0.000482 & \multirow{2}{*}{-0.0313} & \multirow{2}{*}{0.000970} & \multirow{2}{*}{0.000441} \\
				\cline{4-5}
				& & & $\left|j=\frac{1}{2}\right\rangle$ & 0.000609 & & & \\
				\hline 
				\multirow{2}{*}{Hackenbroich et al.~(Ref.~\onlinecite{Hackenbroich2018})} & \multirow{2}{*}{Fig.~\ref{fig:hackenbroichfit}} & \multirow{2}{*}{8} & $|j=0\rangle$ & 0.00222 & \multirow{2}{*}{-0.0554} & \multirow{2}{*}{0.00698} & \multirow{2}{*}{-0.00211} \\
				\cline{4-5} 
				& & & $\left|j=\frac{1}{2}\right\rangle$ & 0.00136 & & & \\
				\hline 
			\end{tabular}
			\caption{
				The values of the fitting function $R_j\left(\{\beta_i\}\right)$, along with the associated normalized numerical values of the $\tilde{\beta}_i$ (where $\tilde{\beta}_i$ is the parameter corresponding to the $i$th conserved quantity of Table \ref{table:modereps}) that were calculated for the best fits to the $\rm{SU}(2)_1$ data of Sec.~\ref{sec:su21results}. The fits to the data of Hickey et al.~and Hackenbroich et al.~were performed simultaneously in both sectors, minimizing the sum of the $R_j\left(\{\beta_i\}\right)$, and so the parameter values of the $\tilde{\beta}_i$ are the same in both sectors for those fits.
			}
			\label{table:su21fitparams}
		\end{table}
		\begin{table}[hbt]
			\centering
			\begin{tabular}{c|c|c|c|c|c|c|c|c|c|c|c|c|c|c}
					Figure & Sector & $R_j\left(\{\beta_i\}\right)$ & $\tilde{\beta}_0$ & $\tilde{\beta}_2$ & $\tilde{\beta}_3$ & $\tilde{\beta}_4$ & $\tilde{\beta}_5$ & $\tilde{\beta}_6$ & $\tilde{\beta}_7$ & $\tilde{\beta}_8$ & $\tilde{\beta}_9$ & $\tilde{\beta}_{10}$ & $\tilde{\beta}_{11}$ \\
					\hline
					Fig.~\ref{fig:chenmmfit_bosons}  & $\left|j=\frac{1}{2}\right\rangle$ & 0.00727 & --- & --- & 0.0045 & -0.03 & \multicolumn{7}{c|}{---}  \\
					\hline 
					Fig.~\ref{fig:chenmmfit_fermions}  & $\left|j=\frac{1}{2}\right\rangle$ & 0.00055 & 0.17 & -0.018 & --- & --- & \multicolumn{7}{c|}{---}  \\
					\hline
					\multirow{2}{*}{Fig.~\ref{fig:chen5opsfit}}  & $|j=0\rangle$ & 0.00043 & --- & --- & \multirow{2}{*}{-0.0061} & \multirow{2}{*}{-0.022} & \multicolumn{7}{c|}{\multirow{2}{*}{---}} \\
					\cline{2-5}
					& $\left|j=\frac{1}{2}\right\rangle$ & 0.00013 & 0.17 & -0.033 &  &  & \multicolumn{7}{c|}{}  \\
					\hline
					\multirow{2}{*}{Fig.~\ref{fig:chenallopsfit}}  & $|j=0\rangle$ &  0.00142 & --- & --- &  \multirow{2}{*}{-0.049} & \multirow{2}{*}{-0.036} & --- & --- & --- &  \multirow{2}{*}{0.006} & \multirow{2}{*}{0.013} & \multirow{2}{*}{0.016} & \multirow{2}{*}{-0.009}\\
					\cline{2-5}\cline{8-10}
					& $\left|j=\frac{1}{2}\right\rangle$ & 0.00174 & 0.19 & -0.064 &  &  & 0\footnote{We set $\tilde{\beta}_5 = 0$ due to the exclusion of $\tilde{H}^{(5)}$ by the $\mathcal{RT}$ symmetry.} & 0.01 & 0.008 &  &  &  &  \\
					\hline
			\end{tabular}
			\caption{
				The values of the fitting function $R_j\left(\{\beta_i\}\right)$, along with the associated normalized numerical values of the $\tilde{\beta}_i$ (where $\tilde{\beta}_i$ is the parameter corresponding to the $i$th conserved quantity of Table \ref{table:modereps}) that were calculated for the best fits to the $\rm{SU}(2)_2$ data of Sec.~\ref{sec:su22results}. Note the substantial improvement in $R_j\left(\{\beta_i\}\right)$ of Fig.~\ref{fig:chenmmfit_fermions} relative to Fig.~\ref{fig:chenmmfit_bosons} that was achieved by fitting with the half-integer dimensional conserved quantities $\tilde{H}_0$ and $\tilde{H}_2$ instead of the integer dimensional conserved quantities $\tilde{H}_3$ and $\tilde{H}_4$. The fits of Figs.~\ref{fig:chen5opsfit} and \ref{fig:chenallopsfit} were performed simultaneously in both sectors, minimizing the sum of the $R_j\left(\{\beta_i\}\right)$, and so the parameter values of the $\tilde{\beta}_i$ corresponding to integer dimensional conserved quantities are the same in both sectors for those fits. Half-integer dimensional conserved quantities were only used for the $|j=1/2\rangle$ sector, so the corresponding $\tilde{\beta}_i$ are marked by --- in the $|j=0\rangle$ sector.
				The data used for these fits is from Chen et al.~(Ref.~\protect\onlinecite{Chen2018}), with size $\ell = 6$.
			}
			\label{table:su22fitparams}
		\end{table}

		\newpage
		
		\bibliography{su2_fits_bibliography}

\begin{thebibliography}{80}%
\makeatletter
\providecommand \@ifxundefined [1]{%
 \@ifx{#1\undefined}
}%
\providecommand \@ifnum [1]{%
 \ifnum #1\expandafter \@firstoftwo
 \else \expandafter \@secondoftwo
 \fi
}%
\providecommand \@ifx [1]{%
 \ifx #1\expandafter \@firstoftwo
 \else \expandafter \@secondoftwo
 \fi
}%
\providecommand \natexlab [1]{#1}%
\providecommand \enquote  [1]{``#1''}%
\providecommand \bibnamefont  [1]{#1}%
\providecommand \bibfnamefont [1]{#1}%
\providecommand \citenamefont [1]{#1}%
\providecommand \href@noop [0]{\@secondoftwo}%
\providecommand \href [0]{\begingroup \@sanitize@url \@href}%
\providecommand \@href[1]{\@@startlink{#1}\@@href}%
\providecommand \@@href[1]{\endgroup#1\@@endlink}%
\providecommand \@sanitize@url [0]{\catcode `\\12\catcode `\$12\catcode
  `\&12\catcode `\#12\catcode `\^12\catcode `\_12\catcode `\%12\relax}%
\providecommand \@@startlink[1]{}%
\providecommand \@@endlink[0]{}%
\providecommand \url  [0]{\begingroup\@sanitize@url \@url }%
\providecommand \@url [1]{\endgroup\@href {#1}{\urlprefix }}%
\providecommand \urlprefix  [0]{URL }%
\providecommand \Eprint [0]{\href }%
\providecommand \doibase [0]{http://dx.doi.org/}%
\providecommand \selectlanguage [0]{\@gobble}%
\providecommand \bibinfo  [0]{\@secondoftwo}%
\providecommand \bibfield  [0]{\@secondoftwo}%
\providecommand \translation [1]{[#1]}%
\providecommand \BibitemOpen [0]{}%
\providecommand \bibitemStop [0]{}%
\providecommand \bibitemNoStop [0]{.\EOS\space}%
\providecommand \EOS [0]{\spacefactor3000\relax}%
\providecommand \BibitemShut  [1]{\csname bibitem#1\endcsname}%
\let\auto@bib@innerbib\@empty
\bibitem [{\citenamefont {Wen}(1991)}]{Wen1991}%
  \BibitemOpen
  \bibfield  {author} {\bibinfo {author} {\bibfnamefont {X.-G.}\ \bibnamefont
  {Wen}},\ }\href {\doibase 10.1142/S0217979291001541} {\bibfield  {journal}
  {\bibinfo  {journal} {Int. J. Mod. Phys. B}\ }\textbf {\bibinfo {volume}
  {5}},\ \bibinfo {pages} {1641} (\bibinfo {year} {1991})}\BibitemShut
  {NoStop}%
\bibitem [{\citenamefont {Kitaev}(2003)}]{Kitaev2003}%
  \BibitemOpen
  \bibfield  {author} {\bibinfo {author} {\bibfnamefont {A.}~\bibnamefont
  {Kitaev}},\ }\href {\doibase https://doi.org/10.1016/S0003-4916(02)00018-0}
  {\bibfield  {journal} {\bibinfo  {journal} {Ann. Phys. (NY)}\ }\textbf
  {\bibinfo {volume} {303}},\ \bibinfo {pages} {2} (\bibinfo {year}
  {2003})}\BibitemShut {NoStop}%
\bibitem [{\citenamefont {Wen}(2004)}]{Wen2004}%
  \BibitemOpen
  \bibfield  {author} {\bibinfo {author} {\bibfnamefont {X.-G.}\ \bibnamefont
  {Wen}},\ }\href@noop {} {\emph {\bibinfo {title} {Quantum Field Theory of
  Many-Body Systems}}}\ (\bibinfo  {publisher} {Oxford University Press,
  Oxford},\ \bibinfo {year} {2004})\BibitemShut {NoStop}%
\bibitem [{\citenamefont {Kitaev}(2006)}]{Kitaev2006}%
  \BibitemOpen
  \bibfield  {author} {\bibinfo {author} {\bibfnamefont {A.}~\bibnamefont
  {Kitaev}},\ }\href {\doibase https://doi.org/10.1016/j.aop.2005.10.005}
  {\bibfield  {journal} {\bibinfo  {journal} {Ann. Phys. (NY)}\ }\textbf
  {\bibinfo {volume} {321}},\ \bibinfo {pages} {2} (\bibinfo {year}
  {2006})}\BibitemShut {NoStop}%
\bibitem [{\citenamefont {Nayak}\ \emph {et~al.}(2008)\citenamefont {Nayak},
  \citenamefont {Simon}, \citenamefont {Stern}, \citenamefont {Freedman},\ and\
  \citenamefont {Das~Sarma}}]{Nayak2008}%
  \BibitemOpen
  \bibfield  {author} {\bibinfo {author} {\bibfnamefont {C.}~\bibnamefont
  {Nayak}}, \bibinfo {author} {\bibfnamefont {S.~H.}\ \bibnamefont {Simon}},
  \bibinfo {author} {\bibfnamefont {A.}~\bibnamefont {Stern}}, \bibinfo
  {author} {\bibfnamefont {M.}~\bibnamefont {Freedman}}, \ and\ \bibinfo
  {author} {\bibfnamefont {S.}~\bibnamefont {Das~Sarma}},\ }\href {\doibase
  10.1103/RevModPhys.80.1083} {\bibfield  {journal} {\bibinfo  {journal} {Rev.
  Mod. Phys.}\ }\textbf {\bibinfo {volume} {80}},\ \bibinfo {pages} {1083}
  (\bibinfo {year} {2008})}\BibitemShut {NoStop}%
\bibitem [{\citenamefont {Halperin}(1982)}]{Halperin1982}%
  \BibitemOpen
  \bibfield  {author} {\bibinfo {author} {\bibfnamefont {B.~I.}\ \bibnamefont
  {Halperin}},\ }\href {\doibase 10.1103/PhysRevB.25.2185} {\bibfield
  {journal} {\bibinfo  {journal} {Phys. Rev. B}\ }\textbf {\bibinfo {volume}
  {25}},\ \bibinfo {pages} {2185} (\bibinfo {year} {1982})}\BibitemShut
  {NoStop}%
\bibitem [{\citenamefont {Witten}(1989)}]{Witten1989}%
  \BibitemOpen
  \bibfield  {author} {\bibinfo {author} {\bibfnamefont {E.}~\bibnamefont
  {Witten}},\ }\href {\doibase 10.1007/BF01217730} {\bibfield  {journal}
  {\bibinfo  {journal} {Commun. Math. Phys.}\ }\textbf {\bibinfo {volume}
  {121}},\ \bibinfo {pages} {351} (\bibinfo {year} {1989})}\BibitemShut
  {NoStop}%
\bibitem [{\citenamefont {Wen}(1990)}]{Wen1990}%
  \BibitemOpen
  \bibfield  {author} {\bibinfo {author} {\bibfnamefont {X.-G.}\ \bibnamefont
  {Wen}},\ }\href {\doibase 10.1103/PhysRevB.41.12838} {\bibfield  {journal}
  {\bibinfo  {journal} {Phys. Rev. B}\ }\textbf {\bibinfo {volume} {41}},\
  \bibinfo {pages} {12838} (\bibinfo {year} {1990})}\BibitemShut {NoStop}%
\bibitem [{\citenamefont {Li}\ and\ \citenamefont {Haldane}(2008)}]{Li2008}%
  \BibitemOpen
  \bibfield  {author} {\bibinfo {author} {\bibfnamefont {H.}~\bibnamefont
  {Li}}\ and\ \bibinfo {author} {\bibfnamefont {F.~D.~M.}\ \bibnamefont
  {Haldane}},\ }\href {\doibase 10.1103/PhysRevLett.101.010504} {\bibfield
  {journal} {\bibinfo  {journal} {Phys. Rev. Lett.}\ }\textbf {\bibinfo
  {volume} {101}},\ \bibinfo {pages} {010504} (\bibinfo {year}
  {2008})}\BibitemShut {NoStop}%
\bibitem [{\citenamefont {Qi}\ \emph {et~al.}(2012)\citenamefont {Qi},
  \citenamefont {Katsura},\ and\ \citenamefont {Ludwig}}]{Qi2012}%
  \BibitemOpen
  \bibfield  {author} {\bibinfo {author} {\bibfnamefont {X.-L.}\ \bibnamefont
  {Qi}}, \bibinfo {author} {\bibfnamefont {H.}~\bibnamefont {Katsura}}, \ and\
  \bibinfo {author} {\bibfnamefont {A.~W.~W.}\ \bibnamefont {Ludwig}},\ }\href
  {\doibase 10.1103/PhysRevLett.108.196402} {\bibfield  {journal} {\bibinfo
  {journal} {Phys. Rev. Lett.}\ }\textbf {\bibinfo {volume} {108}},\ \bibinfo
  {pages} {196402} (\bibinfo {year} {2012})}\BibitemShut {NoStop}%
\bibitem [{\citenamefont {Peschel}\ and\ \citenamefont
  {Chung}(2011)}]{Peschel2011}%
  \BibitemOpen
  \bibfield  {author} {\bibinfo {author} {\bibfnamefont {I.}~\bibnamefont
  {Peschel}}\ and\ \bibinfo {author} {\bibfnamefont {M.-C.}\ \bibnamefont
  {Chung}},\ }\href {\doibase 10.1209/0295-5075/96/50006} {\bibfield  {journal}
  {\bibinfo  {journal} {Europhys. Lett.}\ }\textbf {\bibinfo {volume} {96}},\
  \bibinfo {pages} {50006} (\bibinfo {year} {2011})}\BibitemShut {NoStop}%
\bibitem [{\citenamefont {Chandran}\ \emph {et~al.}(2011)\citenamefont
  {Chandran}, \citenamefont {Hermanns}, \citenamefont {Regnault},\ and\
  \citenamefont {Bernevig}}]{Chandran2011}%
  \BibitemOpen
  \bibfield  {author} {\bibinfo {author} {\bibfnamefont {A.}~\bibnamefont
  {Chandran}}, \bibinfo {author} {\bibfnamefont {M.}~\bibnamefont {Hermanns}},
  \bibinfo {author} {\bibfnamefont {N.}~\bibnamefont {Regnault}}, \ and\
  \bibinfo {author} {\bibfnamefont {B.~A.}\ \bibnamefont {Bernevig}},\ }\href
  {\doibase 10.1103/PhysRevB.84.205136} {\bibfield  {journal} {\bibinfo
  {journal} {Phys. Rev. B}\ }\textbf {\bibinfo {volume} {84}},\ \bibinfo
  {pages} {205136} (\bibinfo {year} {2011})}\BibitemShut {NoStop}%
\bibitem [{\citenamefont {Dubail}\ \emph {et~al.}(2012)\citenamefont {Dubail},
  \citenamefont {Read},\ and\ \citenamefont {Rezayi}}]{Dubail2012}%
  \BibitemOpen
  \bibfield  {author} {\bibinfo {author} {\bibfnamefont {J.}~\bibnamefont
  {Dubail}}, \bibinfo {author} {\bibfnamefont {N.}~\bibnamefont {Read}}, \ and\
  \bibinfo {author} {\bibfnamefont {E.~H.}\ \bibnamefont {Rezayi}},\ }\href
  {\doibase 10.1103/PhysRevB.86.245310} {\bibfield  {journal} {\bibinfo
  {journal} {Phys. Rev. B}\ }\textbf {\bibinfo {volume} {86}},\ \bibinfo
  {pages} {245310} (\bibinfo {year} {2012})}\BibitemShut {NoStop}%
\bibitem [{\citenamefont {Swingle}\ and\ \citenamefont
  {Senthil}(2012)}]{Swingle2012}%
  \BibitemOpen
  \bibfield  {author} {\bibinfo {author} {\bibfnamefont {B.}~\bibnamefont
  {Swingle}}\ and\ \bibinfo {author} {\bibfnamefont {T.}~\bibnamefont
  {Senthil}},\ }\href {\doibase 10.1103/PhysRevB.86.045117} {\bibfield
  {journal} {\bibinfo  {journal} {Phys. Rev. B}\ }\textbf {\bibinfo {volume}
  {86}},\ \bibinfo {pages} {045117} (\bibinfo {year} {2012})}\BibitemShut
  {NoStop}%
\bibitem [{\citenamefont {Bauer}\ \emph {et~al.}(2014)\citenamefont {Bauer},
  \citenamefont {Cincio}, \citenamefont {Keller}, \citenamefont {Dolfi},
  \citenamefont {Vidal}, \citenamefont {Trebst},\ and\ \citenamefont
  {Ludwig}}]{Bauer2014}%
  \BibitemOpen
  \bibfield  {author} {\bibinfo {author} {\bibfnamefont {B.}~\bibnamefont
  {Bauer}}, \bibinfo {author} {\bibfnamefont {L.}~\bibnamefont {Cincio}},
  \bibinfo {author} {\bibfnamefont {B.~P.}\ \bibnamefont {Keller}}, \bibinfo
  {author} {\bibfnamefont {M.}~\bibnamefont {Dolfi}}, \bibinfo {author}
  {\bibfnamefont {G.}~\bibnamefont {Vidal}}, \bibinfo {author} {\bibfnamefont
  {S.}~\bibnamefont {Trebst}}, \ and\ \bibinfo {author} {\bibfnamefont
  {A.~W.~W.}\ \bibnamefont {Ludwig}},\ }\href
  {https://doi.org/10.1038/ncomms6137 http://10.0.4.14/ncomms6137
  https://www.nature.com/articles/ncomms6137{\#}supplementary-information}
  {\bibfield  {journal} {\bibinfo  {journal} {Nat. Commun.}\ }\textbf {\bibinfo
  {volume} {5}},\ \bibinfo {pages} {5137} (\bibinfo {year} {2014})}\BibitemShut
  {NoStop}%
\bibitem [{\citenamefont {Hickey}\ \emph {et~al.}(2016)\citenamefont {Hickey},
  \citenamefont {Cincio}, \citenamefont {Papi\ifmmode~\acute{c}\else
  \'{c}\fi{}},\ and\ \citenamefont {Paramekanti}}]{Hickey2016}%
  \BibitemOpen
  \bibfield  {author} {\bibinfo {author} {\bibfnamefont {C.}~\bibnamefont
  {Hickey}}, \bibinfo {author} {\bibfnamefont {L.}~\bibnamefont {Cincio}},
  \bibinfo {author} {\bibfnamefont {Z.}~\bibnamefont
  {Papi\ifmmode~\acute{c}\else \'{c}\fi{}}}, \ and\ \bibinfo {author}
  {\bibfnamefont {A.}~\bibnamefont {Paramekanti}},\ }\href {\doibase
  10.1103/PhysRevLett.116.137202} {\bibfield  {journal} {\bibinfo  {journal}
  {Phys. Rev. Lett.}\ }\textbf {\bibinfo {volume} {116}},\ \bibinfo {pages}
  {137202} (\bibinfo {year} {2016})}\BibitemShut {NoStop}%
\bibitem [{\citenamefont {Hackenbroich}\ \emph {et~al.}(2018)\citenamefont
  {Hackenbroich}, \citenamefont {Sterdyniak},\ and\ \citenamefont
  {Schuch}}]{Hackenbroich2018}%
  \BibitemOpen
  \bibfield  {author} {\bibinfo {author} {\bibfnamefont {A.}~\bibnamefont
  {Hackenbroich}}, \bibinfo {author} {\bibfnamefont {A.}~\bibnamefont
  {Sterdyniak}}, \ and\ \bibinfo {author} {\bibfnamefont {N.}~\bibnamefont
  {Schuch}},\ }\href {\doibase 10.1103/PhysRevB.98.085151} {\bibfield
  {journal} {\bibinfo  {journal} {Phys. Rev. B}\ }\textbf {\bibinfo {volume}
  {98}},\ \bibinfo {pages} {085151} (\bibinfo {year} {2018})}\BibitemShut
  {NoStop}%
\bibitem [{\citenamefont {Chen}\ \emph {et~al.}(2018)\citenamefont {Chen},
  \citenamefont {Vanderstraeten}, \citenamefont {Capponi},\ and\ \citenamefont
  {Poilblanc}}]{Chen2018}%
  \BibitemOpen
  \bibfield  {author} {\bibinfo {author} {\bibfnamefont {J.-Y.}\ \bibnamefont
  {Chen}}, \bibinfo {author} {\bibfnamefont {L.}~\bibnamefont
  {Vanderstraeten}}, \bibinfo {author} {\bibfnamefont {S.}~\bibnamefont
  {Capponi}}, \ and\ \bibinfo {author} {\bibfnamefont {D.}~\bibnamefont
  {Poilblanc}},\ }\href {\doibase 10.1103/PhysRevB.98.184409} {\bibfield
  {journal} {\bibinfo  {journal} {Phys. Rev. B}\ }\textbf {\bibinfo {volume}
  {98}},\ \bibinfo {pages} {184409} (\bibinfo {year} {2018})}\BibitemShut
  {NoStop}%
\bibitem [{\citenamefont {Zaletel}\ and\ \citenamefont
  {Mong}(2012)}]{Zaletel2012}%
  \BibitemOpen
  \bibfield  {author} {\bibinfo {author} {\bibfnamefont {M.~P.}\ \bibnamefont
  {Zaletel}}\ and\ \bibinfo {author} {\bibfnamefont {R.~S.~K.}\ \bibnamefont
  {Mong}},\ }\href {\doibase 10.1103/PhysRevB.86.245305} {\bibfield  {journal}
  {\bibinfo  {journal} {Phys. Rev. B}\ }\textbf {\bibinfo {volume} {86}},\
  \bibinfo {pages} {245305} (\bibinfo {year} {2012})}\BibitemShut {NoStop}%
\bibitem [{\citenamefont {Davenport}\ \emph {et~al.}(2015)\citenamefont
  {Davenport}, \citenamefont {Rodr{\'{i}}guez}, \citenamefont {Slingerland},\
  and\ \citenamefont {Simon}}]{Davenport2015}%
  \BibitemOpen
  \bibfield  {author} {\bibinfo {author} {\bibfnamefont {S.~C.}\ \bibnamefont
  {Davenport}}, \bibinfo {author} {\bibfnamefont {I.~D.}\ \bibnamefont
  {Rodr{\'{i}}guez}}, \bibinfo {author} {\bibfnamefont {J.~K.}\ \bibnamefont
  {Slingerland}}, \ and\ \bibinfo {author} {\bibfnamefont {S.~H.}\ \bibnamefont
  {Simon}},\ }\href {\doibase 10.1103/PhysRevB.92.115155} {\bibfield  {journal}
  {\bibinfo  {journal} {Phys. Rev. B}\ }\textbf {\bibinfo {volume} {92}},\
  \bibinfo {pages} {115155} (\bibinfo {year} {2015})}\BibitemShut {NoStop}%
\bibitem [{\citenamefont {Huang}\ \emph {et~al.}(2022)\citenamefont {Huang},
  \citenamefont {Zhu}, \citenamefont {Gong}, \citenamefont {Jiang},\ and\
  \citenamefont {Sheng}}]{Huang2021}%
  \BibitemOpen
  \bibfield  {author} {\bibinfo {author} {\bibfnamefont {Y.}~\bibnamefont
  {Huang}}, \bibinfo {author} {\bibfnamefont {W.}~\bibnamefont {Zhu}}, \bibinfo
  {author} {\bibfnamefont {S.-S.}\ \bibnamefont {Gong}}, \bibinfo {author}
  {\bibfnamefont {H.-C.}\ \bibnamefont {Jiang}}, \ and\ \bibinfo {author}
  {\bibfnamefont {D.~N.}\ \bibnamefont {Sheng}},\ }\href {\doibase
  10.1103/PhysRevB.105.155104} {\bibfield  {journal} {\bibinfo  {journal}
  {Phys. Rev. B}\ }\textbf {\bibinfo {volume} {105}},\ \bibinfo {pages}
  {155104} (\bibinfo {year} {2022})}\BibitemShut {NoStop}%
\bibitem [{\citenamefont {Cardy}(2016)}]{Cardy2016}%
  \BibitemOpen
  \bibfield  {author} {\bibinfo {author} {\bibfnamefont {J.}~\bibnamefont
  {Cardy}},\ }\href {http://stacks.iop.org/1742-5468/2016/i=2/a=023103}
  {\bibfield  {journal} {\bibinfo  {journal} {J. Stat. Mech.: Theory Exp.}\
  }\textbf {\bibinfo {volume} {2016}},\ \bibinfo {pages} {023103} (\bibinfo
  {year} {2016})}\BibitemShut {NoStop}%
\bibitem [{\citenamefont {Cho}\ \emph {et~al.}(2017)\citenamefont {Cho},
  \citenamefont {Ludwig},\ and\ \citenamefont {Ryu}}]{ChoLudwigRyu-PRB2017}%
  \BibitemOpen
  \bibfield  {author} {\bibinfo {author} {\bibfnamefont {G.~Y.}\ \bibnamefont
  {Cho}}, \bibinfo {author} {\bibfnamefont {A.~W.~W.}\ \bibnamefont {Ludwig}},
  \ and\ \bibinfo {author} {\bibfnamefont {S.}~\bibnamefont {Ryu}},\ }\href
  {\doibase 10.1103/PhysRevB.95.115122} {\bibfield  {journal} {\bibinfo
  {journal} {Phys. Rev. B}\ }\textbf {\bibinfo {volume} {95}},\ \bibinfo
  {pages} {115122} (\bibinfo {year} {2017})}\BibitemShut {NoStop}%
\bibitem [{\citenamefont {Cardy}(2017)}]{Cardy2017}%
  \BibitemOpen
  \bibfield  {author} {\bibinfo {author} {\bibfnamefont {J.}~\bibnamefont
  {Cardy}},\ }\href {\doibase 10.21468/SciPostPhys.3.2.011} {\bibfield
  {journal} {\bibinfo  {journal} {SciPost Phys.}\ }\textbf {\bibinfo {volume}
  {3}},\ \bibinfo {pages} {011} (\bibinfo {year} {2017})}\BibitemShut {NoStop}%
\bibitem [{\citenamefont {Kalmeyer}\ and\ \citenamefont
  {Laughlin}(1987)}]{Kalmeyer1987}%
  \BibitemOpen
  \bibfield  {author} {\bibinfo {author} {\bibfnamefont {V.}~\bibnamefont
  {Kalmeyer}}\ and\ \bibinfo {author} {\bibfnamefont {R.~B.}\ \bibnamefont
  {Laughlin}},\ }\href {\doibase 10.1103/PhysRevLett.59.2095} {\bibfield
  {journal} {\bibinfo  {journal} {Phys. Rev. Lett.}\ }\textbf {\bibinfo
  {volume} {59}},\ \bibinfo {pages} {2095} (\bibinfo {year}
  {1987})}\BibitemShut {NoStop}%
\bibitem [{Note1()}]{Note1}%
  \BibitemOpen
  \bibinfo {note} {See for instance Ref.~\protect \rev@citealpnum {Huang2021}
  for recent numerical work on real-space entanglement spectra of this chiral
  spin liquid, as well as Ref.~\protect \rev@citealpnum {Chen2018}, a PEPS
  study that we consider in further detail below. Both works present the
  real-space entanglement spectrum in the spin-1/2 sector of $\protect \rm
  {SU}(2)_2$, which is of particular interest in our present
  paper.}\BibitemShut {Stop}%
\bibitem [{\citenamefont {Cirac}\ \emph {et~al.}(2021)\citenamefont {Cirac},
  \citenamefont {P\'erez-Garc\'{\i}a}, \citenamefont {Schuch},\ and\
  \citenamefont {Verstraete}}]{Cirac2021}%
  \BibitemOpen
  \bibfield  {author} {\bibinfo {author} {\bibfnamefont {J.~I.}\ \bibnamefont
  {Cirac}}, \bibinfo {author} {\bibfnamefont {D.}~\bibnamefont
  {P\'erez-Garc\'{\i}a}}, \bibinfo {author} {\bibfnamefont {N.}~\bibnamefont
  {Schuch}}, \ and\ \bibinfo {author} {\bibfnamefont {F.}~\bibnamefont
  {Verstraete}},\ }\href {\doibase 10.1103/RevModPhys.93.045003} {\bibfield
  {journal} {\bibinfo  {journal} {Rev. Mod. Phys.}\ }\textbf {\bibinfo {volume}
  {93}},\ \bibinfo {pages} {045003} (\bibinfo {year} {2021})}\BibitemShut
  {NoStop}%
\bibitem [{\citenamefont {Poilblanc}\ \emph {et~al.}(2016)\citenamefont
  {Poilblanc}, \citenamefont {Schuch},\ and\ \citenamefont
  {Affleck}}]{Poilblanc2016}%
  \BibitemOpen
  \bibfield  {author} {\bibinfo {author} {\bibfnamefont {D.}~\bibnamefont
  {Poilblanc}}, \bibinfo {author} {\bibfnamefont {N.}~\bibnamefont {Schuch}}, \
  and\ \bibinfo {author} {\bibfnamefont {I.}~\bibnamefont {Affleck}},\ }\href
  {\doibase 10.1103/PhysRevB.93.174414} {\bibfield  {journal} {\bibinfo
  {journal} {Phys. Rev. B}\ }\textbf {\bibinfo {volume} {93}},\ \bibinfo
  {pages} {174414} (\bibinfo {year} {2016})}\BibitemShut {NoStop}%
\bibitem [{\citenamefont {Dubail}\ and\ \citenamefont
  {Read}(2015)}]{Dubail2015}%
  \BibitemOpen
  \bibfield  {author} {\bibinfo {author} {\bibfnamefont {J.}~\bibnamefont
  {Dubail}}\ and\ \bibinfo {author} {\bibfnamefont {N.}~\bibnamefont {Read}},\
  }\href {\doibase 10.1103/PhysRevB.92.205307} {\bibfield  {journal} {\bibinfo
  {journal} {Phys. Rev. B}\ }\textbf {\bibinfo {volume} {92}},\ \bibinfo
  {pages} {205307} (\bibinfo {year} {2015})}\BibitemShut {NoStop}%
\bibitem [{\citenamefont {Wahl}\ \emph {et~al.}(2013)\citenamefont {Wahl},
  \citenamefont {Tu}, \citenamefont {Schuch},\ and\ \citenamefont
  {Cirac}}]{Wahl2013}%
  \BibitemOpen
  \bibfield  {author} {\bibinfo {author} {\bibfnamefont {T.~B.}\ \bibnamefont
  {Wahl}}, \bibinfo {author} {\bibfnamefont {H.-H.}\ \bibnamefont {Tu}},
  \bibinfo {author} {\bibfnamefont {N.}~\bibnamefont {Schuch}}, \ and\ \bibinfo
  {author} {\bibfnamefont {J.~I.}\ \bibnamefont {Cirac}},\ }\href {\doibase
  10.1103/PhysRevLett.111.236805} {\bibfield  {journal} {\bibinfo  {journal}
  {Phys. Rev. Lett.}\ }\textbf {\bibinfo {volume} {111}},\ \bibinfo {pages}
  {236805} (\bibinfo {year} {2013})}\BibitemShut {NoStop}%
\bibitem [{Note2()}]{Note2}%
  \BibitemOpen
  \bibinfo {note} {Ref.~\protect \rev@citealpnum {Haegeman2017} sees a gapless
  chiral entanglement spectrum (``dispersion law'') of the PEPS of
  Ref.~\protect \rev@citealpnum {Hackenbroich2018} in the thermodynamic limit,
  as depicted in their Figure 27.\cite {SchuchPriv}}\BibitemShut {NoStop}%
\bibitem [{\citenamefont {Poilblanc}\ \emph {et~al.}(2015)\citenamefont
  {Poilblanc}, \citenamefont {Cirac},\ and\ \citenamefont
  {Schuch}}]{Poilblanc2015}%
  \BibitemOpen
  \bibfield  {author} {\bibinfo {author} {\bibfnamefont {D.}~\bibnamefont
  {Poilblanc}}, \bibinfo {author} {\bibfnamefont {J.~I.}\ \bibnamefont
  {Cirac}}, \ and\ \bibinfo {author} {\bibfnamefont {N.}~\bibnamefont
  {Schuch}},\ }\href {\doibase 10.1103/PhysRevB.91.224431} {\bibfield
  {journal} {\bibinfo  {journal} {Phys. Rev. B}\ }\textbf {\bibinfo {volume}
  {91}},\ \bibinfo {pages} {224431} (\bibinfo {year} {2015})}\BibitemShut
  {NoStop}%
\bibitem [{\citenamefont {Poilblanc}(2017)}]{Poilblanc2017}%
  \BibitemOpen
  \bibfield  {author} {\bibinfo {author} {\bibfnamefont {D.}~\bibnamefont
  {Poilblanc}},\ }\href {\doibase 10.1103/PhysRevB.96.121118} {\bibfield
  {journal} {\bibinfo  {journal} {Phys. Rev. B}\ }\textbf {\bibinfo {volume}
  {96}},\ \bibinfo {pages} {121118(R)} (\bibinfo {year} {2017})}\BibitemShut
  {NoStop}%
\bibitem [{\citenamefont {Hastings}(2004)}]{Hastings2004}%
  \BibitemOpen
  \bibfield  {author} {\bibinfo {author} {\bibfnamefont {M.~B.}\ \bibnamefont
  {Hastings}},\ }\href {\doibase 10.1103/PhysRevB.69.104431} {\bibfield
  {journal} {\bibinfo  {journal} {Phys. Rev. B}\ }\textbf {\bibinfo {volume}
  {69}},\ \bibinfo {pages} {104431} (\bibinfo {year} {2004})}\BibitemShut
  {NoStop}%
\bibitem [{\citenamefont {Hastings}(2010)}]{Hastings2010}%
  \BibitemOpen
  \bibfield  {author} {\bibinfo {author} {\bibfnamefont {M.~B.}\ \bibnamefont
  {Hastings}},\ }\href@noop {} {\enquote {\bibinfo {title} {Locality in quantum
  systems},}\ } (\bibinfo {year} {2010}),\ \Eprint
  {http://arxiv.org/abs/1008.5137} {arXiv:1008.5137 [math-ph]} \BibitemShut
  {NoStop}%
\bibitem [{\citenamefont {Knizhnik}\ and\ \citenamefont
  {Zamolodchikov}(1984)}]{Knizhnik1984}%
  \BibitemOpen
  \bibfield  {author} {\bibinfo {author} {\bibfnamefont {V.}~\bibnamefont
  {Knizhnik}}\ and\ \bibinfo {author} {\bibfnamefont {A.}~\bibnamefont
  {Zamolodchikov}},\ }\href {\doibase
  https://doi.org/10.1016/0550-3213(84)90374-2} {\bibfield  {journal} {\bibinfo
   {journal} {Nucl. Phys. B}\ }\textbf {\bibinfo {volume} {247}},\ \bibinfo
  {pages} {83 } (\bibinfo {year} {1984})}\BibitemShut {NoStop}%
\bibitem [{\citenamefont {Calabrese}\ and\ \citenamefont
  {Cardy}(2006)}]{Calabrese2006}%
  \BibitemOpen
  \bibfield  {author} {\bibinfo {author} {\bibfnamefont {P.}~\bibnamefont
  {Calabrese}}\ and\ \bibinfo {author} {\bibfnamefont {J.}~\bibnamefont
  {Cardy}},\ }\href {\doibase 10.1103/PhysRevLett.96.136801} {\bibfield
  {journal} {\bibinfo  {journal} {Phys. Rev. Lett.}\ }\textbf {\bibinfo
  {volume} {96}},\ \bibinfo {pages} {136801} (\bibinfo {year}
  {2006})}\BibitemShut {NoStop}%
\bibitem [{\citenamefont {Calabrese}\ and\ \citenamefont
  {Cardy}(2007)}]{Calabrese2007}%
  \BibitemOpen
  \bibfield  {author} {\bibinfo {author} {\bibfnamefont {P.}~\bibnamefont
  {Calabrese}}\ and\ \bibinfo {author} {\bibfnamefont {J.}~\bibnamefont
  {Cardy}},\ }\href {http://stacks.iop.org/1742-5468/2007/i=06/a=P06008}
  {\bibfield  {journal} {\bibinfo  {journal} {J. Stat. Mech.: Theory Exp.}\
  }\textbf {\bibinfo {volume} {2007}},\ \bibinfo {pages} {P06008} (\bibinfo
  {year} {2007})}\BibitemShut {NoStop}%
\bibitem [{Note3()}]{Note3}%
  \BibitemOpen
  \bibinfo {note} {That correlation length goes to zero as the fixed point is
  approached when $\tau _0\to 0$, in line with the expectation of the absence
  of a scale at a fixed point.}\BibitemShut {Stop}%
\bibitem [{Note4()}]{Note4}%
  \BibitemOpen
  \bibinfo {note} {These states are maximally entangled states between the
  left- and the right-moving descendants of a primary state in the bulk CFT
  Hilbert space\cite {Ishibashi1989}.}\BibitemShut {Stop}%
\bibitem [{Note5()}]{Note5}%
  \BibitemOpen
  \bibinfo {note} {I.e., the additional boundary operators should be such that
  they are boundary limits of both purely chiral (holomorphic $\Phi _i(z)$ or
  anti-holomorphic $\protect \bar {\Phi }_i(\protect \bar {z})$) bulk
  operators. (This is the case, e.g., for the energy momentum tensor: $\Phi
  _{i=1}=T(x)$ and $\protect \bar {\Phi }_{i=1}=\protect \bar {T}(x)$ in the
  notation of Table \ref {table:modereps} of Appendix \ref {app:details}. These
  are equal, i.e.,~$T(x)=\protect \bar {T}(x)$, at a conformally invariant
  boundary such as the one under consideration.)}\BibitemShut {NoStop}%
\bibitem [{Note6()}]{Note6}%
  \BibitemOpen
  \bibinfo {note} {The actual ground state on the surface of the cylinder of
  any such two-dimensional chiral topological system can be fully represented
  in this manner, upon inclusion of a sufficient number of operators $\Phi
  _i(x)$ and $\protect \overline {\Phi }_i(x)$.}\BibitemShut {Stop}%
\bibitem [{Note7()}]{Note7}%
  \BibitemOpen
  \bibinfo {note} {As an aside, we remark that in the usual quantum quench
  problem one wishes to describe the expectation value of a product of a finite
  number of local bulk operators of the CFT, each of which consists of a left-
  and a right-moving part, after waiting a sufficiently long (real) time. Such
  information is contained in the reduced density matrix for a bipartition of
  position space into a compact spatial interval $A$ containing the locations
  of the finite number of operators, and its complement $\protect \bar {A}$.
  Under (real) time-evolution, such a density matrix relaxes\cite
  {Cardy2016,WenRyuLudwigJStatMech2018} into the thermal density matrix or GGE
  density matrix on the interval $A$. The bipartition in the situation we
  consider here, in contrast, is that between left- and right-moving degrees of
  freedom (in contrast to that between the two spatial regions $A$ and
  $\protect \bar {A}$ above), and the reduced density matrix in our
  Eqs.~\protect \textup {\hbox {\mathsurround \z@ \protect \normalfont
  (\ignorespaces \ref {LabelEq-rho-L-a}\unskip \@@italiccorr )}} and \protect
  \textup {\hbox {\mathsurround \z@ \protect \normalfont (\ignorespaces \ref
  {eq:realrhol}\unskip \@@italiccorr )}} is obtained by performing a trace over
  the right-moving degrees of freedom. This density matrix reflects the
  left-right entanglement of the boundary state $|{\protect \bf G}_a\rangle
  $.}\BibitemShut {Stop}%
\bibitem [{\citenamefont {Sasaki}\ and\ \citenamefont
  {Yamanaka}(1988)}]{Sasaki1988}%
  \BibitemOpen
  \bibfield  {author} {\bibinfo {author} {\bibfnamefont {R.}~\bibnamefont
  {Sasaki}}\ and\ \bibinfo {author} {\bibfnamefont {I.}~\bibnamefont
  {Yamanaka}},\ }\href {\doibase 10.2969/aspm/01610271} {\bibfield  {journal}
  {\bibinfo  {journal} {Adv. Stud. Pure Math.}\ }\textbf {\bibinfo {volume}
  {16}},\ \bibinfo {pages} {271} (\bibinfo {year} {1988})}\BibitemShut
  {NoStop}%
\bibitem [{\citenamefont {Bazhanov}\ \emph {et~al.}(1996)\citenamefont
  {Bazhanov}, \citenamefont {Lukyanov},\ and\ \citenamefont
  {Zamolodchikov}}]{Bazhanov1996}%
  \BibitemOpen
  \bibfield  {author} {\bibinfo {author} {\bibfnamefont {V.~V.}\ \bibnamefont
  {Bazhanov}}, \bibinfo {author} {\bibfnamefont {S.~L.}\ \bibnamefont
  {Lukyanov}}, \ and\ \bibinfo {author} {\bibfnamefont {A.~B.}\ \bibnamefont
  {Zamolodchikov}},\ }\href {\doibase 10.1007/BF02101898} {\bibfield  {journal}
  {\bibinfo  {journal} {Commun. Math. Phys.}\ }\textbf {\bibinfo {volume}
  {177}},\ \bibinfo {pages} {381} (\bibinfo {year} {1996})}\BibitemShut
  {NoStop}%
\bibitem [{\citenamefont {Kulish}\ and\ \citenamefont
  {Zeitlin}(2005)}]{Kulish2005}%
  \BibitemOpen
  \bibfield  {author} {\bibinfo {author} {\bibfnamefont {P.~P.}\ \bibnamefont
  {Kulish}}\ and\ \bibinfo {author} {\bibfnamefont {A.~M.}\ \bibnamefont
  {Zeitlin}},\ }\href {\doibase
  https://doi.org/10.1016/j.nuclphysb.2004.12.031} {\bibfield  {journal}
  {\bibinfo  {journal} {Nucl. Phys. B}\ }\textbf {\bibinfo {volume} {709}},\
  \bibinfo {pages} {578 } (\bibinfo {year} {2005})}\BibitemShut {NoStop}%
\bibitem [{\citenamefont {Mussardo}\ \emph {et~al.}(1988)\citenamefont
  {Mussardo}, \citenamefont {Sotkov},\ and\ \citenamefont
  {Stanishkov}}]{Mussardo1988}%
  \BibitemOpen
  \bibfield  {author} {\bibinfo {author} {\bibfnamefont {G.}~\bibnamefont
  {Mussardo}}, \bibinfo {author} {\bibfnamefont {G.}~\bibnamefont {Sotkov}}, \
  and\ \bibinfo {author} {\bibfnamefont {M.}~\bibnamefont {Stanishkov}},\
  }\href {\doibase https://doi.org/10.1016/0550-3213(88)90686-4} {\bibfield
  {journal} {\bibinfo  {journal} {Nucl. Phys. B}\ }\textbf {\bibinfo {volume}
  {305}},\ \bibinfo {pages} {69 } (\bibinfo {year} {1988})}\BibitemShut
  {NoStop}%
\bibitem [{\citenamefont {Friedan}\ \emph {et~al.}(1985)\citenamefont
  {Friedan}, \citenamefont {Qiu},\ and\ \citenamefont
  {Shenker}}]{FriedanQiuShenkerPhysLettB-SUSY-1985}%
  \BibitemOpen
  \bibfield  {author} {\bibinfo {author} {\bibfnamefont {D.}~\bibnamefont
  {Friedan}}, \bibinfo {author} {\bibfnamefont {Z.}~\bibnamefont {Qiu}}, \ and\
  \bibinfo {author} {\bibfnamefont {S.}~\bibnamefont {Shenker}},\ }\href
  {\doibase 10.1016/0370-2693(85)90819-6} {\bibfield  {journal} {\bibinfo
  {journal} {Phys. Lett. B}\ }\textbf {\bibinfo {volume} {151}},\ \bibinfo
  {pages} {37} (\bibinfo {year} {1985})}\BibitemShut {NoStop}%
\bibitem [{\citenamefont {Dixon}\ \emph {et~al.}(1988)\citenamefont {Dixon},
  \citenamefont {Ginsparg},\ and\ \citenamefont
  {Harvey}}]{DixonGinspargHarvey-SUSY-NPB1988}%
  \BibitemOpen
  \bibfield  {author} {\bibinfo {author} {\bibfnamefont {L.}~\bibnamefont
  {Dixon}}, \bibinfo {author} {\bibfnamefont {P.}~\bibnamefont {Ginsparg}}, \
  and\ \bibinfo {author} {\bibfnamefont {J.}~\bibnamefont {Harvey}},\ }\href
  {\doibase 10.1016/0550-3213(88)90011-9} {\bibfield  {journal} {\bibinfo
  {journal} {Nucl. Phys. B}\ }\textbf {\bibinfo {volume} {306}},\ \bibinfo
  {pages} {470} (\bibinfo {year} {1988})}\BibitemShut {NoStop}%
\bibitem [{Note8()}]{Note8}%
  \BibitemOpen
  \bibinfo {note} {The constant below satisfies ``${\protect \rm const.}=
  \protect \qopname \relax o{ln}Z_a $'', where $1/Z_a$ is the factor by which
  the right hand side of \protect \textup {\hbox {\mathsurround \z@ \protect
  \normalfont (\ignorespaces \ref {eq:realrhol}\unskip \@@italiccorr )}} has to
  be multiplied to ensure the proper normalization of the density matrix on the
  left hand side. In general, this constant has a complicated dependence on the
  coefficients $\beta _i$. However, if in the limit of large system size $\ell
  $ the entanglement Hamiltonian is dominated by the Hamiltonian ${\protect
  \tilde H}_L$ of the CFT [compare \protect \textup {\hbox {\mathsurround \z@
  \protect \normalfont (\ignorespaces \ref {eq:lincomb-SizeDependence}\unskip
  \@@italiccorr )}}], then this constant is known\cite
  {KitaevPreskillPRL2006,Qi2012,AffleckLudwig1991} to take the form
  ``${\protect \rm const.} =\alpha \ell - \gamma _a$'', where $\alpha $ is
  nonuniversal and $\gamma _a$ is related to the quantum dimension of the
  topological excitation $a$.}\BibitemShut {Stop}%
\bibitem [{Note9()}]{Note9}%
  \BibitemOpen
  \bibinfo {note} {See also Appendix \ref
  {LabelSubsectionRTSymatryConservedQuantities}}\BibitemShut {NoStop}%
\bibitem [{Note10()}]{Note10}%
  \BibitemOpen
  \bibinfo {note} {The modes $G_n$ of the superconformal current, together with
  the Virasoro modes $L_n$ of the energy-momentum tensor, obey the $N = 1$
  superconformal algebra\cite {Cohn1988,FriedanQiuShenkerPhysLettB-SUSY-1985}
  at central charge $c=3/2$. In particular, $\{G_m, G_n\} =2 L_{m+n} +\protect
  \frac {c}{3} \left (m^2-\protect \frac {1}{4}\right ) \delta _{m, -n}$, so in
  particular, $G_0^2 = \protect \frac {\{G_0,G_0\}}{2} = L_0 - \protect \frac
  {c}{24}$. Therefore the eigenvalues of $G_0$ must in every case be one of
  $\pm \protect \sqrt {L_0 - \protect \frac {c}{24}}$.}\BibitemShut {Stop}%
\bibitem [{\citenamefont {Haegeman}\ and\ \citenamefont
  {Verstraete}(2017)}]{Haegeman2017}%
  \BibitemOpen
  \bibfield  {author} {\bibinfo {author} {\bibfnamefont {J.}~\bibnamefont
  {Haegeman}}\ and\ \bibinfo {author} {\bibfnamefont {F.}~\bibnamefont
  {Verstraete}},\ }\href {\doibase 10.1146/annurev-conmatphys-031016-025507}
  {\bibfield  {journal} {\bibinfo  {journal} {Annu. Rev. Conden. Ma. P.}\
  }\textbf {\bibinfo {volume} {8}},\ \bibinfo {pages} {355} (\bibinfo {year}
  {2017})}\BibitemShut {NoStop}%
\bibitem [{Note11()}]{Note11}%
  \BibitemOpen
  \bibinfo {note} {As will be seen later, the $\protect \rm {SU}(2)_2$
  entanglement spectrum data available to us from Ref.~\protect \rev@citealpnum
  {Chen2018} only includes clearly observable countings for the $|j=0\rangle $
  and $|j=1/2\rangle $ primary sectors, so those will be the only sectors of
  $\protect \rm {SU}(2)_2$ we fit in Sec.~\ref {sec:su22results}. This does not
  in any way preclude our use of the integrals of \protect \textit {operators}
  that correspond to descendant states in the $|j=1\rangle $ primary
  sector.}\BibitemShut {Stop}%
\bibitem [{\citenamefont {Zamolodchikov}\ and\ \citenamefont
  {Fateev}(1986)}]{ZamolodchikovFateevSovJNuclPhys1986}%
  \BibitemOpen
  \bibfield  {author} {\bibinfo {author} {\bibfnamefont {A.~B.}\ \bibnamefont
  {Zamolodchikov}}\ and\ \bibinfo {author} {\bibfnamefont {V.~A.}\ \bibnamefont
  {Fateev}},\ }\href@noop {} {\bibfield  {journal} {\bibinfo  {journal} {Sov.
  J. Nucl. Phys.}\ }\textbf {\bibinfo {volume} {43}},\ \bibinfo {pages} {657}
  (\bibinfo {year} {1986})}\BibitemShut {NoStop}%
\bibitem [{\citenamefont {Goddard}\ and\ \citenamefont
  {Olive}(1986)}]{Goddard1986}%
  \BibitemOpen
  \bibfield  {author} {\bibinfo {author} {\bibfnamefont {P.}~\bibnamefont
  {Goddard}}\ and\ \bibinfo {author} {\bibfnamefont {D.}~\bibnamefont
  {Olive}},\ }\href {\doibase 10.1142/S0217751X86000149} {\bibfield  {journal}
  {\bibinfo  {journal} {Int. J. Mod. Phys. A}\ }\textbf {\bibinfo {volume}
  {01}},\ \bibinfo {pages} {303} (\bibinfo {year} {1986})}\BibitemShut
  {NoStop}%
\bibitem [{Note12()}]{Note12}%
  \BibitemOpen
  \bibinfo {note} {In contrast to the $\psi ^a(x)$ and their antichiral
  counterparts $\protect \bar {\psi }^a(x)$, which anticommute with each other,
  $\phi ^a(x)$ and $\protect \bar {\phi }^a(x)$ actually commute with each
  other. Within the chiral (antichiral) theory, though, $\phi ^a(x)$ and $\psi
  ^a(x)$ ($\protect \bar {\phi }^a(x)$ and $\protect \bar {\psi }^a(x)$) will
  behave the same way in correlation functions (see, e.g., Ref.~\protect
  \rev@citealpnum {Maldacena1997}.)}\BibitemShut {NoStop}%
\bibitem [{\citenamefont {Cohn}\ and\ \citenamefont
  {Friedan}(1988)}]{Cohn1988}%
  \BibitemOpen
  \bibfield  {author} {\bibinfo {author} {\bibfnamefont {J.}~\bibnamefont
  {Cohn}}\ and\ \bibinfo {author} {\bibfnamefont {D.}~\bibnamefont {Friedan}},\
  }\href {\doibase https://doi.org/10.1016/0550-3213(88)90398-7} {\bibfield
  {journal} {\bibinfo  {journal} {Nucl. Phys. B}\ }\textbf {\bibinfo {volume}
  {296}},\ \bibinfo {pages} {779} (\bibinfo {year} {1988})}\BibitemShut
  {NoStop}%
\bibitem [{\citenamefont {Cincio}\ and\ \citenamefont
  {Vidal}(2013)}]{Cincio2013}%
  \BibitemOpen
  \bibfield  {author} {\bibinfo {author} {\bibfnamefont {L.}~\bibnamefont
  {Cincio}}\ and\ \bibinfo {author} {\bibfnamefont {G.}~\bibnamefont {Vidal}},\
  }\href {\doibase 10.1103/PhysRevLett.110.067208} {\bibfield  {journal}
  {\bibinfo  {journal} {Phys. Rev. Lett.}\ }\textbf {\bibinfo {volume} {110}},\
  \bibinfo {pages} {067208} (\bibinfo {year} {2013})}\BibitemShut {NoStop}%
\bibitem [{Note13()}]{Note13}%
  \BibitemOpen
  \bibinfo {note} {Attempts to fit both sectors simultaneously resulted in some
  of the very highest-energy multiplets, in the highest descendant level
  considered, not agreeing with our expectations based on the set of parameters
  that successfully fit the low-energy part of the spectrum.}\BibitemShut
  {Stop}%
\bibitem [{Note14()}]{Note14}%
  \BibitemOpen
  \bibinfo {note} {The states in this sector can be thought of possessing odd
  fermion parity in $\protect \rm {SU}(2)_2$, as opposed to the states in the
  $|j=0\rangle $ sector (see, e.g., the discussion in Sec.~\ref
  {sec:locconquant2}). Perhaps this might be related to the fact that they do
  not seem to be visible in the data.}\BibitemShut {Stop}%
\bibitem [{Note15()}]{Note15}%
  \BibitemOpen
  \bibinfo {note} {Observe that, for example, the first method is not able to
  lift the degeneracy of even the lowest excited momentum state at descendant
  level $K=1$, while the second method achieves that goal with
  ease.}\BibitemShut {Stop}%
\bibitem [{\citenamefont {Read}\ and\ \citenamefont {Rezayi}(1999)}]{Read1999}%
  \BibitemOpen
  \bibfield  {author} {\bibinfo {author} {\bibfnamefont {N.}~\bibnamefont
  {Read}}\ and\ \bibinfo {author} {\bibfnamefont {E.}~\bibnamefont {Rezayi}},\
  }\href {\doibase 10.1103/PhysRevB.59.8084} {\bibfield  {journal} {\bibinfo
  {journal} {Phys. Rev. B}\ }\textbf {\bibinfo {volume} {59}},\ \bibinfo
  {pages} {8084} (\bibinfo {year} {1999})}\BibitemShut {NoStop}%
\bibitem [{\citenamefont {Wang}(2010)}]{Wang2010}%
  \BibitemOpen
  \bibfield  {author} {\bibinfo {author} {\bibfnamefont {Z.}~\bibnamefont
  {Wang}},\ }\href@noop {} {\emph {\bibinfo {title} {Topological Quantum
  Computation}}},\ \bibinfo {series} {CBMS Regional Conference Series in
  Mathematics}\ No.\ \bibinfo {number} {112}\ (\bibinfo  {publisher} {American
  Mathematical Society, Providence, RI},\ \bibinfo {year} {2010})\BibitemShut
  {NoStop}%
\bibitem [{\citenamefont {Ahn}\ \emph {et~al.}(1990)\citenamefont {Ahn},
  \citenamefont {Bernard},\ and\ \citenamefont
  {Leclair}}]{AhnBernardLeClairNPB1990}%
  \BibitemOpen
  \bibfield  {author} {\bibinfo {author} {\bibfnamefont {C.}~\bibnamefont
  {Ahn}}, \bibinfo {author} {\bibfnamefont {D.}~\bibnamefont {Bernard}}, \ and\
  \bibinfo {author} {\bibfnamefont {A.}~\bibnamefont {Leclair}},\ }\href
  {\doibase https://doi.org/10.1016/0550-3213(90)90287-N} {\bibfield  {journal}
  {\bibinfo  {journal} {Nucl. Phys. B}\ }\textbf {\bibinfo {volume} {346}},\
  \bibinfo {pages} {409} (\bibinfo {year} {1990})}\BibitemShut {NoStop}%
\bibitem [{\citenamefont {Kass}\ \emph {et~al.}(1990)\citenamefont {Kass},
  \citenamefont {Moody}, \citenamefont {Patera},\ and\ \citenamefont
  {Slansky}}]{Kass1990}%
  \BibitemOpen
  \bibfield  {author} {\bibinfo {author} {\bibfnamefont {S.}~\bibnamefont
  {Kass}}, \bibinfo {author} {\bibfnamefont {R.}~\bibnamefont {Moody}},
  \bibinfo {author} {\bibfnamefont {J.}~\bibnamefont {Patera}}, \ and\ \bibinfo
  {author} {\bibfnamefont {R.}~\bibnamefont {Slansky}},\ }\href@noop {} {\emph
  {\bibinfo {title} {Affine Lie Algebras, Weight Multiplicities, and Branching
  Rules}}},\ Vol.~\bibinfo {volume} {2}\ (\bibinfo  {publisher} {University of
  California Press, Berkeley},\ \bibinfo {year} {1990})\BibitemShut {NoStop}%
\bibitem [{Note16()}]{Note16}%
  \BibitemOpen
  \bibinfo {note} {Strictly, we ignore the constant terms in the expressions of
  Table \ref {table:modereps}, because each will shift every state in the
  spectrum by the exact same amount, and therefore none of these terms will
  affect the splittings.}\BibitemShut {Stop}%
\bibitem [{\citenamefont {Di~Francesco}\ \emph {et~al.}(1997)\citenamefont
  {Di~Francesco}, \citenamefont {Mathieu},\ and\ \citenamefont
  {S{\'e}n{\'e}chal}}]{DiFrancesco1997}%
  \BibitemOpen
  \bibfield  {author} {\bibinfo {author} {\bibfnamefont {P.}~\bibnamefont
  {Di~Francesco}}, \bibinfo {author} {\bibfnamefont {P.}~\bibnamefont
  {Mathieu}}, \ and\ \bibinfo {author} {\bibfnamefont {D.}~\bibnamefont
  {S{\'e}n{\'e}chal}},\ }\href@noop {} {\emph {\bibinfo {title} {Conformal
  Field Theory}}},\ Graduate Texts in Contemporary Physics\ (\bibinfo
  {publisher} {Springer-Verlag New York},\ \bibinfo {year} {1997})\BibitemShut
  {NoStop}%
\bibitem [{Note17()}]{Note17}%
  \BibitemOpen
  \bibinfo {note} {We can see this by considering the effect of acting with any
  Virasoro mode $L_n$ for $n > 0$ on a state $|\phi \rangle $ of the multiplet.
  Because the mode $L_n$ is a singlet under $\protect \rm {SU}(2)$, $L_n|\phi
  \rangle $ must be a state with the same $j^z$ quantum number in a multiplet
  of the same dimension, but at a lower level. As there is no such state, it
  must be the case that $L_n|\phi \rangle = 0$. In the $\protect \rm {SU}(2)_1$
  theory, the 1-1-2-3-5 counting of states in the subtowers of the
  highest-weight states of these Virasoro primary multiplets and their
  descendants also guarantees the uniqueness of such multiplets within the
  descendant levels in which they appear (see, e.g., Fig.~\ref
  {fig:baueretalplot}.)}\BibitemShut {NoStop}%
\bibitem [{\citenamefont {Mambrini}\ \emph {et~al.}(2016)\citenamefont
  {Mambrini}, \citenamefont {Or\'us},\ and\ \citenamefont
  {Poilblanc}}]{Mambrini2016}%
  \BibitemOpen
  \bibfield  {author} {\bibinfo {author} {\bibfnamefont {M.}~\bibnamefont
  {Mambrini}}, \bibinfo {author} {\bibfnamefont {R.}~\bibnamefont {Or\'us}}, \
  and\ \bibinfo {author} {\bibfnamefont {D.}~\bibnamefont {Poilblanc}},\ }\href
  {\doibase 10.1103/PhysRevB.94.205124} {\bibfield  {journal} {\bibinfo
  {journal} {Phys. Rev. B}\ }\textbf {\bibinfo {volume} {94}},\ \bibinfo
  {pages} {205124} (\bibinfo {year} {2016})}\BibitemShut {NoStop}%
\bibitem [{Note18()}]{Note18}%
  \BibitemOpen
  \bibinfo {note} {Expanding Eq.~\protect \textup {\hbox {\mathsurround \z@
  \protect \normalfont (\ignorespaces \ref {eq:RTLop}\unskip \@@italiccorr )}}
  in terms of modes, we obtain $\DOTSB \sum@ \slimits@ _{n=-\infty }^\infty
  (\protect \mathcal {R T})L_{n}(\protect \mathcal {R T})^{-1}e^{2\pi inx/\ell
  } = \DOTSB \sum@ \slimits@ _{n=-\infty }^\infty L_{n}e^{2\pi inx/\ell }$, and
  since $(\protect \mathcal {R T})e^{2\pi inx/\ell }(\protect \mathcal {R
  T})^{-1} = e^{2\pi inx/\ell }$, this means that $(\protect \mathcal {R
  T})L_{n}(\protect \mathcal {R T})^{-1} =L_{n}$.}\BibitemShut {Stop}%
\bibitem [{Note19()}]{Note19}%
  \BibitemOpen
  \bibinfo {note} {The normalization can be found by comparing the commutation
  relations of the Pauli matrices $\sigma ^a$, $[\sigma ^a,\sigma ^b] = 2i
  \epsilon _{abc}\sigma ^c$, with Eq.~\protect \textup {\hbox {\mathsurround
  \z@ \protect \normalfont (\ignorespaces \ref {eq:jcommutation}\unskip
  \@@italiccorr )}}.}\BibitemShut {Stop}%
\bibitem [{Note20()}]{Note20}%
  \BibitemOpen
  \bibinfo {note} {The $\phi ^a_n$ modes must satisfy the anti-commutation
  relations $\{\phi ^a_n,\phi ^a_m\} = \delta _{n+m,0}.$ The normalization of
  the action of the $\phi ^a_0$ modes on the $|j=1/2\rangle $ primary state may
  be found by comparison with the Pauli anti-commutation relations $\{\sigma
  ^a,\sigma ^b\} = 2\delta _{ab}I$.}\BibitemShut {Stop}%
\bibitem [{Note21()}]{Note21}%
  \BibitemOpen
  \bibinfo {note} {In terms of the operators $\phi ^a(x)$, we have $\protect
  \mathcal {R} \phi ^a(x) \protect \mathcal {R}^{-1} = \protect \bar {\phi
  }^a(\ell - x)$ and $\protect \mathcal {R} \protect \bar {\phi }^a(x) \protect
  \mathcal {R}^{-1} = \phi ^a(\ell - x)$, analogous to Eq.~\protect \textup
  {\hbox {\mathsurround \z@ \protect \normalfont (\ignorespaces \ref
  {eq:RLop}\unskip \@@italiccorr )}}. In the Ramond sector, where $\phi ^a(x) =
  \phi ^a(\ell +x)$ and $\protect \bar {\phi }^a(x) = \protect \bar {\phi
  }^a(\ell +x)$ are periodic, this works out to $\protect \mathcal {R} \phi
  ^a(x) \protect \mathcal {R}^{-1} = \protect \bar {\phi }^a(- x)$ and
  $\protect \mathcal {R} \protect \bar {\phi }^a(x) \protect \mathcal {R}^{-1}
  = \phi ^a(-x)$, as in Eq.~\protect \textup {\hbox {\mathsurround \z@ \protect
  \normalfont (\ignorespaces \ref {eq:RLop}\unskip \@@italiccorr )}}. In the
  Neveu-Schwarz case, on the other hand, $\phi ^a(x) = -\phi ^a(\ell +x)$ and
  $\protect \bar {\phi }^a(x) = -\protect \bar {\phi }^a(\ell +x)$ are
  anti-periodic, so $\protect \mathcal {R} \phi ^a(x) \protect \mathcal
  {R}^{-1} = -\protect \bar {\phi }^a(- x)$ and $\protect \mathcal {R} \protect
  \bar {\phi }^a(x) \protect \mathcal {R}^{-1} = -\phi ^a(- x)$, leading to the
  relative minus sign of Eq.~\protect \textup {\hbox {\mathsurround \z@
  \protect \normalfont (\ignorespaces \ref {eq:phiz12transformation}\unskip
  \@@italiccorr )}}.}\BibitemShut {Stop}%
\bibitem [{\citenamefont {Schuch}()}]{SchuchPriv}%
  \BibitemOpen
  \bibfield  {author} {\bibinfo {author} {\bibfnamefont {N.}~\bibnamefont
  {Schuch}},\ }\href@noop {} {}\bibinfo {howpublished} {private
  communication}\BibitemShut {NoStop}%
\bibitem [{\citenamefont {Ishibashi}(1989)}]{Ishibashi1989}%
  \BibitemOpen
  \bibfield  {author} {\bibinfo {author} {\bibfnamefont {N.}~\bibnamefont
  {Ishibashi}},\ }\href {\doibase 10.1142/S0217732389000320} {\bibfield
  {journal} {\bibinfo  {journal} {Mod. Phys. Lett. A}\ }\textbf {\bibinfo
  {volume} {04}},\ \bibinfo {pages} {251} (\bibinfo {year} {1989})}\BibitemShut
  {NoStop}%
\bibitem [{\citenamefont {Wen}\ \emph {et~al.}(2018)\citenamefont {Wen},
  \citenamefont {Ryu},\ and\ \citenamefont
  {Ludwig}}]{WenRyuLudwigJStatMech2018}%
  \BibitemOpen
  \bibfield  {author} {\bibinfo {author} {\bibfnamefont {X.}~\bibnamefont
  {Wen}}, \bibinfo {author} {\bibfnamefont {S.}~\bibnamefont {Ryu}}, \ and\
  \bibinfo {author} {\bibfnamefont {A.~W.~W.}\ \bibnamefont {Ludwig}},\ }\href
  {\doibase 10.1088/1742-5468/aae84e} {\bibfield  {journal} {\bibinfo
  {journal} {J. Stat. Mech.: Theory Exp.}\ }\textbf {\bibinfo {volume}
  {2018}},\ \bibinfo {pages} {113103} (\bibinfo {year} {2018})}\BibitemShut
  {NoStop}%
\bibitem [{\citenamefont {Kitaev}\ and\ \citenamefont
  {Preskill}(2006)}]{KitaevPreskillPRL2006}%
  \BibitemOpen
  \bibfield  {author} {\bibinfo {author} {\bibfnamefont {A.}~\bibnamefont
  {Kitaev}}\ and\ \bibinfo {author} {\bibfnamefont {J.}~\bibnamefont
  {Preskill}},\ }\href {\doibase 10.1103/PhysRevLett.96.110404} {\bibfield
  {journal} {\bibinfo  {journal} {Phys. Rev. Lett.}\ }\textbf {\bibinfo
  {volume} {96}},\ \bibinfo {pages} {110404} (\bibinfo {year}
  {2006})}\BibitemShut {NoStop}%
\bibitem [{\citenamefont {Affleck}\ and\ \citenamefont
  {Ludwig}(1991)}]{AffleckLudwig1991}%
  \BibitemOpen
  \bibfield  {author} {\bibinfo {author} {\bibfnamefont {I.}~\bibnamefont
  {Affleck}}\ and\ \bibinfo {author} {\bibfnamefont {A.~W.~W.}\ \bibnamefont
  {Ludwig}},\ }\href {\doibase 10.1103/PhysRevLett.67.161} {\bibfield
  {journal} {\bibinfo  {journal} {Phys. Rev. Lett.}\ }\textbf {\bibinfo
  {volume} {67}},\ \bibinfo {pages} {161} (\bibinfo {year} {1991})}\BibitemShut
  {NoStop}%
\bibitem [{\citenamefont {Maldacena}\ and\ \citenamefont
  {Ludwig}(1997)}]{Maldacena1997}%
  \BibitemOpen
  \bibfield  {author} {\bibinfo {author} {\bibfnamefont {J.~M.}\ \bibnamefont
  {Maldacena}}\ and\ \bibinfo {author} {\bibfnamefont {A.~W.~W.}\ \bibnamefont
  {Ludwig}},\ }\href {\doibase https://doi.org/10.1016/S0550-3213(97)00596-8}
  {\bibfield  {journal} {\bibinfo  {journal} {Nucl. Phys. B}\ }\textbf
  {\bibinfo {volume} {506}},\ \bibinfo {pages} {565} (\bibinfo {year}
  {1997})}\BibitemShut {NoStop}%
\end{thebibliography}%
\end{document}